\documentclass[aps,prx,article,twocolumn,preprintnumbers,superscriptaddress,floatfix]{revtex4-2}

\date{\today}

\usepackage[normalem]{ulem}
\usepackage{placeins}
\usepackage{layout}
\usepackage{float}
\usepackage{amsmath, amssymb, amsthm}
\usepackage{tensor}
\usepackage{braket}
\usepackage{bm}
\usepackage{IEEEtrantools}
\usepackage[svgnames]{xcolor}
\usepackage{hyperref}
\hypersetup{
    colorlinks=true,
    citecolor=NavyBlue,
    filecolor=black,
    linkcolor=NavyBlue,
    urlcolor=NavyBlue,
    linktocpage=true
}

\usepackage[capitalize]{cleveref}
\Crefname{figure}{Fig.}{Figs.}
\Crefname{equation}{Eq.}{Eqs.}

\usepackage{tikz-cd}
\usepackage{siunitx}
\DeclareSIUnit\year{yr}
\usepackage{booktabs}
\newcolumntype{L}[1]{>{\raggedright\let\newline\\\arraybackslash\hspace{0pt}}p{#1}}
\newcolumntype{C}[1]{>{\centering\let\newline\\\arraybackslash\hspace{0pt}}p{#1}}
\newcolumntype{R}[1]{>{\raggedleft\let\newline\\\arraybackslash\hspace{0pt}}p{#1}}

\makeatletter
\def\l@subsection#1#2{}
\def\l@subsubsection#1#2{}
\makeatother

\usepackage[acronym,symbols,nogroupskip,nomain,nonumberlist,nopostdot,toc]{glossaries}
\glsdisablehyper
\newacronym{vqa}{VQA}{variational quantum algorithm}
\newacronym{vqe}{VQE}{a variational quantum eigensolver}
\newacronym{tdse}{TDSE}{time-dependent Schrödinger equation}
\newacronym{tise}{TISE}{time-independent Schrödinger equation}
\newacronym{qpe}{QPE}{quantum phase estimation}
\newacronym{qmc}{QMC}{Quantum Monte Carlo}
\newacronym{vte}{VTE}{variational time evolution}
\newacronym{pf}{PF}{product formula}
\newacronym{vha}{VHA}{Variational Hamiltonian Ansatz}
\newacronym[firstplural=degrees of freedom (DOFs)]{dof}{DOF}{degree of freedom}
\newacronym{oqs}{OQS}{open quantum systems}
\newacronym{kzm}{KZM}{Kibble-Zurek mechanism}
\newacronym{qkzm}{QKZM}{quantum Kibble-Zurek mechanism}
\newacronym{kz}{KZ}{Kibble-Zurek}
\newacronym{lz}{LZ}{Landau-Zener}
\newacronym{ftlz}{FTLZ}{finite-time Landau-Zener}
\newacronym{tfim}{TFIM}{transverse field Ising model}
\newacronym{cp}{CP}{critical point}
\newacronym{gs}{GS}{ground state}
\newacronym{qa}{QA}{quantum annealing}
\newacronym{sa}{SA}{simulated annealing}
\newacronym{sqa}{SQA}{simulated quantum annealing}
\newacronym{cpt}{CPT}{classical phase transition}
\newacronym{qpt}{QPT}{quantum phase transition}
\newacronym{qv}{QV}{quantum volume}
\newacronym{rb}{RB}{randomized benchmarking}
\newacronym{xeb}{XEB}{cross-entropy benchmarking}
\newacronym{qec}{QEC}{quantum error correction}
\newacronym{em}{EM}{error mitigation}
\newacronym{ems}{EMS}{error mitigation and suppression}
\newacronym{rem}{REM}{readout error mitigation}
\newacronym{trex}{TREX}{twirled readout error mitigation}
\newacronym{m3}{M3}{matrix-free measurement mitigation}
\newacronym{dd}{DD}{dynamical decoupling}
\newacronym{pea}{PEA}{probabilistic error amplification}
\newacronym{ecr}{ECR}{echoed cross resonance}
\newacronym{qaoa}{QAOA}{the quantum approximate optimization algorithm}

\newcommand{\sherbrooke}[0]{\texttt{ibm\_sherbrooke}}
\newcommand{\torino}[0]{\texttt{ibm\_torino}}
\newcommand{\auckland}[0]{\texttt{ibm\_auckland}}
\newcommand{\cz}[0]{$CZ$}
\newcommand{\qiskit}[0]{\textsc{Qiskit}}
\newcommand{\estimator}[0]{\textsc{Qiskit}'s Estimator}

\newcommand{\diff}[1]{\mathrm{d} #1 \, }

\newcommand{\sx}{\sigma{\mathstrut}^x}

\newcommand{\sz}{\sigma{\mathstrut}^z}

\def\bea{\begin{IEEEeqnarray}}
\def\eea{\end{IEEEeqnarray}}
\def\tf{t_\mathrm{f}}
\def\HP{H_\mathrm{P}}
\def\HM{H_\mathrm{M}}

\def\ndef{n_\mathrm{def}}
\def\Ne{N_\mathrm{e}}
\def\rzz{R_{ZZ}}
\def\cz{CZ}
\def\Eres{E_\mathrm{res}}
\def\tone{\mathrm{T}_1}
\def\ttwo{\mathrm{T}_2}

\begin{document}

\title{Benchmarking digital quantum simulations above hundreds of qubits using quantum critical dynamics}

\author{Alexander Miessen}
\email[\href{mailto:lex@zurich.ibm.com}{lex@zurich.ibm.com}]{}
\affiliation{IBM Quantum, IBM Research -- Zurich, 8803 Rüschlikon, Switzerland}
\affiliation{Institute for Computational Science, University of Zurich, Winterthurerstrasse 190, 8057 Zurich, Switzerland}
\author{Daniel J. Egger}
\affiliation{IBM Quantum, IBM Research -- Zurich, 8803 Rüschlikon, Switzerland}
\author{Ivano Tavernelli}
\affiliation{IBM Quantum, IBM Research -- Zurich, 8803 Rüschlikon, Switzerland}
\author{Guglielmo Mazzola}
\email[\href{mailto:guglielmo.mazzola@uzh.ch}{guglielmo.mazzola@uzh.ch}]{}
\affiliation{Institute for Computational Science, University of Zurich, Winterthurerstrasse 190, 8057 Zurich, Switzerland}

\date{\today}

\begin{abstract}
The real-time simulation of large many-body quantum systems is a formidable task, that may only be achievable with a genuine quantum computational platform. 
Currently, quantum hardware with a number of qubits sufficient to make classical emulation challenging is available.
This condition is necessary for the pursuit of a so-called quantum advantage, but it also makes verifying the results very difficult.
In this manuscript, we flip the perspective and utilize known theoretical results about many-body quantum critical dynamics to benchmark quantum hardware and various error mitigation techniques on up to 133 qubits.
In particular, we benchmark against known universal scaling laws in the Hamiltonian simulation of a time-dependent transverse field Ising Hamiltonian.
Incorporating only basic error mitigation and suppression methods, our study shows reliable control up to a two-qubit gate depth of 28, featuring a maximum of 1396 two-qubit gates, before noise becomes prevalent.
These results are transferable to applications such as Hamiltonian simulation, variational algorithms, optimization, or quantum machine learning.
We demonstrate this on the example of digitized quantum annealing for optimization and identify an optimal working point in terms of both circuit depth and time step on a 133-site optimization problem.
\end{abstract}

\maketitle

\section{Introduction}
\label{sec: intro}

Quantum critical dynamics occur when a quantum system reaches a 
\gls{cp}, characterized by a non-analytic change in the system's ground state energy as a function of a Hamiltonian parameter.
The system's correlation length and relaxation time are maximal at the \gls{cp} and diverge in the thermodynamic limit.
Crossing it, the system undergoes a \gls{qpt}.
As a result, system details no longer affect macroscopic quantities, causing the emergence of universal behavior, a key property of critical phenomena~\cite{stanley1971phase,sachdev_2011,sondhi1997continuous}.

\Glspl{qpt} are experimentally realized by changing the control parameters of the system Hamiltonian over time.
However, understanding the real-time dynamics of many-body quantum systems close to the critical point is a formidable task.
Many believe that only quantum simulators~\cite{Feynman1982}, i.e., controllable quantum systems that can emulate others~\cite{Georgescu2014}, can tackle this problem at scale~\cite{Altmann2021, Monroe2021}.
Quantum simulators were first realized with ultracold dilute gas in an optical lattice~\cite{Jaksch1998} and have since been implemented on a variety of platforms~\cite{Altmann2021, Monroe2021,lewenstein2012ultracold,bernien2017probing,scholl2021quantum}.
Most of these platforms are \textit{analog} quantum simulators, which are subject to calibration errors and decoherence.

A parallel approach is to perform the simulation on \textit{digital} quantum computers using suitable algorithms compiled to the native basis gate set of the hardware~\cite{Miessen2023perspective, abrams1999quantum, Kim2023utility, Mi2022timeCrystals, Keenan2023KPZscaling, Miessen2021}.
However, current digital machines are prone to errors, such as calibration errors, cross-talk, and decoherence.
In the future, quantum error correction could potentially enable fault-tolerant simulations of many-body quantum systems on digital machines~\cite{Bravyi2024}.
Ref.~\citenum{daley2022practical} gives a perspective on the relative strengths and weaknesses of analog and digital platforms.

Crucially, quantum critical dynamics have implications beyond condensed matter and statistical physics.
For example, \Glspl{qpt} occur ubiquitously in quantum optimization, where a quantum algorithm helps solve a classical optimization problem.
Such applications are among the most anticipated and economically impactful use cases for quantum computers~\cite{Abbas2023optimizationWorkingGroup}.
\Gls{qa} is an algorithm to find ground states that can be used to solve combinatorial optimization problems~\cite{Albash2018annealingRMP}.
It evolves an easy-to-prepare ground state of one Hamiltonian to the unknown ground state of another problem Hamiltonian, which corresponds to the classical optimization problem to solve.
If the evolution is adiabatically slow, the system follows the instantaneous ground state of the time-dependent Hamiltonian and ends in the solution of the optimization problem.
Particularly for \gls{qa}, \glspl{qpt} create algorithmic bottlenecks~\cite{kadowaki1998quantum,knysh2016zero,Albash2018annealingRMP}, as they imply an energy gap between the ground and the first excited state that closes in the thermodynamic limit.
For finite annealing times $\tf$, the evolution may therefore not be adiabatic due to a diverging relaxation time at the \gls{cp}.
This mechanism produces defects that carry through to the final state~\cite{Zurek2005}.

In a noise-free, closed system, the density of defects is a non-increasing function of $\tf$, and several scaling regimes can be identified as $\tf$ increases~\cite{Zurek2005,Zeng2023universal}.
The \gls{qkzm}~\cite{Zurek2005,kibble1980some,zurek1985cosmological} quantifies the relationship between $\tf$ and the number of defects produced during the annealing run in a regime of sufficiently slow, yet finite, $\tf$.
It predicts a universal scaling, a power-law decay, of the density of defects in the final solution as a function of $\tf$ and the system's critical exponents~\cite{Bando2020}, i.e., a set of numbers characterizing the system's behavior near its \gls{cp}.
It has been the subject of a range of largely analog experimental studies~\cite{Li2023probingLongRangeKZM, keesling2019quantum, king2022coherent, King2023spinGlass,ebadi2021quantum}.

We first present an application-oriented benchmarking method that utilizes the predictability of known universal scaling laws, such as the \gls{kz} scaling.
Those attributes make it ideal for an intuitive and easily scalable metric to assess the quality of large-scale quantum simulations.
This is increasingly important as digital quantum computing devices and algorithms have left the infantile stage of a few tens of qubits with error rates prohibiting more than a handful of two-qubit gates~\cite{Bravyi2022, vazquez2024scaling}.
Today's quantum devices can exceed 100 qubits at error rates that, combined with \gls{ems} techniques~\cite{Cai2023em}, allow a coherent simulation of thousands of two-qubit gates~\cite{Kim2023utility, shtanko2023uncovering}.

Many benchmarking methods employ randomized circuits.
This makes them more comparable, objective, and device-agnostic while accounting for, e.g., qubit connectivity and basis gate set~\cite{Amico2023defining}.
Moreover, randomness helps collect many of the error sources into the same benchmarking process.
Examples of such methods span an entire field of research, including variants of randomized benchmarking~\cite{Magesan2012rb, Proctor2022mirrorRB, Hines2023}, cross-entropy benchmarking~\cite{Boixo2018xeb}, and quantum volume~\cite{Cross2019qv}.
These methods are crucial to assess gate fidelities and characterize different kinds of noise.
However, they do not appropriately represent the capabilities of current quantum hardware to run a specific application.
Indeed, applications for noisy quantum devices have circuits with highly structured repeated layers that are often tailored to the connectivity and the basis gate set of the hardware.
It is therefore of little surprise that such simulations achieve much higher gate counts and qubit numbers than one would predict from generic benchmarks.
Even though there are very promising new benchmarking methods designed to overcome these issues~\cite{Mckay2023layerfidelity}, they still only provide a fidelity without relating to applications.
Pure application benchmarks are the other extreme.
Either by testing against a set of well-understood problems and, possibly, corresponding solutions~\cite{Lubinski2023, Wu2023}, or based on specific applications such as quantum optimization~\cite{Santra2024} or discrete time crystals~\cite{zhang2023characterizing}.

\begin{figure*}[t]
    \centering
    \includegraphics{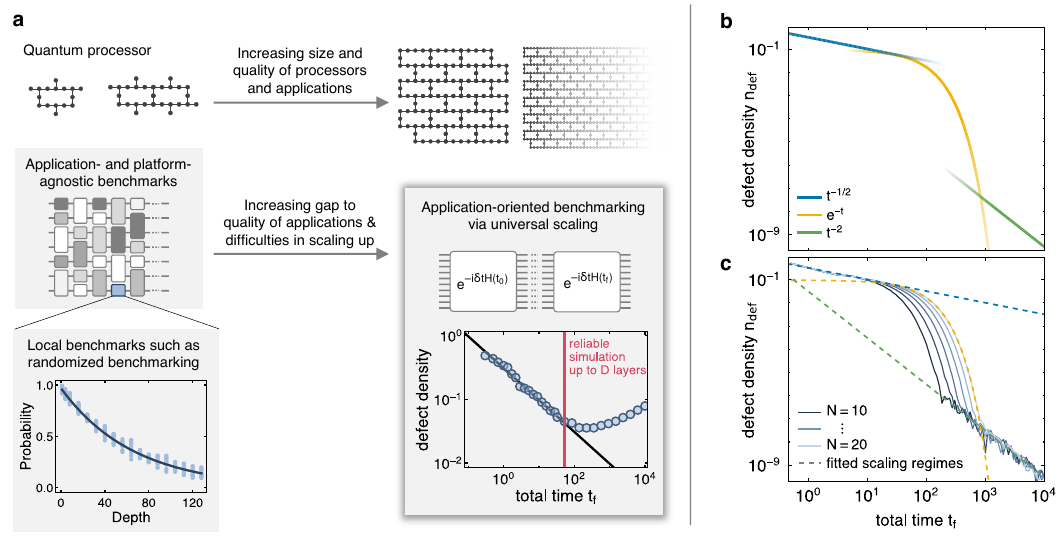}
    \caption{\textbf{Application-oriented benchmarking of quantum simulations.} 
    \textbf{a} Benchmarking experiments give detailed insight into device characteristics such as error rates and decoherence times.
    As devices scale, however, it is desirable to have a benchmarking method that resembles the applications that are executed.
    By simulating critical dynamics and measuring an observable that follows a known universal scaling, we track for how many time steps, which correspond directly to the number of circuit layers, we can reproduce the expected universal scaling.
    Here, we measure the density of defects produced in quantum annealing, predicted by the Kibble-Zurek scaling.
    This provides a direct and concrete metric predicting how many layers of two-qubit gates can be reliably simulated given, e.g., a certain device, qubit subset, and error mitigation technique.
    Note that \textbf{a} contains only schematic plots and shows no actual data.
    \textbf{b} For finite system sizes away from the thermodynamic limit, the density of defects exhibits two scaling regimes beyond the Kibble-Zurek scaling that are theoretically known (see main text).
    These can benchmark deep circuits.
    \textbf{c} The departure from Kibble-Zurek scaling is determined by the system size and all three scaling regimes can be identified in statevector simulations with $N=10, \ldots, 20$ spins (solid lines).
    Shown here are results from Trotterized simulations with $\Delta t = 0.5$.
    }
    \label{fig: schematic and statevector KZM}
\end{figure*}

We go a step further and propose a concrete, application-oriented benchmarking scheme to assess how many two-qubit gate layers of a given application circuit can be reliably simulated given a specific device and, importantly, \gls{ems} method.
A schematic of the principle is shown in \Cref{fig: schematic and statevector KZM}\textbf{a}.
The result is not an abstract fidelity but simply the number of circuit layers that can be reliably simulated, which can be directly transferred to other applications.
Here, we focus on the \gls{kz} scaling, though other universal scalings, e.g. at short times, can also be used~\cite{Zeng2023universal}.

This benchmark is transferrable to applications in time evolution, optimization, quantum machine learning, and variational algorithms.
To showcase this, we apply the benchmark to combinatorial optimization specifically as
the underlying circuits directly implement digitized \gls{qa} or \gls{qaoa}.
QAOA is an annealing-inspired variational ansatz, with variational layers representing time steps in digitized \gls{qa}, used in conjunction with \gls{vqe}~\cite{farhi2014quantum}.
Recent studies indicate that solving hard optimization problems requires many \gls{qaoa} layers~\cite{Zhou2020, farhi2020quantum}.
At the same time, the optimally converged parameter values of \gls{qaoa} in the large-layer limit reproduce annealing schedules, meaning \gls{qaoa} implements digitized \gls{qa} with variationally found annealing schedules.
It is therefore important to better understand digitized \gls{qa}.
A relevant question is whether the algorithmic error stemming from a finite time step is detrimental or possibly even beneficial, similar to what happens in simulated \gls{qa}~\cite{Heim2015spinGlass}.
Here, we identify an optimal working point with respect to the time step and number of circuit layers to minimize the residual energy given the finite hardware resources and how it depends on the system's minimum energy gap.

In summary, we explore two directions -- benchmarking and optimization -- and show how the former can be used to guide the design of the latter.
\Cref{sec: quantum annealing} provides the necessary technical background to \gls{qa}, the \gls{qkzm}, and Trotterized time evolution.
In \Cref{sec: benchmarking}, we introduce and implement our application-oriented benchmark of quantum simulation against universal behavior.
Here, we present the main results of this work -- a comparison of different levels of \gls{ems} on two quantum processors with up to 133 qubits using our \gls{qkzm}-based benchmarking.
\Cref{sec: optimization} studies digitized \gls{qa}, using the same underlying circuits as in the benchmarking experiments, for solving optimization problems and showcases how results from \Cref{sec: benchmarking} are transferable to related applications.
We conclude in \Cref{sec: outlook}.

\section{Digitized quantum annealing and defect production}
\label{sec: quantum annealing}

Given a (mixing) Hamiltonian $\HM$ with an easy-to-prepare ground state and a problem Hamiltonian $\HP$ whose ground state we wish to compute, we construct a time-dependent combination of the two,
\begin{equation}
    H(s) = A(s) \HM + B(s) \HP \ .
\label{eq: hamiltonian annealing undefined}
\end{equation}
Here, $s = t / \tf$ is the time $t \in [0, \tf]$ rescaled by the total annealing time $\tf$.
$A(s)$ and $B(s)$ are the annealing schedules such that $H(0) = \HM$ and $H(1) = \HP$.
The system is initially prepared in the ground state of $\HM$ and evolved for time $\tf$.
If this evolution is adiabatically slow, i.e., for large enough $\tf$, the system remains in the ground state of the instantaneous Hamiltonian and ends in the desired ground state of $\HP$ at $t = \tf$~\cite{Albash2018annealingRMP}.

The prototypical Hamiltonian considered in \gls{qa} is an $N$-site \gls{tfim} with nearest-neighbor interactions,
\begin{equation}
    \HM = - \sum_{i} \sx_i
    \ , \quad
    \HP = - \sum_{\langle i,j \rangle} J_{ij} \sz_i \sz_{j} \ ,
\label{eq: hamiltonian ising}
\end{equation}
with Pauli matrices $\sx_i$ and $\sz_i$ acting on site $i$ and couplings $J_{ij}$ between neighboring sites indicated by $\langle i,j \rangle$.
The ground state of $H(0) = \HM$ is $\ket{+}^N$.
We apply linear schedules $A(s) = (1-s)$ and $B(s) = s$ for simplicity, resulting in
\begin{equation}
    H(s) = - A(s) \sum_{i} \sx_i - B(s) \sum_{\langle i,j \rangle} J_{ij} \sz_i \sz_j \ .
\label{eq: hamiltonian ising annealing}
\end{equation}
Unless otherwise stated, we consider an ordered Ising model with uniform couplings $J_{ij} \equiv J$.

The \gls{tfim} is the simplest model exhibiting a \gls{qpt}~\cite{sachdev_2011} in the thermodynamic limit $N \rightarrow \infty$.
In the simple case of uniform couplings $J>0$ ($J<0$), the \gls{cp} at $J_c = A / B$ separates the paramagnetic phase ($A \gg J B$) from the doubly-degenerate (anti-)ferromagnetic phase ($A \ll JB$) with all spins (anti-)aligned.
In the disordered case, the latter is substituted by a glassy phase, i.e., an energy landscape with many deep local minima~\cite{young2010first, PhysRevLett.104.207206, knysh2016zero}.

\subsection{The Kibble-Zurek mechanism}

The \gls{qkzm} describes the non-equilibrium dynamics of the systems in the realistic setting of crossing this \gls{cp} with finite annealing time $\tf$, leading to defects in the final solution.
Ref.~\citenum{Zurek2005} provides an excellent account of the \gls{qkzm} and the resulting density of defects scaling in a \gls{tfim}, both by employing the original (classical) reasoning of the \gls{kzm} and in terms of a fully quantum description based on \gls{lz} theory.
Here, a defect refers to the wrong orientation of a spin and the density of defects to the number of defects averaged over all sites.
In the ferromagnetic phase specifically, defects are domain walls between anti-aligned spins.
We now summarize the classical description of the \gls{kzm} following Ref.~\citenum{Zurek2005}.
Classically, the \gls{cp} is characterized by a diverging correlation length $\xi$ and relaxation time $\tau$.
Far away from the \gls{cp}, $\tau$ is small enough for the system to instantaneously relax to equilibrium.
As the system approaches the \gls{cp}, $\tau$ becomes equal to and, eventually, grows beyond the time scale on which $H$ changes.
When this happens, the system can no longer relax to its instantaneous ground state, with reactions to changes slowing down until they halt completely.
This is usually referred to as the critical slowdown and subsequent freezing out.
Past the \gls{cp}, as $\tau$ decreases again, the system unfreezes and, crucially, continues evolving from approximately the frozen-out state, resulting in defects in the final state.
In summary, the growth of the system's relaxation time at the \gls{cp}, which is finite for finite system sizes, determines the necessary rate of change of the system $\diff s / \diff t = 1/\tf$ required to track the instantaneous ground state.
This is the essence of the \gls{kzm}.
It predicts the density of defects as a function of a finite annealing time $\tf$  with a power-law decay~\cite{Bando2020,keesling2019quantum}
\begin{equation}
    \ndef \propto \tf^{ - \frac{d \nu}{1 + z \nu}} \ ,
\label{eq:kzm defect density}
\end{equation}
fully determined by the system dimension $d$ and its critical exponents $\nu$ and $z$.
In 1D, with $d=1, z=1, \nu=1$, this becomes
\begin{equation}
     \ndef \propto \tf^{ - 1/2} \ .
\label{eq: kz scaling 1D}
\end{equation}

In the quantum description, the existence of a \gls{qpt} and all its related quantities originate from the closing of the system's energy gap at the \gls{cp}.
For finite system sizes $N < \infty$, the minimum gap between the ground state and the first excited state is small but finite and decreases as $\propto 1/N$.
It closes in the thermodynamic limit $N \rightarrow \infty$, causing both the correlation length $\xi$ and relaxation time $\tau$ to diverge.
Assuming annealing times fast enough to produce at least one defect on average, \gls{lz} theory yields the same \gls{kz} scaling in a quantum setting~\cite{Zurek2005}.

For uniform couplings $J$, defects in the ferromagnetic solution after annealing appear as domain walls between anti-aligned spins and can be measured by $\sigma_i^z\sigma_j^z$ correlators over all $\Ne$ edges of the spin-lattice, i.e., all nearest-neighbor correlators.
The density of defects is measured as the density of domain walls,
\begin{equation}
    \ndef = \frac{1}{2 \Ne} \sum_{\langle i, j \rangle}^{\Ne} \bigl( 1 - \sigma_i^z \sigma_j^z \bigr) \ .
\label{eq: defect density pauli op}
\end{equation}
For example, an open chain of $N$ spins has $\Ne=N-1$ edges and a periodic chain has $\Ne=N$ edges.

\subsection{Density of defect-scaling across various regimes}
\label{subsec: beyond kzm}

As mentioned above, the \gls{qkzm} is not expected to hold for every possible annealing time $\tf$.
Deviations from the \gls{kz} scaling occur at both very short and very long $\tf$~\cite{Zeng2023universal, Schmitt2022,chandran2012kibble}.
For finite system sizes and very slow anneals, i.e., large $\tf$, Ref.~\citenum{Zurek2005} describes an exponential drop in defect density following the regime of \gls{kz} scaling that is captured by \gls{lz} dynamics.
In that case, when the anneal is slow enough to never freeze out and to produce less than one defect in the chain on average, the number of defects is proportional to the \gls{lz} probability $p_\mathrm{LZ}$ of exciting the system,
\begin{equation}
    \ndef \propto p_\mathrm{LZ} \approx \exp \Bigl( - b\frac{\tf}{N^2} \Bigr) \ ,
\label{eq: lz scaling}
\end{equation}
with $b$ a constant.
Following Refs.~\citenum{Morita2008,Caneva2008}, for adiabatically slow anneals, i.e., even larger $\tf$, the scaling reads
\begin{equation}
    \ndef
    \approx p_\mathrm{LZ}(\tf) + \frac{1 - 2 p_\mathrm{LZ}(\tf)}{ a \tf^2},
\label{eq: ftlz scaling}
\end{equation}
where $a$ is a constant that depends on the derivatives of the annealing schedules~\cite{Morita2008}.
The expressions of constants $a$ and $b$ can be found in Refs.~\citenum{Zurek2005,Caneva2008}.

The theoretical predictions for the three scaling regimes in \Cref{eq: kz scaling 1D,eq: lz scaling,eq: ftlz scaling} are shown in \Cref{fig: schematic and statevector KZM}\textbf{b}.
Density of defects scalings obtained from ideal statevector simulations in \Cref{fig: schematic and statevector KZM}\textbf{c} with system sizes up to $N=20$ spins show that the beyond-\gls{kzm} scaling regimes, i.e., \Cref{eq: lz scaling,eq: ftlz scaling} are finite-size effects and that their onset is controlled by the system size.

\subsection{Product formulas for time-dependent Hamiltonians}

Ultimately, \gls{qa} implements the time evolution under a time-dependent Hamiltonian $H(t)$.
The unitary time evolution operator obtained by integrating the time-dependent Schrödinger equation reads
\begin{equation}
    U(t_f,t_0) = \mathcal{T} \exp \Biggl( -i \int_{t_0}^{t_f} \diff{t'} H(t') \Biggr) \ .
\label{eq: time evo op t-dep H}
\end{equation}
The time-ordering $\mathcal{T}$ accounts for any non-commutativity of the Hamiltonian with itself at different times, $[H(t), H(t')] \neq 0$.

To implement \gls{qa} on a digital quantum computer, we must decompose \Cref{eq: time evo op t-dep H} into quantum gates~\cite{Miessen2023perspective}.
The simplest digitized implementation of \Cref{eq: time evo op t-dep H} for a Hamiltonian $H(t) = \sum_{l=1}^L a_l(t) H_l$ is via a first order \gls{pf}, which approximates the integral as a Riemann sum, $\int_0^{t_f} \diff{t'} H(t') = \lim_{n\rightarrow \infty} \sum_{m=1}^n H(m \Delta t) \Delta t$ with $\Delta t = t_f/n$~\cite{Poulin2011,Childs2021theoryTrotterError}.
Importantly, the time ordering in the Trotterized $U(t_f,t_0)$ is enforced with a ``right-to-left ordering'', resulting in
\begin{equation}
    U \approx \prod_{m=n}^1 e^{ -i \Delta t \sum_l a_l(m \Delta t) H_l} \ .
\end{equation}
Therefore, in contrast to the case of a time-independent Hamiltonian, splitting the exponent into discrete time steps introduces a first approximation.
Typically,  a digital quantum computer cannot natively implement $e^{-it H}$.
Therefore, the Hamiltonian at time step $m \Delta t$ is further decomposed using a first order \gls{pf}, or any higher-order \gls{pf},
\begin{equation}
    U \approx \prod_{m=n}^1 \prod_{l=1}^L e^{-i a_l(m \Delta t) H_l \Delta t} \ .
\label{eq:full-trotter1-tdep}
\end{equation}
Each of these exponentials can be directly represented by the hardware native basis gates of the digital machine.
Note that, in principle, any other quantum algorithm for quantum dynamics could be used to decompose the time evolution operator as long as it allows for time-dependent Hamiltonians~\cite{Miessen2023perspective, Rajput2021}.

\section{Application-oriented benchmarking through universal scaling}
\label{sec: benchmarking}

Here, we suggest a direct approach to benchmark close to applications.
We simulate the time evolution under a time-dependent Hamiltonian and benchmark the accuracy with which a known universal scaling is replicated.
A schematic of our method is shown in \Cref{fig: schematic and statevector KZM}\textbf{a}.
We propose to measure, for example, the density of defects after digitized \gls{qa} as a function of total annealing time $\tf$.
Given a fixed time step, the time $\tf$ directly corresponds to the number of circuit layers in the Trotterized time evolution circuit.
In the \gls{kz} regime, the density of defects will decrease according to the predicted scaling up to a certain threshold $\tf$, after which hardware noise becomes dominant and leads to a deviation from the \gls{kz} scaling.
The number of circuit layers for which the expected scaling of $\ndef$ is observed is the number of circuit layers for which reliable simulation was achieved.

In particular, decoherence of the system will lead to a deviation of the defect density from the predicted \gls{kz} scaling.
This has been previously confirmed in several numerical and analog experiments that studied the effect of dissipation on the \gls{kz} scaling~\cite{Dutta2016anti,Arceci2018,Bando2020,king2022coherent}.
These studies show that different kinds of noise can cause decoherence of the system.
However, the behavior of the density of defects as noise accumulates, and whether it increases or decreases, depends on the noise.

In this regard, it is important to note that observing a $\tf^{-1/2}$ scaling may not be a sufficient condition to claim coherent quantum dynamics. 
For instance, a purely classical diffusion model may reproduce the same density of defects scaling~\cite{Mayo2021,king2022coherent}.
This is particularly important for analog quantum annealers. 
In the presence of sufficiently strong thermal noise that destroys coherence, the density of defects could still decrease with  $\tf$, since the machine may behave like a classical thermal annealer (at low temperature) or as a machine featuring incoherent quantum tunneling events~\cite{boixo2014evidence,PhysRevX.6.031015,isakov2016understanding}. These processes could still lower the density of defects.
Indeed, this effect has been observed in Ref.~\citenum{king2022coherent}, necessitating additional analysis to reasonably prove the existence of coherent annealing~\cite{Dziarmaga2022kinkcorrelations}. 

In our digital quantum setting, the situation is different: the presence of hardware noise never yields a behavior that resembles a classical thermal limit.
To prove this, in \Cref{app: noisy simulations}, we study the impact of simulated hardware noise with varying strength.
We observe that introducing hardware noise qualitatively changes $\ndef$ as a function of $\tf$.
Even for large $\tf$ and accumulating noise, we do not recover a decreasing $\ndef(\tf)$, contrary to the noisy analog quantum annealing case~\cite{king2022coherent}.
Therefore, the density of defects following the expected scaling constitutes evidence of a noise-free digital simulation up to a threshold $\tf$.
This argument is further corroborated by the fact that the noise model is in very good agreement with hardware results (\Cref{app: noisy simulations}).

The proposed benchmark has several advantages.
First and foremost, it yields a concrete, intuitive, and scalable metric; the number of simulable circuit layers as they can be found in countless applications.
Even more so since Hamiltonian simulation is a prime application for quantum computers and a building block for many other applications, for example, quantum phase estimation~\cite{Motta2022} and sampling algorithms~\cite{layden2023quantum}.
Second, our method can benchmark hardware and \gls{ems} algorithms separately or in combination.
Third, our method is scalable since no classical verification is required.
Fourth, since the \gls{kz} scaling is defined in the thermodynamic limit $N \rightarrow \infty$, it is particularly well-suited to benchmark large digital quantum computers without scaling issues.
Finally, since finite-size scaling regimes are also well-understood (cf. \Cref{subsec: beyond kzm} and \Cref{fig: schematic and statevector KZM}\textbf{b},\textbf{c}), the method is applicable in the setting of scaling to large circuit depths at qubit counts far below the thermodynamic limit.

\subsection{Experimental setup}
\label{subsec: benchmarking experimental setup}

We now discuss benchmarking different levels of \gls{ems} on two quantum processors through digitized \gls{qa} and measuring defect densities.
An $N$-spin Ising model with uniform couplings $J=1$ and linear schedules (cf. \Cref{sec: quantum annealing}) is mapped to $N$ qubits.
The qubit register is initialized as $\psi(t=0) = \ket{+}^N$, the ground state of $H(t=0) = \HM$, and time evolved under the time-dependent Hamiltonian in \Cref{eq: hamiltonian ising annealing}.
The time evolution is implemented using a first-order \gls{pf} as in \Cref{eq:full-trotter1-tdep} with a time step of $\Delta t$.
Throughout this section, we use a time step of $\Delta t = 0.5$.
We justify the choice of this time step in \Cref{app: time step}, showing results of noiseless statevector simulations with different time steps and comparing them to continuum results.
Although smaller time steps would yield higher accuracy, more Trotter steps, i.e., deeper circuits, would be required to explore the relevant \gls{kz} regime.
Our results show that $\Delta t = 0.5$ is a reasonable compromise to avoid significant algorithmic errors and allow for sufficiently large annealing times.
Hardware improvements will eventually allow for smaller time steps and thereby increase the overall accuracy of the results, particularly at short time scales.
After annealing for a total time $\tf$, we measure the defect density via \Cref{eq: defect density pauli op} to compare to the expected scaling in \Cref{eq: kz scaling 1D}.
Expectation values of observables are estimated from measurements with \estimator{} primitive~\cite{Qiskit}.

\begin{figure}[t]
    \includegraphics[]{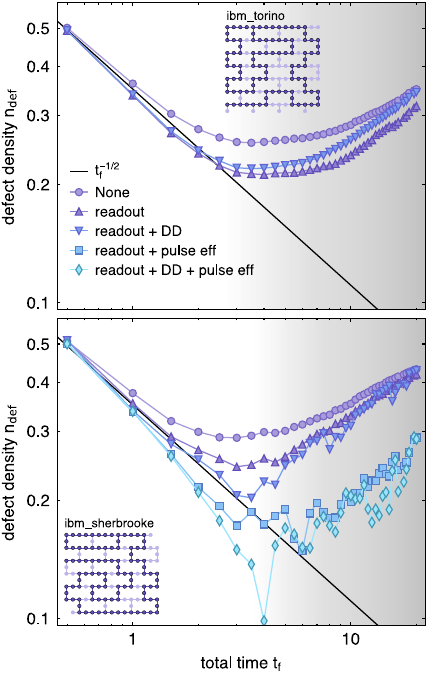}
    \caption{
        \textbf{Density of defects scaling in 1D from 100-qubit circuits.} Hardware results comparing the density of defects scaling on subsets of 100 qubits on \torino{} (top) and \sherbrooke{} (bottom) employing different levels of EMS.
        Each point corresponds to one additional Trotter layer or time step of $\Delta t = 0.5$.
        Raw simulation results with no EMS (None) are compared to results using combinations of REM, DD error suppression, and pulse-efficient transpilation.
        Insets show the qubit layout of the respective processor with the chosen qubit subset highlighted.
        The grey shading indicates decoherence and deviation from the expected KZ scaling shown by the black solid line.
    }
    \label{fig: kzm device 1D}
\end{figure}

\begin{figure*}[ht]
    \includegraphics[]{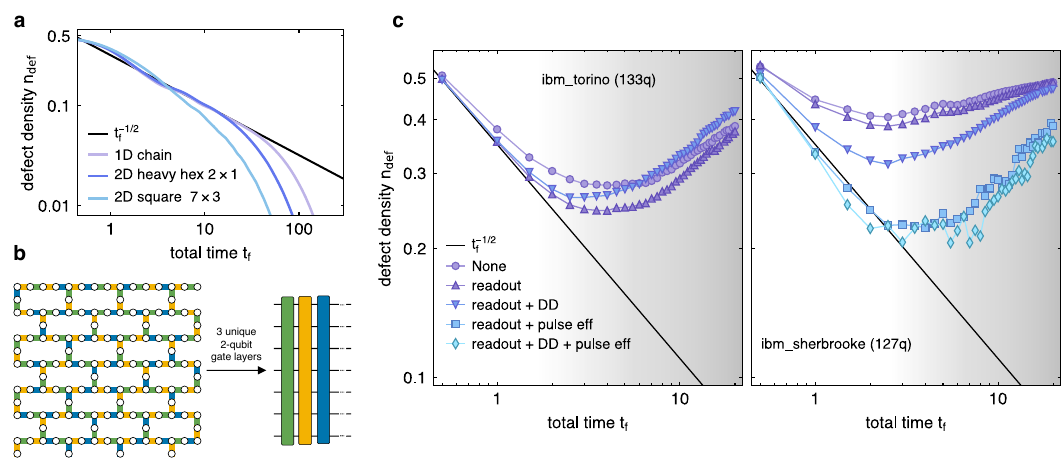}
    \caption{
        \textbf{Density of defects scaling on a heavy-hexagonal lattice from 133- and 127-qubit circuits.}
        \textbf{a} Statevector results comparing the density of defects scaling in a periodic chain, a heavy-hexagonal lattice consisting of two heavy-hex cells ($2 \times 1$), and a $7 \times 3$ square lattice, all with $N=21$ spins.
        The heavy-hexagonal lattice exhibits the same $t^{-1/2}$ scaling as predicted for 1D systems, whereas defects in a true 2D lattice show a steeper scaling.
        \textbf{b} The couplings on the heavy-hexagonal lattice of \torino{} and \sherbrooke{} are grouped such that each Trotter layer consists of three layers of $R_{ZZ}$ gates.
        \textbf{c} Hardware results comparing the density of defects scaling on \torino{} (left) and \sherbrooke{} (right) employing different levels of EMS.
        Each point corresponds to one additional Trotter layer or time step of $\Delta t = 0.5$.
        Raw simulation results with no EMS (None) are compared to results using combinations of REM, DD error suppression, and pulse-efficient transpilation.
        The grey shading indicates decoherence and deviation from the expected KZ scaling shown by the black solid line.
    }
    \label{fig: kzm device heavy-hex}
\end{figure*}

We employ the IBM Heron processor \torino{} and the IBM Eagle processor \sherbrooke{} for our simulations, with 133 and 127 qubits, respectively, and heavy-hexagonal qubit connectivity~\cite{IBMQuantumPlatform}.
Hardware characteristics such as error rates are given in \Cref{app: hardware properties}.
Eagle and Heron processors differ in their native two-qubit gates and how they are physically realized.
\sherbrooke{}'s native two-qubit gate is an \gls{ecr} gate implemented via a dispersive coupling mediated by a fixed-frequency resonator~\cite{Rigetti2010ecrGate, Sheldon2016ecrGate}.
By contrast, \torino{}'s native two-qubit gate is a \cz{} gate realized through tunable-frequency couplers.
In a tunable coupler architecture, the coupling element is frequency tunable, and driving it can create different two-qubit gates~\cite{Mckay2016universal, Ganzhorn2020benchmarking}.
Most importantly, this results in \torino{} having a roughly $6 \times$ shorter two-qubit gate time compared to \sherbrooke{} and a higher two-qubit gate quality~\cite{Mckay2023layerfidelity}.

On both devices, we compare simulations with no \gls{ems} to \gls{rem}~\cite{Berg2022trex,Nation2021m3} as well as to \gls{rem} combined with \gls{dd}~\cite{Barron2023qaoa}.
In addition, \sherbrooke{} allows for pulse-efficient transpilation of our circuits, another method of error suppression~\cite{Earnest2023pulse}.
Pulse-efficient transpilation scales hardware native cross-resonance pulses and their echoes (combined, this makes up an \gls{ecr} gate on \sherbrooke{}) and is thus not compatible with \torino{}'s native \cz{} gates.
\Cref{app: error mitigation} gives a brief technical summary of each method.

\subsection{Results: 1D chain}
\label{subsec: benchmarking 1D results}

The Trotter circuit for the time evolution of a spin chain requires at least two layers of $\rzz$ gates per Trotter layer, or time step.
Since each $\rzz$ gate is transpiled to two hardware native two-qubit gates (ECR or CZ), the final circuit for a 1D Ising model has a two-qubit gate depth of $4 \times N_t$ where $N_t$ is the number of time steps.
\Cref{fig: kzm device 1D} shows the density of defects scaling obtained from digitized \gls{qa} on subsets of 100 qubits connected through an open line (see insets), which was chosen to minimize cumulative two-qubit gate errors along the line (cf. \Cref{app: hardware properties}).
Owing to noise, the experimental results differ from the noiseless statevector results in \Cref{fig: schematic and statevector KZM}\textbf{c}, which show a monotonic decrease in defect density with increasing $\tf$.
Hardware errors only make it possible to follow the KZ scaling and subsequent scaling regimes up to a certain threshold time.
Beyond this point (indicated by grey shading), the defect density deviates from the $t^{-1/2}$ scaling and eventually increases, as the noise accumulates with increasing circuit depth as predicted by the noisy emulation of the algorithm.

This threshold time depends on the hardware and the \gls{ems} employed.
The top panel of \Cref{fig: kzm device 1D} shows the results for \torino{} comparing no \gls{ems} with only \gls{rem}, and with \gls{rem} combined with \gls{dd}.
While we can simulate two Trotter layers with a two-qubit gate depth of 8 and a total of 396 CZ gates without any level of \gls{ems}, adding \gls{rem} already significantly improves the results.
We can reliably simulate up to five Trotter layers with a two-qubit gate depth of 20 and 990 CZ gates in total.
Assuming \torino{}'s median \cz{} gate time of $84 \, \mathrm{ns}$ for all two-qubit gates, this corresponds to a circuit execution time of $\qty{1.7}{\us}$.
This is remarkable given that \gls{rem} is the simplest and easiest-to-implement form of \gls{em}.
We attribute the negligible impact of the \gls{dd} sequence on the \torino{} results to the reduced cross-talk of the device compared to \sherbrooke{}.

The picture is different on \sherbrooke{}, shown in the bottom panel of \Cref{fig: kzm device 1D}.
We compare the same \gls{ems} methods as on \torino{} in addition to pulse-efficiently transpiled circuits.
The results without any \gls{ems} and only \gls{rem} are slightly worse than on \torino{}.
However, adding \gls{dd} to counter static $ZZ$ cross-talk results in substantial gains on this device.
According to our metric, we achieve a reliable simulation of seven circuit layers with a two-qubit gate depth of 28 and 1386 ECR gates in total.
Assuming \sherbrooke{}'s median ECR gate time of $\qty{533}{\ns}$~\cite{IBMQuantumPlatform} for all two-qubit gates, this equates to a circuit execution time of roughly $\qty{15}{\us}$.
Moreover, we show in \Cref{app: kink-kink correlator} that the correlations between defects show the characteristic non-monotonic fingerprint of a genuine quantum \gls{kzm}~\cite{Nowak2021quantum,Dziarmaga2022kinkcorrelations}.
Pulse-efficient transpilation fairs similarly, though we observe that the curve dips below the expected scaling.
We attribute this to rotation errors in the scaled two-qubit pulses at small angles~\cite{Earnest2023pulse}, see \Cref{app: error mitigation}.
As the two-qubit gate angles $\theta = -2J B(m \Delta t / \tf) \Delta t = -2J m \Delta t^2 / \tf$, with $m\Delta t = t$, are proportional to the annealing schedule, increasing $\tf$ means smaller and smaller angles at the beginning of an anneal.
This causes a growing disparity between the target operation of the circuit before transpilation and the final sequence of gates with growing $\tf$.
Fixing these rotation errors requires custom calibrations~\cite{Vazquez2023wellConditioned}, which are difficult at scale through a cloud-based quantum computing service.
Since our goal is to compare out-of-the-box \gls{ems} methods, we did not conduct any custom calibrations.

\subsection{Results: heavy-hexagonal lattice}
\label{subsec: benchmarking heavy hex results}

Our approach is not limited to one-dimensional spin systems.
In fact, such critical behavior and the resulting universal scaling manifests in lattice geometries beyond 1D~\cite{Schmitt2022} and can hence benchmark any qubit connectivity for which the critical exponents and scaling are known.
The processors we use have a heavy-hexagonal qubit connectivity for which we need to infer the predicted \gls{kz} scaling.
The degree of connectivity, i.e., the number of edges over the number of sites, in a heavy-hex graph is $N_\mathrm{e} / N_\mathrm{s} = 6/5=1.2$.
In terms of connectivity, it is hence much closer to a 1D system with $N_\mathrm{e} / N_\mathrm{s} = 1$ than to a 2D square lattice with $N_\mathrm{e} / N_\mathrm{s} = 2$.
We therefore conclude that the density of defects on a heavy-hex lattice should follow a scaling similar to the 1D scaling of $t^{-1/2}$.
This is further supported by noiseless statevector simulations shown in \Cref{fig: kzm device heavy-hex}\textbf{a} comparing the density of defects scaling of a 1D periodic chain, a heavy-hex lattice, and a 2D square lattice with $N=21$ sites each.
The data shows that the density of defects in 1D and heavy-hex lattices coincide on the $t^{-1/2}$ scaling before the exponential drop-off at larger $\tf$, attributed to finite-size effects, while the 2D square lattice shows a steeper scaling.
This is why we reasonably consider the $t^{-1/2}$ scaling also as the benchmark line for the heavy-hex lattice, even though deviations from the exact $t^{-1/2}$ scaling are possible for this geometry and the critical exponents should be further investigated.

The Trotter circuit for a Hamiltonian \Cref{eq: hamiltonian ising annealing} on a heavy-hex lattice is made depth-optimal by grouping the couplings according to the edge-coloring shown in \Cref{fig: kzm device heavy-hex}\textbf{b}.
Each Trotter layer in the circuit therefore consists of three layers of $\rzz$ gates.
Again, each $\rzz$ gate is transpiled into two hardware-native two-qubit gates, resulting in a total two-qubit gate depth of $6 \times N_t$.

Utilizing all 133 and 127 qubits of \torino{} and \sherbrooke{}, respectively, we see an even starker difference between both chips in defect density (\Cref{fig: kzm device heavy-hex}\textbf{c}).
We can reliably simulate up to three Trotter layers on \torino{} employing only \gls{rem}.
With 150 edges, this equals a two-qubit gate depth of 18 and 900 CZ gates in total.
Again, assuming \torino{}'s median \cz{} gate time of $\qty{84}{ns}$ for all two-qubit gates, this implies a circuit execution time of $\qty{1.5}{\us}$ using merely \gls{rem}.
\sherbrooke{}, on the other hand, requires more involved \gls{ems}, which becomes even more apparent when using all qubits of the device.
Although \gls{dd} again provides sizable improvements, we do not reproduce the expected \gls{kz} scaling.
However, pulse-efficient transpilation combined with \gls{rem} extends the circuit depth up to five Trotter layers before noise becomes prevalent.
This is equivalent to a Trotter circuit with a CNOT gate depth of 30 and 1440 CNOT gates in total.
In this case, the circuit duration reduces to roughly $\qty{6}{\us}$ due to the pulse-efficient transpilation.

\section{Digitized quantum annealing for optimization}
\label{sec: optimization}

\begin{figure*}[t]
    \includegraphics{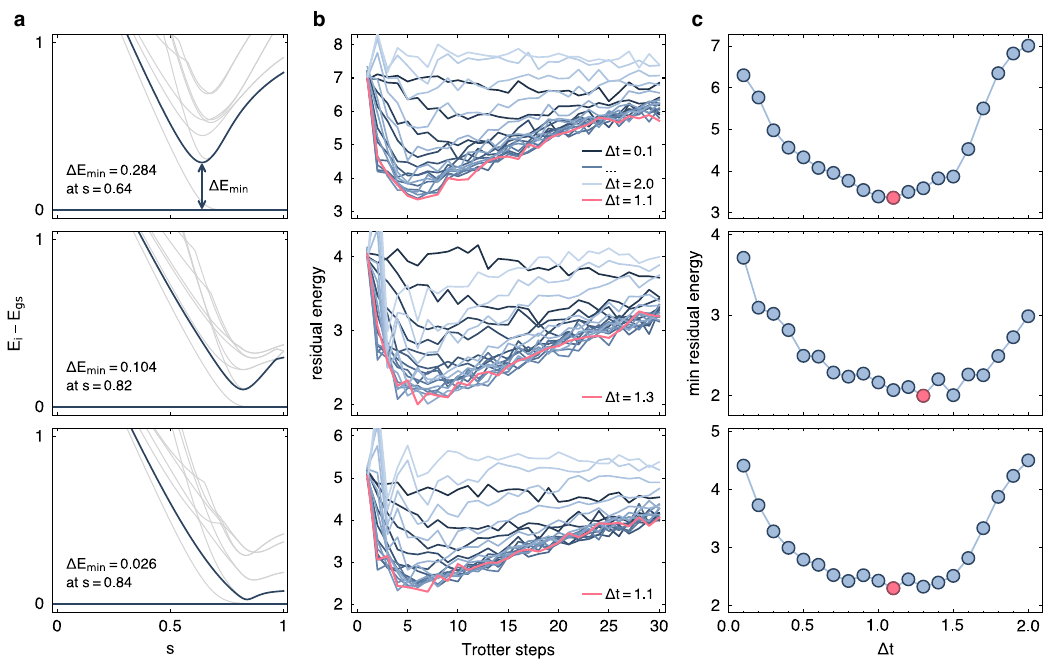}
    \caption{
        \textbf{Hardware results of the residual energy dependence on time step and spectral gap.}
        \textbf{a} Spectra relative to the ground state of three different instances of disorder in a 12-qubit periodic chain with coupling coefficients uniformly sampled from $J_{ij} \in [-1, 1]$.
        The instances were specifically chosen with three different minimum energy gaps, differing by an order of magnitude between the largest (top) and the smallest gap (bottom).
        \textbf{b} Residual energy averaged over $400$ samples obtained from QA of the system in \textbf{a} on \auckland{} using only REM with fixed time steps $\Delta t \in \{ 0.1, \ldots, 2.0 \}$ as a function of circuit depth (number of time steps).
        \textbf{c} Minimum residual energy from \textbf{b} as a function of the time step, i.e., each point corresponds to the minimum over one curve in \textbf{b}.
    }
    \label{fig: residual energy auckland}
\end{figure*}

The proposed benchmark provides a reasonable prediction of hardware and error mitigation capabilities for a set of problems broader than the non-equilibrium quantum dynamics of a ferromagnetic Ising model.
Estimating how many circuit layers can be reliably executed before noise prevails is useful in many applications.
For instance, variational algorithms~\cite{Wu2023,astrakhantsev2023phenomenological} for many-body spin systems feature circuits very similar to the ones considered in the previous section.
Circuits with the same structure may also be used in quantum machine learning~\cite{havlivcek2019supervised, Melo2023pulseefficient,abbas2021power}.
Finally, the entangling blocks used previously can be adapted to \gls{qaoa} or digitized \gls{qa} of disordered spin models.

We now turn to combinatorial optimization, where the couplings $J_{ij}$ in \Cref{eq: hamiltonian ising} can be arbitrary in magnitude and sign. 
Optimization problems defined by this type of Hamiltonian may fall within the NP-complete complexity class, depending on the connectivity of the underlying graph, and are frequently employed as benchmarks for quantum optimization~\cite{ronnow2014defining,Albash2018scalingAdvantageAnnealing}.
It is therefore of interest to determine whether the results obtained from benchmarking the \gls{qkzm} and identifying an optimal working point are transferable to an optimization context.
This is not straightforward since the \gls{qkzm} is connected with second-order phase transitions, while the bottlenecks in quantum annealing of hard optimization problems are due to first-order transitions in the glassy phase~\cite{young2010first, PhysRevLett.104.207206, knysh2016zero}.

\Gls{qa} has been little explored on digital quantum computers~\cite{keever2023adiabatic}.
Instead, the focus has been mostly on \gls{qaoa}, an annealing-inspired variational ansatz and the most popular near-term algorithm for quantum optimization~\cite{farhi2014quantum, Barron2023qaoa, Sack2024qaoa, Hadfield2019}.
However, optimizing the variational parameters in \gls{qaoa} is costly in itself~\cite{Scriva2024qaoa} and, once found, optimal variational parameters reduce \gls{qaoa} to \gls{qa} in the large-layer limit~\cite{Zhou2020}.
This does not mean that the optimal \gls{qaoa} parameters can be directly derived from annealing schedules.
Rather, the variational optimization results in optimized annealing schedules.
Nonetheless, precisely because of this relationship, annealing schedules can be used to initialize \gls{qaoa} parameters, thereby reducing the training cost of the variational ansatz~\cite{Scriva2024qaoa,Sack2019qaoa}.
It is therefore sensible to study digitized \gls{qa} in more detail.
Previous experimental realizations of digitized \gls{qa} featured only up to nine qubits~\cite{barends2016digitized}.

Since digitized \gls{qa} requires discretizing the time evolution, a central question is what effect does the choice of the time step have on the success of the annealing.
In principle, a large time-step error could, counterintuitively, even improve the performance of digitized \gls{qa}.
This effect has been observed in classical path-integral simulations of \gls{qa}, where an unconverged, finite Trotter step is beneficial for tunneling between configurations~\cite{Heim2015spinGlass}, reconciling earlier expectations of quantum speed-up~\cite{santoro2002theory} with observations~\cite{ronnow2014defining}.
While investigating this aspect
we identify an optimal time step for digitized \gls{qa} given a fixed number of circuit layers or time steps $N_t$.
Fixing $N_t$ is meaningful since, in reality, one can only access a limited amount of resources such as the expendable number of gates, determining both the computational cost and, more importantly before achieving fault-tolerance, the amount of error introduced by hardware noise.
The success of \gls{qa} relies on as large as possible annealing times $\tf$, which is achieved by increasing the time step $\Delta t$ for a fixed number of time steps $N_t$.
On the other hand, the errors of time evolution algorithms typically scale with $\Delta t$~\cite{Miessen2023perspective}.
\Glspl{pf} are derived on the assumption of small time steps $\Delta t \ll 1$ and algorithmic errors scale with $\mathcal{O}(\Delta t^2)$~\cite{Childs2021theoryTrotterError}.
This naturally introduces a trade-off between \gls{qa} performance and the algorithmic error due to $\Delta t$~\cite{Sack2019qaoa}.
Note that, despite this trade-off, having a converged time step is no requirement here.
Instead, the sole objective is to obtain the best possible solution, i.e., the lowest energy.
This differs from \Cref{sec: benchmarking} where we chose the largest possible time step that is still reasonably close to the continuum while allowing us to reach as-large-as-possible $\tf$.

Since the solution to a classical optimization problem is a single bitstring, we are usually interested in individual measurement samples rather than expectation values.
Therefore, we use \qiskit{}'s Sampler primitive~\cite{Qiskit} throughout this section to obtain individual measurement samples, i.e., bitstrings of measurement outcomes.
Many \gls{em} techniques apply only to expectation values.
By contrast, error suppression methods such as \gls{dd} and pulse-efficient transpilation apply to sampling.
However, as discussed in \Cref{sec: benchmarking}, \gls{dd} does not yield substantial improvements on \torino{}. 
Furthermore, IBM Heron processors, such as \torino{}, are based on tunable couplers and thus do not allow for pulse-efficient transpilation which was designed for cross-resonance-based hardware~\cite{Earnest2023pulse}.
To make results comparable across different devices, we therefore only use \gls{rem} in this section.
Furthermore, we will compute the residual energy,
\begin{equation}
    \Eres = E(\tf) - E_0 \ ,
\end{equation}
where $E(\tf) = \braket{\psi(\tf) | H(\tf) | \psi(\tf)}$ and $E_0$ is the exact ground state obtained through exact diagonalization for small system sizes and using CPLEX~\cite{cplex} for large systems.

\subsection{Dependence of the residual energy on the time step and spectral gap}
\label{subsec: optimization auckland}

We first study the dependence of the digitized \gls{qa} result on the time step on a small system of a periodic 12-qubit chain.
Three instances of disorder with couplings randomly sampled from $J_{ij} \in [-1, 1]$ are chosen with varying minimum energy gaps between the ground and the first parity-preserving excited state since a smaller gap entails longer annealing times and makes finding a solution generally more difficult.
Note that the Hamiltonian \Cref{eq: hamiltonian ising annealing} and therefore its spectrum is a function of $s$.
The minimum energy gap takes on its smallest value at the critical point.
Furthermore, the first excited state of opposite parity becomes degenerate with the ground state as $s \rightarrow 1$.
The spectra of the three different Hamiltonians as a function of $s$ are shown in \Cref{fig: residual energy auckland}\textbf{a} with minimum gaps ranging across one order of magnitude.
\Cref{fig: residual energy auckland}\textbf{b} displays the residual energy after annealing the respective system from \textbf{a} on \auckland{}~\footnote{\auckland{} is one of IBM's by now retired 27-qubit Falcon chips} with fixed time step $\Delta t \in \{ 0.1, \ldots, 2.0 \}$ as a function of the number of time steps (depth).
Here, we observe that the smallest residual energies are reached after approximately $5-10$ Trotter layers, owing to decoherence.
The observation of a minimum is compatible with both numerical predictions~\cite{Arceci2018} and experimental results from analog simulation~\cite{king2022coherent, ebadi2022quantum}.
Furthermore, it is analogous to what was observed in the benchmarking experiment \Cref{fig: kzm device heavy-hex}\textbf{c}, where we also located a minimum defect density after roughly 5-10 Trotter layers.

For each $\Delta t$, the minimum residual energy achieved over all circuit depths is plotted in \Cref{fig: residual energy auckland}\textbf{c}.
For all three systems, the lowest residual energy is obtained with a time step of $\Delta t > 1$, specifically $1.0 < \Delta t < 1.5$.
Noiseless statevector simulations confirm that this is not merely a consequence of hardware noise, as seen in \Cref{app: optimization statevector}, yielding an optimal time step of $1.2 < \Delta t < 1.4$.
However, in statevector simulation, the existence of a finite optimal time step is likely due to the finite range of depths chosen.
Therefore, choosing an infinitesimal time step seems to be both (i) inefficient in noise-free simulations and (ii) impractical in real experiments.
On the other hand, a too-large time step implies algorithmic errors.

To summarize, we observe a trade-off between realizing the largest possible annealing times while keeping the algorithmic error under control.
Choosing an unconverged, large time step is indeed advantageous when using digitized \gls{qa} for optimization.
This is of further relevance when initializing \gls{qaoa} variational parameters mentioned previously.
In the context of directly studying \gls{qaoa} performance, Ref.~\cite{Sack2019qaoa} finds an optimal time step of $0.75$ for initializing variational parameters.
Increasing $\Delta t$ even further, however, does not provide any benefits as shown by the statevector results.
This observation is consequential for practical noisy settings and allows us to optimize the number of Trotter steps needed to reach a target annealing time.

\subsection{Optimization of disordered heavy-hexagonal graph}
\label{subsec: optimization torino}

\begin{figure}[t]
    \includegraphics[]{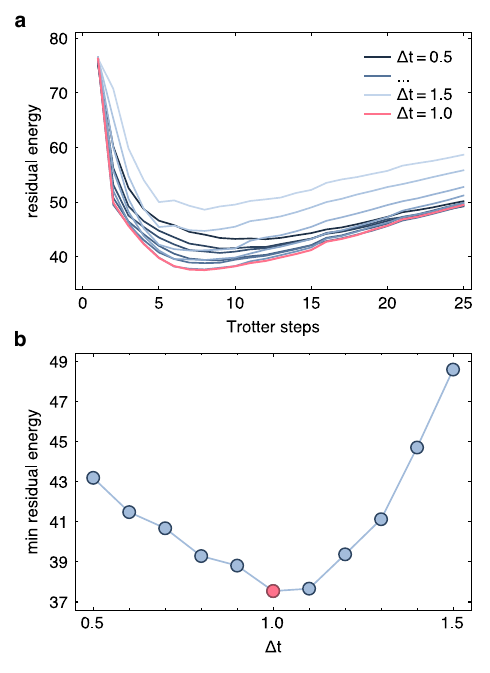}
    \caption{
        \textbf{Hardware results of the residual energy dependence on the time step.}
        \textbf{a} Residual energy of a heavy-hexagonal 133-qubit graph with coupling coefficients uniformly sampled from $J_{ij} \in [-1, 1]$, averaged over $10^5$ samples obtained from QA on \torino{} with fixed time steps $\Delta t \in \{ 0.5, \ldots, 1.5 \}$ as a function of circuit depth (number of time steps).
        \textbf{b} Minimum mean residual energy from \textbf{a} as a function of the time step, i.e., each point corresponds to the minimum over one curve in \textbf{a}.
    }
    \label{fig: residual energy torino}
\end{figure}

We now study a heavy-hexagonal spin-lattice of 133 qubits, i.e. \torino{}'s full chip.
We consider just one realization of disorder with uniformly sampled couplings $J_{ij} \in [-1, 1]$, and compute $\Eres$ as a function of circuit depth and time step as before.
The ground state is again computed using CPLEX.
This time, the system is too large to compute its instantaneous higher excited states, which is why we have no information about the size of its minimum energy gap.
However, the comparison of different systems in the previous section indicates that the dependence of the optimal time step on the spectral gap is small.
\Cref{fig: residual energy torino}\textbf{a} shows the results for time steps $\Delta t \in \{0.5, \ldots 1.5 \}$.
Again, we observe that, for each time step, the minimum residual energy is obtained at around $5-10$ Trotter layers, or time steps, after which decoherence begins to dominate.
\Cref{fig: residual energy torino}\textbf{b} reports the minimum $\Eres$ for each value of the time step.
Also here, we observe a continuous decrease in $\Eres$ with increasing $\Delta t$ up to a time step of $1.0$ to $1.1$, after which $\Eres$ sharply increases.
In conclusion, this suggests an unconverged, large time step of $\Delta t \gtrsim 1.0$ to be optimal for digital \gls{qa} regardless of the system instance and graph connectivity.
Further studies will be required to explore the generality of this empirical finding.

\subsection{Consistency with the KZ benchmark}
\label{subsec: relationship optimization}
Here, we briefly summarize the relation of our large-scale optimization results in \Cref{subsec: optimization torino} to the results of \Cref{sec: benchmarking}.
In \Cref{sec: benchmarking}, the 1D ferromagnetic system (i.e., uniform couplings) simulation on \torino{}, see \Cref{fig: kzm device 1D}, indicates that about 5 Trotter steps can be executed reliably before the density of defects substantially deviates from the expected scaling (referring to the \gls{rem} curve for consistency with the optimization experiments).
Subsequently, at about 8 Trotter steps, the density of defects reaches a minimum before it increases due to the accumulated noise.
Similar values of 3 and 8 Trotter steps, respectively, are obtained on the heavy-hex lattice, see \Cref{fig: kzm device heavy-hex}.

The optimization experiment shares the same connectivity as these benchmarks but has random couplings and different time steps.
Depending on the time step used, the minimum energy we measure (which is the disordered analog to the defect density) lies again in the range of 5-10 Trotter steps, see \Cref{fig: residual energy torino}\textbf{a}.
This finding is consistent with our results from \Cref{sec: benchmarking}.
For applications such as optimization or variational methods, the location of the minimum in $\ndef(\tf)$ is more relevant than the point where it starts deviating from the expected scaling.

This demonstrates the usefulness of our benchmarking routine in an application context.
Suppose, for example, that one is interested in using \gls{qaoa} rather than digitized \gls{qa}.
Since \gls{qaoa} features the same circuit structure in a closed-loop classical parameter optimization, it is expensive to discover the optimal circuit depth empirically.
Running the proposed benchmark beforehand allows one to directly select the optimal \gls{qaoa} depth according to the hardware capabilities.

\section{Discussion and Outlook}
\label{sec: outlook}

This work presents an application-oriented method to jointly benchmark the hardware and the algorithms.
Developing such a method to predict simulation quality, and, importantly, more closely resemble device and algorithm capabilities, is critical.
Particularly so in times of rapid algorithmic and hardware advancements.
The simple figure of merit that we introduce is the number of Trotter steps, i.e., the threshold depth, before which errors induce deviations from a universal scaling law.
While we propose to use digitized \gls{qa} and the universal \gls{kz} scaling in our work, we emphasize the central idea: benchmark against the prediction of a known universal behavior.
Other scaling laws, model Hamiltonians, or lattice geometries could be employed as well.

We observe that different machines, equipped with different \gls{ems}, provide different threshold depths.
Importantly, the simple benchmark we propose can measure the continuous improvements of hardware and algorithms in the near future.
Moreover, it is tailored to Hamiltonian simulation, which is among the most anticipated applications with provable quantum advantage in physics and a building block in many other applications.

We test our method on hardware native geometries and the resulting circuits contain only dense layers of $\rzz$ gates.
However, this does not limit its predictive power since all two-qubit gates are eventually transpiled to the same hardware-native two-qubit gate.
One can therefore directly transfer results in terms of two-qubit-gate depth to the simulation of more complicated models.
Nonetheless, future research could devise more varied models exhibiting universal behavior that could be used analogously to the \gls{kz} scaling.
For instance, circuits with interleaved two-qubit gates of different kinds that do not commute with each other such that they cannot be arranged in a dense layer of gates, as for example encountered in fermionic models.
Furthermore, our method could be used to select best-qubit subsets.
By measuring all nearest-neighbor $\sz_i \sz_j$ correlators on a given quantum processor, the defect densities of different qubit-subsets can be reconstructed in post-processing.
For example, computing the density of defects for all qubit-subsets of size $N$ in this way would allow selecting the subset that best matches the \gls{kz} scaling.
Moreover, the known finite-size scaling regimes for large annealing times could one day be used to benchmark very deep quantum circuits once they become accessible with advanced \gls{ems} techniques or, eventually, \gls{qec}.

Our benchmark extends beyond quantum many-body simulations to quantum optimization and variational algorithms.
We observe consistency between the threshold number of Trotter steps before the noise becomes prevalent, identified using the \gls{kz} scaling, and the optimal depth 
of digitized \gls{qa} for combinatorial optimization.
We demonstrate that, counter-intuitively, the time step should indeed be chosen at a value $\gtrsim 1.0$, resulting in significantly improved residual energy values after annealing.
This means in practice that digitized \gls{qa} could be a competitive quantum optimization framework as it (i) avoids a classical optimization loop and (ii) greatly reduces the runtime of \gls{qaoa}~\cite{Weidenfeller2022scalingofquantum}, avoiding the costly iterative optimization, affected by hardware and shot noise~\cite{Scriva2024qaoa,mazzola2024quantum}.
More generally, our results suggest that current hardware with simple \gls{ems} may support variational ansatze (for a similar class of Hamiltonians) comprising about ten Trotter-circuit layers.
Our results seem to be robust against several instances of disorder with varying minimum spectral gaps as well as on systems of different sizes and connectivities.
Nonetheless, future work should aim at exploring other settings such as solving dense graphs~\cite{Sack2024qaoa}.

In introducing this benchmark, we report some of the largest-scale digital quantum simulations as of today~\cite{Kim2023utility,farrell2023scalable,pelofske2023scaling, chowdhury2024enhancing}.
Compared to Ref.~\citenum{Kim2023utility}, we move closer to a physical system by incorporating Trotter error and time-dependent, general two-qubit gate angles, requiring a full gate decomposition of the $\rzz$ gate into hardware native two-qubit gates.
However, we do not introduce sophisticated \gls{ems} such as probabilistic error amplification, which could be a goal of future work.

Finally, we emphasize that the purpose of this paper is not to claim direct quantum advantage with the featured experiments, including the optimization demonstration.
For one, the sparse-graph optimization problem studied here is not hard. 
Other quantum or classical platforms may demonstrate better performance than we do~\cite{Rehfeldt2023}.
Moreover, we do not claim that the digitized quantum annealing method is the best-performing digital algorithm.
Indeed, warm-start methods~\cite{Egger2021warmstart, Tate2023warmstart} and digitized counterdiabatic approaches~\cite{Chandarana2022Counterdiabatic} all improve the performance of quantum approximate optimization.
Instead, the value of our optimization experiments lies in understanding its correlation with the proposed benchmark.
Furthermore, the benchmarks introduced here will be instrumental to verify future quantum simulations on sufficiently large hardware, beyond the reach of approximate classical methods.


\subsection*{Data Availability}
The numerical data that support the findings of this study are available from the corresponding authors upon reasonable request.

\subsection*{Code Availability}
The relevant scripts of this study are available from the corresponding authors upon reasonable request.

\subsection*{Acknowledgments}
We thank Julien Gacon, Elena Peña Tapia, and Almudena Carrera Vazquez for help with the code, and Stefan Wörner, Laurin Fischer, Samuele Piccinelli, and Francesco Tacchino for insightful and inspiring discussions.
We also thank Adolfo Del Campo, Jacek Dziarmaga, and Andrew King for providing useful feedback.
G.\,M. acknowledges financial support from the Swiss 
National Science Foundation (grant PCEFP2\_203455).
This research was supported by the NCCR MARVEL, a National Centre of Competence in Research, funded by the Swiss National Science Foundation (grant number 205602).
We acknowledge the use of IBM Quantum services for this work. The views expressed are those of the authors, and do not reflect the official policy or position of IBM or the IBM Quantum team.

\appendix

\section{Influence of hardware noise}
\label{app: noisy simulations}

\begin{figure}[t]
    \includegraphics{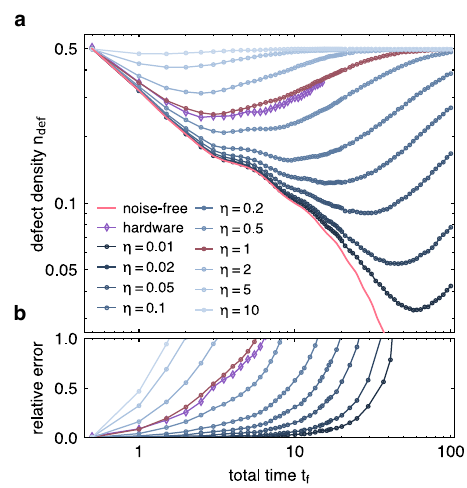}
    \caption{
        \textbf{Density of defects scaling in the presence of simulated hardware noise.}
        \textbf{a} All curves show simulated Trotterized time evolution of a 12-qubit chain with $\Delta t = 0.5$.
        The noise-free reference curve (pink line) represents a statevector simulation.
        The hardware curve (purple diamonds) corresponds to results obtained from \sherbrooke{} using only \gls{rem}.
        All other curves show results of sampling $10^5$ measurements from a simulated circuit and include simulated hardware noise that corresponds to the full noise model of \sherbrooke{}.
        \textbf{b} Error of the density of defects relative to the noise-free results.
    }
    \label{fig: noisy simulations}
\end{figure}

The effects of noise such as a finite temperature or bath-couplings on \gls{qa} have been previously studied both numerically~\cite{Arceci2018, Bando2020} and experimentally~\cite{king2022coherent} in analog settings.
The result is a deviation from the \gls{kz} scaling and a subsequent increase in defect density.

Here, we perform simulations of the Trotter algorithm including a tunable noise model.
Moreover, since noise generally leads to decoherence of the system, this further validates our claim that observing a density of defects that follows the \gls{kz} scaling implies coherent evolution.
Scaling the average device errors across several orders of magnitude allows for a detailed understanding of their influence on the density of defects.
In other words, the noisier the system, the sooner we expect decoherence and, in turn, a deviation from the expected scaling.

The noise model is constructed from error rates of \sherbrooke{} obtained through standard calibrations~\cite{IBMQuantumPlatform}.
To simplify the scaling of the noise, we use error rates obtained from the average of individual gate and qubit properties.
Concretely, our model applies the following noise to all qubits,
\begin{align}
& T_1 = T_1^{\mathrm{ave}} / \eta 
\ , \quad
T_2 = T_2^{\mathrm{ave}} / \eta \\
& e_{\mathrm{1q}} =  \eta e_{\mathrm{1q}}^{\mathrm{ave}}
\ , \quad
e_{\mathrm{2q}} =  \eta e_{\mathrm{2q}}^{\mathrm{ave}}
\ , \quad
e_{\mathrm{ro}} =  \eta e_{\mathrm{ro}}^{\mathrm{ave}} \ . \nonumber
\end{align}
Here, $T_1, T_2$ are relaxation time and dephasing time, $e_{\mathrm{1q}}, e_{\mathrm{2q}}, e_{\mathrm{ro}}$ are single-qubit, two-qubit, and readout errors, respectively, and $\eta$ is a scaling factor.
The average values are $T_1^{\mathrm{ave}} = \qty{266.37}{\mu s}$, $T_2^{\mathrm{ave}} = \qty{178.71}{\us}$, $e_{\mathrm{1q}}^{\mathrm{ave}} = 1.25 \times 10^{-3}$, $e_{\mathrm{2q}}^{\mathrm{ave}} = 1.10 \times 10^{-2}$, and $e_{\mathrm{ro}}^{\mathrm{ave}} = 2.41 \times 10^{-2}$.
All gate and measurement times are left unchanged and correspond to those of \sherbrooke{}.

\Cref{fig: noisy simulations}\textbf{a} shows the density of defects obtained from \gls{qa} of a periodic 12-qubit system with simulated noise scaled by $\eta \in \{ 0.01, \ldots,10 \}$.
The noise-free line is a statevector simulation of the same system.
In addition, we plot real hardware results of simulating the same 12-qubit system on \sherbrooke{}, using only \gls{rem}.
The results of this experiment match the noisy simulations with unscaled noise $\eta=1$, thereby validating our choice of noise model.
Both the noisy and the noise-free simulations as well as the hardware experiment were done with the same time step of $\Delta t = 0.5$ as in \Cref{sec: benchmarking}.

Since the system size that we can simulate exactly with noise is small, the behavior of the density of defects over time appears less smooth compared to larger sizes.
These finite size effects manifest as a drop in the noise-free density of defects after roughly $\tf \approx 10$.
The noise-free simulations indicates a monotonically decreasing density of defects with $\tf$.
In the noisy cases instead, after a threshold time $\tf^*$, the density of defects starts increasing with time. 
As the noise strength decreases, $\tf^*$ increases.
This is further underlined by \Cref{fig: noisy simulations}\textbf{b}, showing the relative error of the density of defects with respect to the noise-free curve, $|(\ndef(\eta) - \tilde{n}_\mathrm{def}) / \tilde{n}_\mathrm{def} |$, with $\tilde{n}_\mathrm{def}$ the noise-free defect density.
While this behavior is compatible with earlier numerical studies on dissipative quantum annealing and analog hardware experiments~\cite{Arceci2018,king2022coherent,Bando2020}, the main difference here is that the density of defects always keeps increasing with time after $\tf^*$. 
In contrast, dissipative quantum dynamics or annealing experiments~\cite{king2022coherent,King2023spinGlass}, indicated that the density of defects may decrease again, possibly with the same $-1/2$ exponent, even beyond the coherence limit.

\section{Trotter error of benchmarking experiments}
\label{app: time step}

\begin{figure}[t]
    \includegraphics{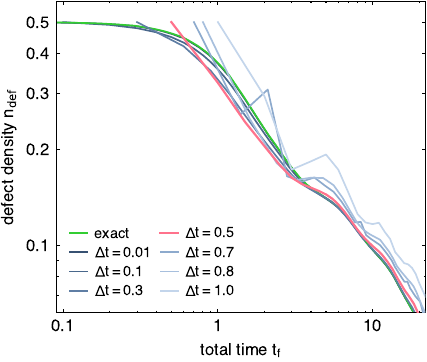}
    \caption{
        \textbf{Density of defects scaling for different discretizations $\Delta t$.}
        Statevector simulations of a periodic 12-qubit chain comparing the exact, i.e., non-discrete time evolution to Trotterized simulations with different time steps $\Delta t$.
        The time step $\Delta t = 0.5$ is highlighted in pink and corresponds to the time step chosen for the hardware experiments in \Cref{sec: benchmarking}.
    }
    \label{fig: kzm time steps}
\end{figure}

Discretization of the time evolution requires setting a time step, which in turn controls the discretization error, in our case the Trotter error.
Ideally, the time step should be sufficiently small to reduce the Trotter error and accurately describe the continuous-time limit of the dynamics.
However, smaller time steps require more Trotter circuit layers to reach the same final time.
In \Cref{sec: benchmarking}, we adopt a practical approach and utilize a small, yet non-vanishing, time step of $\Delta t = 0.5$.

\Cref{fig: kzm time steps}\textbf{a} shows statevector simulations of a periodic 12-qubit chain with time steps varying between $\Delta t = 0.01, \ldots 1.0$ and compares the resulting density of defects to the exact, continuous-time result.
The figure shows that our choice of $\Delta = 0.5$, despite minor deviations from the continuum solution, provides a stable evolution over the entire time range we consider.
This is not the case for larger time steps as can be seen by the increased fluctuations.
Crucially, this time step allows us to experimentally probe relevant regimes of $\tf$.
More concretely, the \gls{kz} scaling we seek to benchmark against is not present for small annealing times $\tf$, and observing the \gls{kzm} requires probing sufficiently large $\tf$.
However, this is not a limitation of the proposed method.
Rather, it is a limitation of the current hardware but as the hardware improves and longer circuits, i.e., more Trotter steps, can be reliably executed, smaller time steps can be adopted to improve the accuracy at smaller $\tf$.

\section{Error mitigation and suppression}
\label{app: error mitigation}

Error suppression refers to methods that actively reduce errors at the circuit level, while error mitigation refers to methods that invert errors in post-processing.
We employ some of the most established techniques and provide brief introductions to each one of them in this appendix.
More detailed descriptions can be found in the respective references.

\subsection{Readout error mitigation}
We employ two methods to mitigate readout errors.
When expectation values are evaluated directly, i.e., through \qiskit{}'s Estimator primitive without evaluating single-measurement bitstrings, \gls{trex} is used~\cite{Berg2022trex}.
\gls{trex} twirls the readout with $X$ gates and averages a measured expectation value over these twirls. 
It can therefore only mitigate readout errors in expectation values, and not on individual bitstrings.
That is why, for optimization problems, where individual bitstrings are sampled and processed (with \qiskit{}'s Sampler primitive), we use \gls{m3} instead of \gls{trex}~\cite{Nation2021m3}.

\subsection{Dynamical decoupling}
\gls{dd} is an error suppression technique that inserts quantum gate sequences during qubit idle times.
\gls{dd} sequences are designed to remove certain system-environment interactions.
There exist a plethora of possible \gls{dd} sequences canceling different errors or canceling errors to different orders~\cite{Ezzell2023dd}.
In all experiments with \gls{dd}, we used a staggered XY4 sequence~\cite{Barron2023qaoa}, $Y_0 \tau Y_1 \tau X_0 \tau X_1 \tau Y_0 \tau Y_1 \tau X_0 \tau X_1 \tau$, with $Y_i$ acting on qubit $i$ and $\tau$ being a delay duration.
The $\mathrm{XY4} = Y\tau X \tau Y \tau X \tau$ sequence universally cancels all single qubit interactions up to first order. 
By staggering them, i.e., introducing a relative shift between the sequences on neighboring qubits, we also cancel static $ZZ$ cross-talk~\cite{Barron2023qaoa}.

\begin{figure}[t]
    \includegraphics{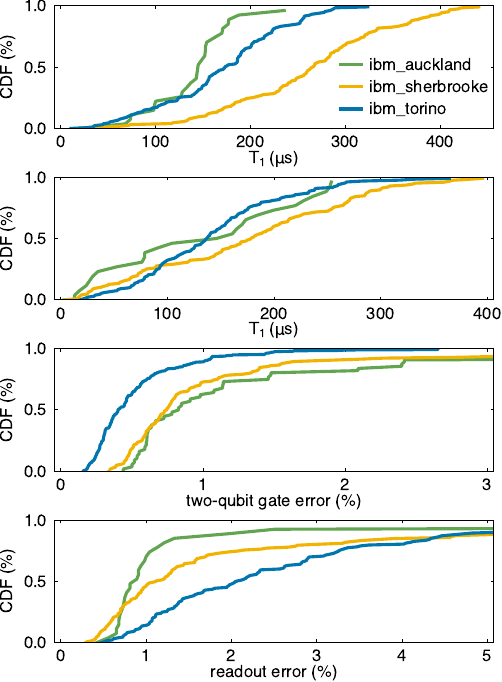}
    \caption{
        \textbf{Backend Properties.}
        Cumulative distribution functions of the decoherence times $\mathrm{T}_1$ and $\mathrm{T}_2$, two-qubit gate and readout errors for the three devices used in this work, \torino{}, \sherbrooke{}, and \auckland{}.
    }
    \label{fig: hardware properties}
\end{figure}

\subsection{Pulse-efficient transpilation}

By default, parameterized two-qubit gates such as $\rzz (\theta)$ are transpiled to a parameterized single-qubit $R_Z(\theta)$ sandwiched between two maximally entangling, hardware-native two-qubit gates.
However, if the backend exposes to the user ECR gates as underlying or being its native two-qubit gate and allows for pulse-level access, such gates can be transpiled to significantly shorter final pulse sequences than obtained with standard transpilation.
This leads to an overall reduction of the circuit duration.
Details can be found in~\cite{Earnest2023pulse} but, in essence, this is achieved by scaling the area of the ECR pulses as a function of the gate parameter $\theta$.
In our application, having time-dependent annealing schedules, two-qubit gate parameters start off being small at the beginning of our Trotter circuit and grow towards the end.
For our circuits, pulse-efficient transpilation yields a reduction in pulse duration of $\sim 40\%$ on average.

When the gate angle and the corresponding pulse area become so small that not only the pulse width but also its amplitude are adjusted, a fine amplitude calibration becomes necessary~\cite{Vazquez2023wellConditioned}.
This is due to non-linearities in the cross-resonance gate with respect to pulse amplitude.

\section{Hardware properties and qubit selection}
\label{app: hardware properties}

\begin{figure}[t]
    \includegraphics{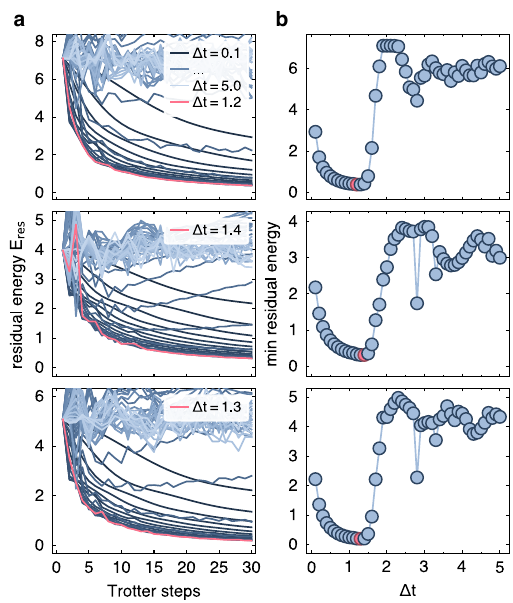}
    \caption{
        \textbf{Statevector simulations of the residual energy dependence on time step and spectral gap.}
        \textbf{a} Residual energy obtained from statevector simulations of QA with fixed time steps $\Delta t \in \{ 0.1, \ldots, 5.0 \}$ as a function of circuit depth, i.e., number of time steps.
        Each row corresponds to the respective spectrum in \Cref{fig: residual energy auckland}.
        \textbf{b} Minimum residual energy from \textbf{a} as a function of the time step with fixed depth, i.e., each point corresponds to the minimum over one curve in \textbf{a}.
    }
    \label{fig: residual energy statevector}
\end{figure}

For completeness, we report the decoherence times $\tone$ and $\ttwo$, the two-qubit gate error, and the readout error as reported for the quantum processors on which we execute the circuits, see \Cref{fig: hardware properties}.
These properties were accessed on 13$^\text{th}$ February 2024 for \torino{}, on 22$^\text{nd}$ February 2024 for \sherbrooke{}, and on 6$^\text{th}$ November 2023 for \auckland{}.
They are indicative of the device's performance of when the corresponding experiments in the main text were executed, even though the data reported in the main text were gathered on several different days
The 100-qubit line in \Cref{fig: kzm device 1D} was chosen by computing the cumulative two-qubit gate error along all 100-qubit lines on the respective processor and choosing the one with the smallest error.

\section{Quantum annealing with very large time steps}
\label{app: optimization statevector}

\begin{figure}[t]
    \includegraphics{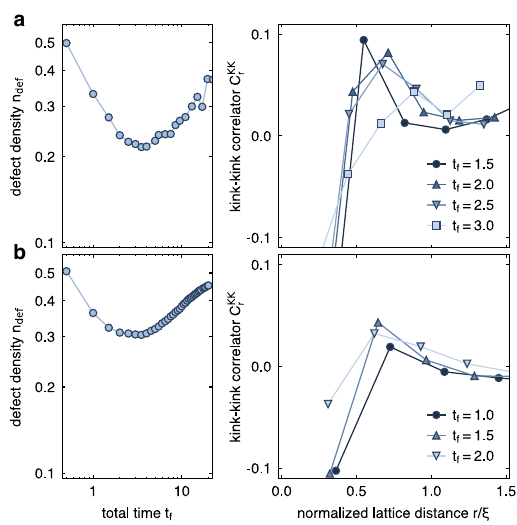}
    \caption{
        \textbf{Kink-kink correlation functions for different final annealing times $\tf$.}
        \textbf{a} Density of defects (left) for a periodic 12-qubit chain on \auckland{} using REM and pulse-efficient transpilation and corresponding kink-kink correlators (right).
        \textbf{b} Density of defects (left) of an open 100-qubit chain on \sherbrooke{} using no EMS and corresponding kink-kink correlators (right).
    }
    \label{fig: kink-kink correlator}
\end{figure}

Here, we confirm that the optimal time step for digitized \gls{qa} for optimization is $\Delta t > 1$ through ideal statevector simulations.
\Cref{fig: residual energy statevector} shows the results for the same systems as in \Cref{fig: residual energy auckland} from the main text, that is, instances of disordered couplings $J_{ij} \in [-1, 1]$.
When considering a fixed number of time steps or circuit layers, the minimum residual energy obtained from up to 30 time steps decreases with growing time step size up to an optimal time step $1.2 < \Delta t < 1.4$ (depending on the system), before sharply increasing, see \Cref{fig: residual energy statevector}.
Increasing the time step even further does not yield any benefit whatsoever and, since it induces significant algorithmic errors, results in randomly fluctuating residual energies with increasing circuit depth.

\section{Kink-kink correlator}
\label{app: kink-kink correlator}

We investigate the correlation between different defects in spin chains after digitized quantum annealing, also termed kink-kink correlation in the literature~\cite{king2022coherent,Cincio2007entropy}.
Given uniform couplings $J$, the solution after annealing is ferromagnetic.
In this setting, defects are misalignments of spins on edges $i$ between lattice sites, and the correlator between defects $i$ and $i + r$ is defined as
\begin{equation}
    C_r^\mathrm{KK} = \frac{1}{N_{\mathrm{e}, r}} \sum_{i=1}^{N_{\mathrm{e}, r}} \frac{\braket{K_i K_{i + r}} - \ndef^2}{\ndef^2} \ ,
\end{equation}
where $K_i = 1 - \sigma_i^z \sigma_{i+1}^z$ measures whether there is a defect, i.e. a spin-flip, between sites $i$ and $i+1$, and $N_{\mathrm{e}, r}$ is the number of edges on the graph between edges $i$ and $i+r$.

\Cref{fig: kink-kink correlator} shows the density of defects (left) of annealing a periodic 12-qubit chain on \auckland{} (\textbf{a)} using \gls{rem} and pulse-efficient transpilation and of a 100-qubit chain on \sherbrooke{} (\textbf{b}) using no \gls{ems}.
The corresponding kink-kink correlators for different $\tf$ are shown in the respective right panel as a function of the normalized lattice distance $r/\xi$ with $1/\xi = \ndef$.
The existence of a positive peak, which we observe between $r/\xi = 0.5$ and $1.0$, is expected from theoretical results\cite{Nowak2021quantum,Dziarmaga2022kinkcorrelations,Roychowdhury2021}.

\bibliography{refs.bib}

\begin{thebibliography}{110}%
\makeatletter
\providecommand \@ifxundefined [1]{%
 \@ifx{#1\undefined}
}%
\providecommand \@ifnum [1]{%
 \ifnum #1\expandafter \@firstoftwo
 \else \expandafter \@secondoftwo
 \fi
}%
\providecommand \@ifx [1]{%
 \ifx #1\expandafter \@firstoftwo
 \else \expandafter \@secondoftwo
 \fi
}%
\providecommand \natexlab [1]{#1}%
\providecommand \enquote  [1]{``#1''}%
\providecommand \bibnamefont  [1]{#1}%
\providecommand \bibfnamefont [1]{#1}%
\providecommand \citenamefont [1]{#1}%
\providecommand \href@noop [0]{\@secondoftwo}%
\providecommand \href [0]{\begingroup \@sanitize@url \@href}%
\providecommand \@href[1]{\@@startlink{#1}\@@href}%
\providecommand \@@href[1]{\endgroup#1\@@endlink}%
\providecommand \@sanitize@url [0]{\catcode `\\12\catcode `\$12\catcode `\&12\catcode `\#12\catcode `\^12\catcode `\_12\catcode `\%12\relax}%
\providecommand \@@startlink[1]{}%
\providecommand \@@endlink[0]{}%
\providecommand \url  [0]{\begingroup\@sanitize@url \@url }%
\providecommand \@url [1]{\endgroup\@href {#1}{\urlprefix }}%
\providecommand \urlprefix  [0]{URL }%
\providecommand \Eprint [0]{\href }%
\providecommand \doibase [0]{https://doi.org/}%
\providecommand \selectlanguage [0]{\@gobble}%
\providecommand \bibinfo  [0]{\@secondoftwo}%
\providecommand \bibfield  [0]{\@secondoftwo}%
\providecommand \translation [1]{[#1]}%
\providecommand \BibitemOpen [0]{}%
\providecommand \bibitemStop [0]{}%
\providecommand \bibitemNoStop [0]{.\EOS\space}%
\providecommand \EOS [0]{\spacefactor3000\relax}%
\providecommand \BibitemShut  [1]{\csname bibitem#1\endcsname}%
\let\auto@bib@innerbib\@empty
\bibitem [{\citenamefont {Stanley}(1971)}]{stanley1971phase}%
  \BibitemOpen
  \bibfield  {author} {\bibinfo {author} {\bibfnamefont {H.~E.}\ \bibnamefont {Stanley}},\ }\href@noop {} {\emph {\bibinfo {title} {Phase transitions and critical phenomena}}},\ Vol.~\bibinfo {volume} {7}\ (\bibinfo  {publisher} {Clarendon Press, Oxford},\ \bibinfo {year} {1971})\BibitemShut {NoStop}%
\bibitem [{\citenamefont {Sachdev}(2011)}]{sachdev_2011}%
  \BibitemOpen
  \bibfield  {author} {\bibinfo {author} {\bibfnamefont {S.}~\bibnamefont {Sachdev}},\ }\href {https://doi.org/10.1017/CBO9780511973765} {\emph {\bibinfo {title} {Quantum Phase Transitions}}},\ \bibinfo {edition} {2nd}\ ed.\ (\bibinfo  {publisher} {Cambridge University Press},\ \bibinfo {year} {2011})\BibitemShut {NoStop}%
\bibitem [{\citenamefont {Sondhi}\ \emph {et~al.}(1997)\citenamefont {Sondhi}, \citenamefont {Girvin}, \citenamefont {Carini},\ and\ \citenamefont {Shahar}}]{sondhi1997continuous}%
  \BibitemOpen
  \bibfield  {author} {\bibinfo {author} {\bibfnamefont {S.~L.}\ \bibnamefont {Sondhi}}, \bibinfo {author} {\bibfnamefont {S.~M.}\ \bibnamefont {Girvin}}, \bibinfo {author} {\bibfnamefont {J.~P.}\ \bibnamefont {Carini}},\ and\ \bibinfo {author} {\bibfnamefont {D.}~\bibnamefont {Shahar}},\ }\bibfield  {title} {\bibinfo {title} {Continuous quantum phase transitions},\ }\href {https://doi.org/10.1103/RevModPhys.69.315} {\bibfield  {journal} {\bibinfo  {journal} {Rev. Mod. Phys.}\ }\textbf {\bibinfo {volume} {69}},\ \bibinfo {pages} {315} (\bibinfo {year} {1997})}\BibitemShut {NoStop}%
\bibitem [{\citenamefont {Feynman}(1982)}]{Feynman1982}%
  \BibitemOpen
  \bibfield  {author} {\bibinfo {author} {\bibfnamefont {R.~P.}\ \bibnamefont {Feynman}},\ }\bibfield  {title} {\bibinfo {title} {Simulating physics with computers},\ }\href {https://doi.org/10.1007/BF02650179} {\bibfield  {journal} {\bibinfo  {journal} {Int. J. Theor. Phys.}\ }\textbf {\bibinfo {volume} {21}},\ \bibinfo {pages} {467} (\bibinfo {year} {1982})}\BibitemShut {NoStop}%
\bibitem [{\citenamefont {Georgescu}\ \emph {et~al.}(2014)\citenamefont {Georgescu}, \citenamefont {Ashhab},\ and\ \citenamefont {Nori}}]{Georgescu2014}%
  \BibitemOpen
  \bibfield  {author} {\bibinfo {author} {\bibfnamefont {I.~M.}\ \bibnamefont {Georgescu}}, \bibinfo {author} {\bibfnamefont {S.}~\bibnamefont {Ashhab}},\ and\ \bibinfo {author} {\bibfnamefont {F.}~\bibnamefont {Nori}},\ }\bibfield  {title} {\bibinfo {title} {Quantum simulation},\ }\href {https://doi.org/10.1103/RevModPhys.86.153} {\bibfield  {journal} {\bibinfo  {journal} {Rev. Mod. Phys.}\ }\textbf {\bibinfo {volume} {86}},\ \bibinfo {pages} {153} (\bibinfo {year} {2014})}\BibitemShut {NoStop}%
\bibitem [{\citenamefont {Altman}\ \emph {et~al.}(2021)\citenamefont {Altman}, \citenamefont {Brown}, \citenamefont {Carleo}, \citenamefont {Carr}, \citenamefont {Demler}, \citenamefont {Chin}, \citenamefont {DeMarco}, \citenamefont {Economou}, \citenamefont {Eriksson}, \citenamefont {Fu} \emph {et~al.}}]{Altmann2021}%
  \BibitemOpen
  \bibfield  {author} {\bibinfo {author} {\bibfnamefont {E.}~\bibnamefont {Altman}}, \bibinfo {author} {\bibfnamefont {K.~R.}\ \bibnamefont {Brown}}, \bibinfo {author} {\bibfnamefont {G.}~\bibnamefont {Carleo}}, \bibinfo {author} {\bibfnamefont {L.~D.}\ \bibnamefont {Carr}}, \bibinfo {author} {\bibfnamefont {E.}~\bibnamefont {Demler}}, \bibinfo {author} {\bibfnamefont {C.}~\bibnamefont {Chin}}, \bibinfo {author} {\bibfnamefont {B.}~\bibnamefont {DeMarco}}, \bibinfo {author} {\bibfnamefont {S.~E.}\ \bibnamefont {Economou}}, \bibinfo {author} {\bibfnamefont {M.~A.}\ \bibnamefont {Eriksson}}, \bibinfo {author} {\bibfnamefont {K.-M.~C.}\ \bibnamefont {Fu}}, \emph {et~al.},\ }\bibfield  {title} {\bibinfo {title} {Quantum simulators: Architectures and opportunities},\ }\href {https://doi.org/10.1103/PRXQuantum.2.017003} {\bibfield  {journal} {\bibinfo  {journal} {PRX Quantum}\ }\textbf {\bibinfo {volume} {2}},\ \bibinfo {pages} {017003} (\bibinfo {year} {2021})}\BibitemShut {NoStop}%
\bibitem [{\citenamefont {Monroe}\ \emph {et~al.}(2021)\citenamefont {Monroe}, \citenamefont {Campbell}, \citenamefont {Duan}, \citenamefont {Gong}, \citenamefont {Gorshkov}, \citenamefont {Hess}, \citenamefont {Islam}, \citenamefont {Kim}, \citenamefont {Linke}, \citenamefont {Pagano} \emph {et~al.}}]{Monroe2021}%
  \BibitemOpen
  \bibfield  {author} {\bibinfo {author} {\bibfnamefont {C.}~\bibnamefont {Monroe}}, \bibinfo {author} {\bibfnamefont {W.~C.}\ \bibnamefont {Campbell}}, \bibinfo {author} {\bibfnamefont {L.-M.}\ \bibnamefont {Duan}}, \bibinfo {author} {\bibfnamefont {Z.-X.}\ \bibnamefont {Gong}}, \bibinfo {author} {\bibfnamefont {A.~V.}\ \bibnamefont {Gorshkov}}, \bibinfo {author} {\bibfnamefont {P.~W.}\ \bibnamefont {Hess}}, \bibinfo {author} {\bibfnamefont {R.}~\bibnamefont {Islam}}, \bibinfo {author} {\bibfnamefont {K.}~\bibnamefont {Kim}}, \bibinfo {author} {\bibfnamefont {N.~M.}\ \bibnamefont {Linke}}, \bibinfo {author} {\bibfnamefont {G.}~\bibnamefont {Pagano}}, \emph {et~al.},\ }\bibfield  {title} {\bibinfo {title} {Programmable quantum simulations of spin systems with trapped ions},\ }\href {https://doi.org/10.1103/RevModPhys.93.025001} {\bibfield  {journal} {\bibinfo  {journal} {Rev. Mod. Phys.}\ }\textbf {\bibinfo {volume} {93}},\ \bibinfo {pages} {025001} (\bibinfo {year} {2021})}\BibitemShut {NoStop}%
\bibitem [{\citenamefont {Jaksch}\ \emph {et~al.}(1998)\citenamefont {Jaksch}, \citenamefont {Bruder}, \citenamefont {Cirac}, \citenamefont {Gardiner},\ and\ \citenamefont {Zoller}}]{Jaksch1998}%
  \BibitemOpen
  \bibfield  {author} {\bibinfo {author} {\bibfnamefont {D.}~\bibnamefont {Jaksch}}, \bibinfo {author} {\bibfnamefont {C.}~\bibnamefont {Bruder}}, \bibinfo {author} {\bibfnamefont {J.~I.}\ \bibnamefont {Cirac}}, \bibinfo {author} {\bibfnamefont {C.~W.}\ \bibnamefont {Gardiner}},\ and\ \bibinfo {author} {\bibfnamefont {P.}~\bibnamefont {Zoller}},\ }\bibfield  {title} {\bibinfo {title} {Cold bosonic atoms in optical lattices},\ }\href {https://doi.org/10.1103/PhysRevLett.81.3108} {\bibfield  {journal} {\bibinfo  {journal} {Phys. Rev. Lett.}\ }\textbf {\bibinfo {volume} {81}},\ \bibinfo {pages} {3108} (\bibinfo {year} {1998})}\BibitemShut {NoStop}%
\bibitem [{\citenamefont {Lewenstein}\ \emph {et~al.}(2012)\citenamefont {Lewenstein}, \citenamefont {Sanpera},\ and\ \citenamefont {Ahufinger}}]{lewenstein2012ultracold}%
  \BibitemOpen
  \bibfield  {author} {\bibinfo {author} {\bibfnamefont {M.}~\bibnamefont {Lewenstein}}, \bibinfo {author} {\bibfnamefont {A.}~\bibnamefont {Sanpera}},\ and\ \bibinfo {author} {\bibfnamefont {V.}~\bibnamefont {Ahufinger}},\ }\href {https://academic.oup.com/book/26218?searchresult=1&utm_source=TrendMD&utm_medium=cpc&utm_campaign=Oxford_Academic_Books_TrendMD_1} {\emph {\bibinfo {title} {Ultracold Atoms in Optical Lattices: Simulating quantum many-body systems}}}\ (\bibinfo  {publisher} {OUP Oxford},\ \bibinfo {year} {2012})\BibitemShut {NoStop}%
\bibitem [{\citenamefont {Bernien}\ \emph {et~al.}(2017)\citenamefont {Bernien}, \citenamefont {Schwartz}, \citenamefont {Keesling}, \citenamefont {Levine}, \citenamefont {Omran}, \citenamefont {Pichler}, \citenamefont {Choi}, \citenamefont {Zibrov}, \citenamefont {Endres}, \citenamefont {Greiner} \emph {et~al.}}]{bernien2017probing}%
  \BibitemOpen
  \bibfield  {author} {\bibinfo {author} {\bibfnamefont {H.}~\bibnamefont {Bernien}}, \bibinfo {author} {\bibfnamefont {S.}~\bibnamefont {Schwartz}}, \bibinfo {author} {\bibfnamefont {A.}~\bibnamefont {Keesling}}, \bibinfo {author} {\bibfnamefont {H.}~\bibnamefont {Levine}}, \bibinfo {author} {\bibfnamefont {A.}~\bibnamefont {Omran}}, \bibinfo {author} {\bibfnamefont {H.}~\bibnamefont {Pichler}}, \bibinfo {author} {\bibfnamefont {S.}~\bibnamefont {Choi}}, \bibinfo {author} {\bibfnamefont {A.~S.}\ \bibnamefont {Zibrov}}, \bibinfo {author} {\bibfnamefont {M.}~\bibnamefont {Endres}}, \bibinfo {author} {\bibfnamefont {M.}~\bibnamefont {Greiner}}, \emph {et~al.},\ }\bibfield  {title} {\bibinfo {title} {Probing many-body dynamics on a 51-atom quantum simulator},\ }\href {https://doi.org/10.1038/nature24622} {\bibfield  {journal} {\bibinfo  {journal} {Nature}\ }\textbf {\bibinfo {volume} {551}},\ \bibinfo {pages} {579} (\bibinfo {year} {2017})}\BibitemShut {NoStop}%
\bibitem [{\citenamefont {Scholl}\ \emph {et~al.}(2021)\citenamefont {Scholl}, \citenamefont {Schuler}, \citenamefont {Williams}, \citenamefont {Eberharter}, \citenamefont {Barredo}, \citenamefont {Schymik}, \citenamefont {Lienhard}, \citenamefont {Henry}, \citenamefont {Lang}, \citenamefont {Lahaye} \emph {et~al.}}]{scholl2021quantum}%
  \BibitemOpen
  \bibfield  {author} {\bibinfo {author} {\bibfnamefont {P.}~\bibnamefont {Scholl}}, \bibinfo {author} {\bibfnamefont {M.}~\bibnamefont {Schuler}}, \bibinfo {author} {\bibfnamefont {H.~J.}\ \bibnamefont {Williams}}, \bibinfo {author} {\bibfnamefont {A.~A.}\ \bibnamefont {Eberharter}}, \bibinfo {author} {\bibfnamefont {D.}~\bibnamefont {Barredo}}, \bibinfo {author} {\bibfnamefont {K.-N.}\ \bibnamefont {Schymik}}, \bibinfo {author} {\bibfnamefont {V.}~\bibnamefont {Lienhard}}, \bibinfo {author} {\bibfnamefont {L.-P.}\ \bibnamefont {Henry}}, \bibinfo {author} {\bibfnamefont {T.~C.}\ \bibnamefont {Lang}}, \bibinfo {author} {\bibfnamefont {T.}~\bibnamefont {Lahaye}}, \emph {et~al.},\ }\bibfield  {title} {\bibinfo {title} {{Quantum simulation of 2D antiferromagnets with hundreds of Rydberg atoms}},\ }\href {https://doi.org/10.1038/s41586-021-03585-1} {\bibfield  {journal} {\bibinfo  {journal} {Nature}\ }\textbf {\bibinfo {volume} {595}},\ \bibinfo {pages} {233} (\bibinfo {year} {2021})}\BibitemShut {NoStop}%
\bibitem [{\citenamefont {Miessen}\ \emph {et~al.}(2023)\citenamefont {Miessen}, \citenamefont {Ollitrault}, \citenamefont {Tacchino},\ and\ \citenamefont {Tavernelli}}]{Miessen2023perspective}%
  \BibitemOpen
  \bibfield  {author} {\bibinfo {author} {\bibfnamefont {A.}~\bibnamefont {Miessen}}, \bibinfo {author} {\bibfnamefont {P.~J.}\ \bibnamefont {Ollitrault}}, \bibinfo {author} {\bibfnamefont {F.}~\bibnamefont {Tacchino}},\ and\ \bibinfo {author} {\bibfnamefont {I.}~\bibnamefont {Tavernelli}},\ }\bibfield  {title} {\bibinfo {title} {Quantum algorithms for quantum dynamics},\ }\href {https://doi.org/10.1038/s43588-022-00374-2} {\bibfield  {journal} {\bibinfo  {journal} {Nat. Comput. Sci.}\ }\textbf {\bibinfo {volume} {3}},\ \bibinfo {pages} {25} (\bibinfo {year} {2023})}\BibitemShut {NoStop}%
\bibitem [{\citenamefont {Abrams}\ and\ \citenamefont {Lloyd}(1999)}]{abrams1999quantum}%
  \BibitemOpen
  \bibfield  {author} {\bibinfo {author} {\bibfnamefont {D.~S.}\ \bibnamefont {Abrams}}\ and\ \bibinfo {author} {\bibfnamefont {S.}~\bibnamefont {Lloyd}},\ }\bibfield  {title} {\bibinfo {title} {Quantum algorithm providing exponential speed increase for finding eigenvalues and eigenvectors},\ }\href {https://doi.org/10.1103/PhysRevLett.83.5162} {\bibfield  {journal} {\bibinfo  {journal} {Phys. Rev. Lett.}\ }\textbf {\bibinfo {volume} {83}},\ \bibinfo {pages} {5162} (\bibinfo {year} {1999})}\BibitemShut {NoStop}%
\bibitem [{\citenamefont {Kim}\ \emph {et~al.}(2023)\citenamefont {Kim}, \citenamefont {Eddins}, \citenamefont {Anand}, \citenamefont {Wei}, \citenamefont {van~den Berg}, \citenamefont {Rosenblatt}, \citenamefont {Nayfeh}, \citenamefont {Wu}, \citenamefont {Zaletel}, \citenamefont {Temme},\ and\ \citenamefont {Kandala}}]{Kim2023utility}%
  \BibitemOpen
  \bibfield  {author} {\bibinfo {author} {\bibfnamefont {Y.}~\bibnamefont {Kim}}, \bibinfo {author} {\bibfnamefont {A.}~\bibnamefont {Eddins}}, \bibinfo {author} {\bibfnamefont {S.}~\bibnamefont {Anand}}, \bibinfo {author} {\bibfnamefont {K.~X.}\ \bibnamefont {Wei}}, \bibinfo {author} {\bibfnamefont {E.}~\bibnamefont {van~den Berg}}, \bibinfo {author} {\bibfnamefont {S.}~\bibnamefont {Rosenblatt}}, \bibinfo {author} {\bibfnamefont {H.}~\bibnamefont {Nayfeh}}, \bibinfo {author} {\bibfnamefont {Y.}~\bibnamefont {Wu}}, \bibinfo {author} {\bibfnamefont {M.}~\bibnamefont {Zaletel}}, \bibinfo {author} {\bibfnamefont {K.}~\bibnamefont {Temme}},\ and\ \bibinfo {author} {\bibfnamefont {A.}~\bibnamefont {Kandala}},\ }\bibfield  {title} {\bibinfo {title} {Evidence for the utility of quantum computing before fault tolerance},\ }\href {https://doi.org/10.1038/s41586-023-06096-3} {\bibfield  {journal} {\bibinfo  {journal} {Nature}\ }\textbf {\bibinfo {volume} {618}},\ \bibinfo {pages} {500} (\bibinfo {year}
  {2023})}\BibitemShut {NoStop}%
\bibitem [{\citenamefont {Mi}\ \emph {et~al.}(2022)\citenamefont {Mi}, \citenamefont {Ippoliti}, \citenamefont {Quintana}, \citenamefont {Greene}, \citenamefont {Chen}, \citenamefont {Gross}, \citenamefont {Arute}, \citenamefont {Arya}, \citenamefont {Atalaya}, \citenamefont {Babbush} \emph {et~al.}}]{Mi2022timeCrystals}%
  \BibitemOpen
  \bibfield  {author} {\bibinfo {author} {\bibfnamefont {X.}~\bibnamefont {Mi}}, \bibinfo {author} {\bibfnamefont {M.}~\bibnamefont {Ippoliti}}, \bibinfo {author} {\bibfnamefont {C.}~\bibnamefont {Quintana}}, \bibinfo {author} {\bibfnamefont {A.}~\bibnamefont {Greene}}, \bibinfo {author} {\bibfnamefont {Z.}~\bibnamefont {Chen}}, \bibinfo {author} {\bibfnamefont {J.}~\bibnamefont {Gross}}, \bibinfo {author} {\bibfnamefont {F.}~\bibnamefont {Arute}}, \bibinfo {author} {\bibfnamefont {K.}~\bibnamefont {Arya}}, \bibinfo {author} {\bibfnamefont {J.}~\bibnamefont {Atalaya}}, \bibinfo {author} {\bibfnamefont {R.}~\bibnamefont {Babbush}}, \emph {et~al.},\ }\bibfield  {title} {\bibinfo {title} {Time-crystalline eigenstate order on a quantum processor},\ }\href {https://doi.org/10.1038/s41586-021-04257-w} {\bibfield  {journal} {\bibinfo  {journal} {Nature}\ }\textbf {\bibinfo {volume} {601}},\ \bibinfo {pages} {531} (\bibinfo {year} {2022})}\BibitemShut {NoStop}%
\bibitem [{\citenamefont {Keenan}\ \emph {et~al.}(2023)\citenamefont {Keenan}, \citenamefont {Robertson}, \citenamefont {Murphy}, \citenamefont {Zhuk},\ and\ \citenamefont {Goold}}]{Keenan2023KPZscaling}%
  \BibitemOpen
  \bibfield  {author} {\bibinfo {author} {\bibfnamefont {N.}~\bibnamefont {Keenan}}, \bibinfo {author} {\bibfnamefont {N.~F.}\ \bibnamefont {Robertson}}, \bibinfo {author} {\bibfnamefont {T.}~\bibnamefont {Murphy}}, \bibinfo {author} {\bibfnamefont {S.}~\bibnamefont {Zhuk}},\ and\ \bibinfo {author} {\bibfnamefont {J.}~\bibnamefont {Goold}},\ }\bibfield  {title} {\bibinfo {title} {Evidence of kardar-parisi-zhang scaling on a digital quantum simulator},\ }\href {https://doi.org/10.1038/s41534-023-00742-4} {\bibfield  {journal} {\bibinfo  {journal} {npj Quantum Inf.}\ }\textbf {\bibinfo {volume} {9}},\ \bibinfo {pages} {72} (\bibinfo {year} {2023})}\BibitemShut {NoStop}%
\bibitem [{\citenamefont {Miessen}\ \emph {et~al.}(2021)\citenamefont {Miessen}, \citenamefont {Ollitrault},\ and\ \citenamefont {Tavernelli}}]{Miessen2021}%
  \BibitemOpen
  \bibfield  {author} {\bibinfo {author} {\bibfnamefont {A.}~\bibnamefont {Miessen}}, \bibinfo {author} {\bibfnamefont {P.~J.}\ \bibnamefont {Ollitrault}},\ and\ \bibinfo {author} {\bibfnamefont {I.}~\bibnamefont {Tavernelli}},\ }\bibfield  {title} {\bibinfo {title} {Quantum algorithms for quantum dynamics: A performance study on the spin-boson model},\ }\href {https://doi.org/10.1103/PhysRevResearch.3.043212} {\bibfield  {journal} {\bibinfo  {journal} {Phys. Rev. Res.}\ }\textbf {\bibinfo {volume} {3}},\ \bibinfo {pages} {043212} (\bibinfo {year} {2021})}\BibitemShut {NoStop}%
\bibitem [{\citenamefont {Bravyi}\ \emph {et~al.}(2024)\citenamefont {Bravyi}, \citenamefont {Cross}, \citenamefont {Gambetta}, \citenamefont {Maslov}, \citenamefont {Rall},\ and\ \citenamefont {Yoder}}]{Bravyi2024}%
  \BibitemOpen
  \bibfield  {author} {\bibinfo {author} {\bibfnamefont {S.}~\bibnamefont {Bravyi}}, \bibinfo {author} {\bibfnamefont {A.~W.}\ \bibnamefont {Cross}}, \bibinfo {author} {\bibfnamefont {J.~M.}\ \bibnamefont {Gambetta}}, \bibinfo {author} {\bibfnamefont {D.}~\bibnamefont {Maslov}}, \bibinfo {author} {\bibfnamefont {P.}~\bibnamefont {Rall}},\ and\ \bibinfo {author} {\bibfnamefont {T.~J.}\ \bibnamefont {Yoder}},\ }\bibfield  {title} {\bibinfo {title} {High-threshold and low-overhead fault-tolerant quantum memory},\ }\href {https://doi.org/10.1038/s41586-024-07107-7} {\bibfield  {journal} {\bibinfo  {journal} {Nature}\ }\textbf {\bibinfo {volume} {627}},\ \bibinfo {pages} {778} (\bibinfo {year} {2024})}\BibitemShut {NoStop}%
\bibitem [{\citenamefont {Daley}\ \emph {et~al.}(2022)\citenamefont {Daley}, \citenamefont {Bloch}, \citenamefont {Kokail}, \citenamefont {Flannigan}, \citenamefont {Pearson}, \citenamefont {Troyer},\ and\ \citenamefont {Zoller}}]{daley2022practical}%
  \BibitemOpen
  \bibfield  {author} {\bibinfo {author} {\bibfnamefont {A.~J.}\ \bibnamefont {Daley}}, \bibinfo {author} {\bibfnamefont {I.}~\bibnamefont {Bloch}}, \bibinfo {author} {\bibfnamefont {C.}~\bibnamefont {Kokail}}, \bibinfo {author} {\bibfnamefont {S.}~\bibnamefont {Flannigan}}, \bibinfo {author} {\bibfnamefont {N.}~\bibnamefont {Pearson}}, \bibinfo {author} {\bibfnamefont {M.}~\bibnamefont {Troyer}},\ and\ \bibinfo {author} {\bibfnamefont {P.}~\bibnamefont {Zoller}},\ }\bibfield  {title} {\bibinfo {title} {Practical quantum advantage in quantum simulation},\ }\href {https://doi.org/10.1038/s41586-022-04940-6} {\bibfield  {journal} {\bibinfo  {journal} {Nature}\ }\textbf {\bibinfo {volume} {607}},\ \bibinfo {pages} {667} (\bibinfo {year} {2022})}\BibitemShut {NoStop}%
\bibitem [{\citenamefont {Abbas}\ \emph {et~al.}(2023)\citenamefont {Abbas} \emph {et~al.}}]{Abbas2023optimizationWorkingGroup}%
  \BibitemOpen
  \bibfield  {author} {\bibinfo {author} {\bibfnamefont {A.}~\bibnamefont {Abbas}} \emph {et~al.},\ }\bibfield  {title} {\bibinfo {title} {Quantum optimization: Potential, challenges, and the path forward},\ }\href {http://arxiv.org/abs/2312.02279} {\bibfield  {journal} {\bibinfo  {journal} {arXiv}\ } (\bibinfo {year} {2023})},\ \Eprint {https://arxiv.org/abs/2312.02279} {arXiv:2312.02279 [quant-ph]} \BibitemShut {NoStop}%
\bibitem [{\citenamefont {Albash}\ and\ \citenamefont {Lidar}(2018{\natexlab{a}})}]{Albash2018annealingRMP}%
  \BibitemOpen
  \bibfield  {author} {\bibinfo {author} {\bibfnamefont {T.}~\bibnamefont {Albash}}\ and\ \bibinfo {author} {\bibfnamefont {D.~A.}\ \bibnamefont {Lidar}},\ }\bibfield  {title} {\bibinfo {title} {Adiabatic quantum computation},\ }\href {https://doi.org/10.1103/RevModPhys.90.015002} {\bibfield  {journal} {\bibinfo  {journal} {Rev. Mod. Phys.}\ }\textbf {\bibinfo {volume} {90}},\ \bibinfo {pages} {015002} (\bibinfo {year} {2018}{\natexlab{a}})}\BibitemShut {NoStop}%
\bibitem [{\citenamefont {Kadowaki}\ and\ \citenamefont {Nishimori}(1998)}]{kadowaki1998quantum}%
  \BibitemOpen
  \bibfield  {author} {\bibinfo {author} {\bibfnamefont {T.}~\bibnamefont {Kadowaki}}\ and\ \bibinfo {author} {\bibfnamefont {H.}~\bibnamefont {Nishimori}},\ }\bibfield  {title} {\bibinfo {title} {Quantum annealing in the transverse ising model},\ }\href {https://doi.org/10.1103/PhysRevE.58.5355} {\bibfield  {journal} {\bibinfo  {journal} {Phys. Rev. E}\ }\textbf {\bibinfo {volume} {58}},\ \bibinfo {pages} {5355} (\bibinfo {year} {1998})}\BibitemShut {NoStop}%
\bibitem [{\citenamefont {Knysh}(2016)}]{knysh2016zero}%
  \BibitemOpen
  \bibfield  {author} {\bibinfo {author} {\bibfnamefont {S.}~\bibnamefont {Knysh}},\ }\bibfield  {title} {\bibinfo {title} {Zero-temperature quantum annealing bottlenecks in the spin-glass phase},\ }\href {https://doi.org/10.1038/ncomms12370} {\bibfield  {journal} {\bibinfo  {journal} {Nat. Commun.}\ }\textbf {\bibinfo {volume} {7}},\ \bibinfo {pages} {12370} (\bibinfo {year} {2016})}\BibitemShut {NoStop}%
\bibitem [{\citenamefont {Zurek}\ \emph {et~al.}(2005)\citenamefont {Zurek}, \citenamefont {Dorner},\ and\ \citenamefont {Zoller}}]{Zurek2005}%
  \BibitemOpen
  \bibfield  {author} {\bibinfo {author} {\bibfnamefont {W.~H.}\ \bibnamefont {Zurek}}, \bibinfo {author} {\bibfnamefont {U.}~\bibnamefont {Dorner}},\ and\ \bibinfo {author} {\bibfnamefont {P.}~\bibnamefont {Zoller}},\ }\bibfield  {title} {\bibinfo {title} {Dynamics of a quantum phase transition},\ }\href {https://doi.org/10.1103/PhysRevLett.95.105701} {\bibfield  {journal} {\bibinfo  {journal} {Phys. Rev. Lett.}\ }\textbf {\bibinfo {volume} {95}},\ \bibinfo {pages} {105701} (\bibinfo {year} {2005})}\BibitemShut {NoStop}%
\bibitem [{\citenamefont {Zeng}\ \emph {et~al.}(2023)\citenamefont {Zeng}, \citenamefont {Xia},\ and\ \citenamefont {del Campo}}]{Zeng2023universal}%
  \BibitemOpen
  \bibfield  {author} {\bibinfo {author} {\bibfnamefont {H.-B.}\ \bibnamefont {Zeng}}, \bibinfo {author} {\bibfnamefont {C.-Y.}\ \bibnamefont {Xia}},\ and\ \bibinfo {author} {\bibfnamefont {A.}~\bibnamefont {del Campo}},\ }\bibfield  {title} {\bibinfo {title} {Universal breakdown of kibble-zurek scaling in fast quenches across a phase transition},\ }\href {https://doi.org/10.1103/PhysRevLett.130.060402} {\bibfield  {journal} {\bibinfo  {journal} {Phys. Rev. Lett.}\ }\textbf {\bibinfo {volume} {130}},\ \bibinfo {pages} {060402} (\bibinfo {year} {2023})}\BibitemShut {NoStop}%
\bibitem [{\citenamefont {Kibble}(1980)}]{kibble1980some}%
  \BibitemOpen
  \bibfield  {author} {\bibinfo {author} {\bibfnamefont {T.~W.}\ \bibnamefont {Kibble}},\ }\bibfield  {title} {\bibinfo {title} {Some implications of a cosmological phase transition},\ }\href {https://doi.org/10.1016/0370-1573(80)90091-5} {\bibfield  {journal} {\bibinfo  {journal} {Phys. Rep.}\ }\textbf {\bibinfo {volume} {67}},\ \bibinfo {pages} {183} (\bibinfo {year} {1980})}\BibitemShut {NoStop}%
\bibitem [{\citenamefont {Zurek}(1985)}]{zurek1985cosmological}%
  \BibitemOpen
  \bibfield  {author} {\bibinfo {author} {\bibfnamefont {W.~H.}\ \bibnamefont {Zurek}},\ }\bibfield  {title} {\bibinfo {title} {Cosmological experiments in superfluid helium?},\ }\href {https://doi.org/10.1038/317505a0} {\bibfield  {journal} {\bibinfo  {journal} {Nature}\ }\textbf {\bibinfo {volume} {317}},\ \bibinfo {pages} {505} (\bibinfo {year} {1985})}\BibitemShut {NoStop}%
\bibitem [{\citenamefont {Bando}\ \emph {et~al.}(2020)\citenamefont {Bando}, \citenamefont {Susa}, \citenamefont {Oshiyama}, \citenamefont {Shibata}, \citenamefont {Ohzeki}, \citenamefont {G\'omez-Ruiz}, \citenamefont {Lidar}, \citenamefont {Suzuki}, \citenamefont {del Campo},\ and\ \citenamefont {Nishimori}}]{Bando2020}%
  \BibitemOpen
  \bibfield  {author} {\bibinfo {author} {\bibfnamefont {Y.}~\bibnamefont {Bando}}, \bibinfo {author} {\bibfnamefont {Y.}~\bibnamefont {Susa}}, \bibinfo {author} {\bibfnamefont {H.}~\bibnamefont {Oshiyama}}, \bibinfo {author} {\bibfnamefont {N.}~\bibnamefont {Shibata}}, \bibinfo {author} {\bibfnamefont {M.}~\bibnamefont {Ohzeki}}, \bibinfo {author} {\bibfnamefont {F.~J.}\ \bibnamefont {G\'omez-Ruiz}}, \bibinfo {author} {\bibfnamefont {D.~A.}\ \bibnamefont {Lidar}}, \bibinfo {author} {\bibfnamefont {S.}~\bibnamefont {Suzuki}}, \bibinfo {author} {\bibfnamefont {A.}~\bibnamefont {del Campo}},\ and\ \bibinfo {author} {\bibfnamefont {H.}~\bibnamefont {Nishimori}},\ }\bibfield  {title} {\bibinfo {title} {{Probing the universality of topological defect formation in a quantum annealer: Kibble-Zurek mechanism and beyond}},\ }\href {https://doi.org/10.1103/PhysRevResearch.2.033369} {\bibfield  {journal} {\bibinfo  {journal} {Phys. Rev. Research}\ }\textbf {\bibinfo {volume} {2}},\ \bibinfo {pages} {033369} (\bibinfo
  {year} {2020})}\BibitemShut {NoStop}%
\bibitem [{\citenamefont {Li}\ \emph {et~al.}(2023)\citenamefont {Li}, \citenamefont {Wu}, \citenamefont {Mei}, \citenamefont {Yao}, \citenamefont {Lian}, \citenamefont {Cai}, \citenamefont {Wang}, \citenamefont {Qi}, \citenamefont {Yao}, \citenamefont {He} \emph {et~al.}}]{Li2023probingLongRangeKZM}%
  \BibitemOpen
  \bibfield  {author} {\bibinfo {author} {\bibfnamefont {B.-W.}\ \bibnamefont {Li}}, \bibinfo {author} {\bibfnamefont {Y.-K.}\ \bibnamefont {Wu}}, \bibinfo {author} {\bibfnamefont {Q.-X.}\ \bibnamefont {Mei}}, \bibinfo {author} {\bibfnamefont {R.}~\bibnamefont {Yao}}, \bibinfo {author} {\bibfnamefont {W.-Q.}\ \bibnamefont {Lian}}, \bibinfo {author} {\bibfnamefont {M.-L.}\ \bibnamefont {Cai}}, \bibinfo {author} {\bibfnamefont {Y.}~\bibnamefont {Wang}}, \bibinfo {author} {\bibfnamefont {B.-X.}\ \bibnamefont {Qi}}, \bibinfo {author} {\bibfnamefont {L.}~\bibnamefont {Yao}}, \bibinfo {author} {\bibfnamefont {L.}~\bibnamefont {He}}, \emph {et~al.},\ }\bibfield  {title} {\bibinfo {title} {Probing critical behavior of long-range transverse-field ising model through quantum {Kibble-Zurek} mechanism},\ }\href {https://doi.org/10.1103/PRXQuantum.4.010302} {\bibfield  {journal} {\bibinfo  {journal} {PRX Quantum}\ }\textbf {\bibinfo {volume} {4}},\ \bibinfo {pages} {010302} (\bibinfo {year} {2023})}\BibitemShut {NoStop}%
\bibitem [{\citenamefont {Keesling}\ \emph {et~al.}(2019)\citenamefont {Keesling}, \citenamefont {Omran}, \citenamefont {Levine}, \citenamefont {Bernien}, \citenamefont {Pichler} \emph {et~al.}}]{keesling2019quantum}%
  \BibitemOpen
  \bibfield  {author} {\bibinfo {author} {\bibfnamefont {A.}~\bibnamefont {Keesling}}, \bibinfo {author} {\bibfnamefont {A.}~\bibnamefont {Omran}}, \bibinfo {author} {\bibfnamefont {H.}~\bibnamefont {Levine}}, \bibinfo {author} {\bibfnamefont {H.}~\bibnamefont {Bernien}}, \bibinfo {author} {\bibfnamefont {H.}~\bibnamefont {Pichler}}, \emph {et~al.},\ }\bibfield  {title} {\bibinfo {title} {Quantum kibble–zurek mechanism and critical dynamics on a programmable rydberg simulator},\ }\href {https://doi.org/10.1038/s41586-019-1070-1} {\bibfield  {journal} {\bibinfo  {journal} {Nature}\ }\textbf {\bibinfo {volume} {568}},\ \bibinfo {pages} {207} (\bibinfo {year} {2019})}\BibitemShut {NoStop}%
\bibitem [{\citenamefont {King}\ \emph {et~al.}(2022)\citenamefont {King}, \citenamefont {Suzuki}, \citenamefont {Raymond}, \citenamefont {Zucca}, \citenamefont {Lanting}, \citenamefont {Altomare}, \citenamefont {Berkley}, \citenamefont {Ejtemaee}, \citenamefont {Hoskinson}, \citenamefont {Huang} \emph {et~al.}}]{king2022coherent}%
  \BibitemOpen
  \bibfield  {author} {\bibinfo {author} {\bibfnamefont {A.~D.}\ \bibnamefont {King}}, \bibinfo {author} {\bibfnamefont {S.}~\bibnamefont {Suzuki}}, \bibinfo {author} {\bibfnamefont {J.}~\bibnamefont {Raymond}}, \bibinfo {author} {\bibfnamefont {A.}~\bibnamefont {Zucca}}, \bibinfo {author} {\bibfnamefont {T.}~\bibnamefont {Lanting}}, \bibinfo {author} {\bibfnamefont {F.}~\bibnamefont {Altomare}}, \bibinfo {author} {\bibfnamefont {A.~J.}\ \bibnamefont {Berkley}}, \bibinfo {author} {\bibfnamefont {S.}~\bibnamefont {Ejtemaee}}, \bibinfo {author} {\bibfnamefont {E.}~\bibnamefont {Hoskinson}}, \bibinfo {author} {\bibfnamefont {S.}~\bibnamefont {Huang}}, \emph {et~al.},\ }\bibfield  {title} {\bibinfo {title} {{Coherent quantum annealing in a programmable 2,000 qubit Ising chain}},\ }\href {https://doi.org/10.1038/s41567-022-01741-6} {\bibfield  {journal} {\bibinfo  {journal} {Nat. Phys.}\ }\textbf {\bibinfo {volume} {18}},\ \bibinfo {pages} {1324} (\bibinfo {year} {2022})},\ \Eprint
  {https://arxiv.org/abs/2202.05847} {2202.05847} \BibitemShut {NoStop}%
\bibitem [{\citenamefont {King}\ \emph {et~al.}(2023)\citenamefont {King}, \citenamefont {Raymond}, \citenamefont {Lanting}, \citenamefont {Harris}, \citenamefont {Zucca}, \citenamefont {Altomare}, \citenamefont {Berkley}, \citenamefont {Boothby}, \citenamefont {Ejtemaee}, \citenamefont {Enderud} \emph {et~al.}}]{King2023spinGlass}%
  \BibitemOpen
  \bibfield  {author} {\bibinfo {author} {\bibfnamefont {A.~D.}\ \bibnamefont {King}}, \bibinfo {author} {\bibfnamefont {J.}~\bibnamefont {Raymond}}, \bibinfo {author} {\bibfnamefont {T.}~\bibnamefont {Lanting}}, \bibinfo {author} {\bibfnamefont {R.}~\bibnamefont {Harris}}, \bibinfo {author} {\bibfnamefont {A.}~\bibnamefont {Zucca}}, \bibinfo {author} {\bibfnamefont {F.}~\bibnamefont {Altomare}}, \bibinfo {author} {\bibfnamefont {A.~J.}\ \bibnamefont {Berkley}}, \bibinfo {author} {\bibfnamefont {K.}~\bibnamefont {Boothby}}, \bibinfo {author} {\bibfnamefont {S.}~\bibnamefont {Ejtemaee}}, \bibinfo {author} {\bibfnamefont {C.}~\bibnamefont {Enderud}}, \emph {et~al.},\ }\bibfield  {title} {\bibinfo {title} {Quantum critical dynamics in a 5,000-qubit programmable spin glass},\ }\href {https://doi.org/10.1038/s41586-023-05867-2} {\bibfield  {journal} {\bibinfo  {journal} {Nature}\ }\textbf {\bibinfo {volume} {617}},\ \bibinfo {pages} {61} (\bibinfo {year} {2023})}\BibitemShut {NoStop}%
\bibitem [{\citenamefont {Ebadi}\ \emph {et~al.}(2021)\citenamefont {Ebadi}, \citenamefont {Wang}, \citenamefont {Levine}, \citenamefont {Keesling}, \citenamefont {Semeghini}, \citenamefont {Omran}, \citenamefont {Bluvstein}, \citenamefont {Samajdar}, \citenamefont {Pichler}, \citenamefont {Ho} \emph {et~al.}}]{ebadi2021quantum}%
  \BibitemOpen
  \bibfield  {author} {\bibinfo {author} {\bibfnamefont {S.}~\bibnamefont {Ebadi}}, \bibinfo {author} {\bibfnamefont {T.~T.}\ \bibnamefont {Wang}}, \bibinfo {author} {\bibfnamefont {H.}~\bibnamefont {Levine}}, \bibinfo {author} {\bibfnamefont {A.}~\bibnamefont {Keesling}}, \bibinfo {author} {\bibfnamefont {G.}~\bibnamefont {Semeghini}}, \bibinfo {author} {\bibfnamefont {A.}~\bibnamefont {Omran}}, \bibinfo {author} {\bibfnamefont {D.}~\bibnamefont {Bluvstein}}, \bibinfo {author} {\bibfnamefont {R.}~\bibnamefont {Samajdar}}, \bibinfo {author} {\bibfnamefont {H.}~\bibnamefont {Pichler}}, \bibinfo {author} {\bibfnamefont {W.~W.}\ \bibnamefont {Ho}}, \emph {et~al.},\ }\bibfield  {title} {\bibinfo {title} {Quantum phases of matter on a 256-atom programmable quantum simulator},\ }\href {https://doi.org/10.1038/s41586-021-03582-4} {\bibfield  {journal} {\bibinfo  {journal} {Nature}\ }\textbf {\bibinfo {volume} {595}},\ \bibinfo {pages} {227} (\bibinfo {year} {2021})}\BibitemShut {NoStop}%
\bibitem [{\citenamefont {Bravyi}\ \emph {et~al.}(2022)\citenamefont {Bravyi}, \citenamefont {Dial}, \citenamefont {Gambetta}, \citenamefont {Gil},\ and\ \citenamefont {Nazario}}]{Bravyi2022}%
  \BibitemOpen
  \bibfield  {author} {\bibinfo {author} {\bibfnamefont {S.}~\bibnamefont {Bravyi}}, \bibinfo {author} {\bibfnamefont {O.}~\bibnamefont {Dial}}, \bibinfo {author} {\bibfnamefont {J.~M.}\ \bibnamefont {Gambetta}}, \bibinfo {author} {\bibfnamefont {D.}~\bibnamefont {Gil}},\ and\ \bibinfo {author} {\bibfnamefont {Z.}~\bibnamefont {Nazario}},\ }\bibfield  {title} {\bibinfo {title} {{The future of quantum computing with superconducting qubits}},\ }\href {https://doi.org/10.1063/5.0082975} {\bibfield  {journal} {\bibinfo  {journal} {J. Appl. Phys}\ }\textbf {\bibinfo {volume} {132}},\ \bibinfo {pages} {160902} (\bibinfo {year} {2022})}\BibitemShut {NoStop}%
\bibitem [{\citenamefont {Vazquez}\ \emph {et~al.}(2024)\citenamefont {Vazquez}, \citenamefont {Tornow}, \citenamefont {Riste}, \citenamefont {Woerner}, \citenamefont {Takita},\ and\ \citenamefont {Egger}}]{vazquez2024scaling}%
  \BibitemOpen
  \bibfield  {author} {\bibinfo {author} {\bibfnamefont {A.~C.}\ \bibnamefont {Vazquez}}, \bibinfo {author} {\bibfnamefont {C.}~\bibnamefont {Tornow}}, \bibinfo {author} {\bibfnamefont {D.}~\bibnamefont {Riste}}, \bibinfo {author} {\bibfnamefont {S.}~\bibnamefont {Woerner}}, \bibinfo {author} {\bibfnamefont {M.}~\bibnamefont {Takita}},\ and\ \bibinfo {author} {\bibfnamefont {D.~J.}\ \bibnamefont {Egger}},\ }\bibfield  {title} {\bibinfo {title} {Scaling quantum computing with dynamic circuits},\ }\href {https://arxiv.org/abs/2402.17833} {\bibfield  {journal} {\bibinfo  {journal} {arXiv}\ } (\bibinfo {year} {2024})},\ \Eprint {https://arxiv.org/abs/2402.17833} {arXiv:2402.17833 [quant-ph]} \BibitemShut {NoStop}%
\bibitem [{\citenamefont {Cai}\ \emph {et~al.}(2023)\citenamefont {Cai}, \citenamefont {Babbush}, \citenamefont {Benjamin}, \citenamefont {Endo}, \citenamefont {Huggins}, \citenamefont {Li}, \citenamefont {McClean},\ and\ \citenamefont {O'Brien}}]{Cai2023em}%
  \BibitemOpen
  \bibfield  {author} {\bibinfo {author} {\bibfnamefont {Z.}~\bibnamefont {Cai}}, \bibinfo {author} {\bibfnamefont {R.}~\bibnamefont {Babbush}}, \bibinfo {author} {\bibfnamefont {S.~C.}\ \bibnamefont {Benjamin}}, \bibinfo {author} {\bibfnamefont {S.}~\bibnamefont {Endo}}, \bibinfo {author} {\bibfnamefont {W.~J.}\ \bibnamefont {Huggins}}, \bibinfo {author} {\bibfnamefont {Y.}~\bibnamefont {Li}}, \bibinfo {author} {\bibfnamefont {J.~R.}\ \bibnamefont {McClean}},\ and\ \bibinfo {author} {\bibfnamefont {T.~E.}\ \bibnamefont {O'Brien}},\ }\bibfield  {title} {\bibinfo {title} {Quantum error mitigation},\ }\href {https://doi.org/10.1103/RevModPhys.95.045005} {\bibfield  {journal} {\bibinfo  {journal} {Rev. Mod. Phys.}\ }\textbf {\bibinfo {volume} {95}},\ \bibinfo {pages} {045005} (\bibinfo {year} {2023})}\BibitemShut {NoStop}%
\bibitem [{\citenamefont {Shtanko}\ \emph {et~al.}(2023)\citenamefont {Shtanko}, \citenamefont {Wang}, \citenamefont {Zhang}, \citenamefont {Harle}, \citenamefont {Seif}, \citenamefont {Movassagh},\ and\ \citenamefont {Minev}}]{shtanko2023uncovering}%
  \BibitemOpen
  \bibfield  {author} {\bibinfo {author} {\bibfnamefont {O.}~\bibnamefont {Shtanko}}, \bibinfo {author} {\bibfnamefont {D.~S.}\ \bibnamefont {Wang}}, \bibinfo {author} {\bibfnamefont {H.}~\bibnamefont {Zhang}}, \bibinfo {author} {\bibfnamefont {N.}~\bibnamefont {Harle}}, \bibinfo {author} {\bibfnamefont {A.}~\bibnamefont {Seif}}, \bibinfo {author} {\bibfnamefont {R.}~\bibnamefont {Movassagh}},\ and\ \bibinfo {author} {\bibfnamefont {Z.}~\bibnamefont {Minev}},\ }\bibfield  {title} {\bibinfo {title} {Uncovering local integrability in quantum many-body dynamics},\ }\href {https://arxiv.org/abs/2307.07552} {\bibfield  {journal} {\bibinfo  {journal} {arXiv}\ } (\bibinfo {year} {2023})},\ \Eprint {https://arxiv.org/abs/2307.07552} {arXiv:2307.07552 [quant-ph]} \BibitemShut {NoStop}%
\bibitem [{\citenamefont {Amico}\ \emph {et~al.}(2023)\citenamefont {Amico}, \citenamefont {Zhang}, \citenamefont {Jurcevic}, \citenamefont {Bishop}, \citenamefont {Nation}, \citenamefont {Wack},\ and\ \citenamefont {McKay}}]{Amico2023defining}%
  \BibitemOpen
  \bibfield  {author} {\bibinfo {author} {\bibfnamefont {M.}~\bibnamefont {Amico}}, \bibinfo {author} {\bibfnamefont {H.}~\bibnamefont {Zhang}}, \bibinfo {author} {\bibfnamefont {P.}~\bibnamefont {Jurcevic}}, \bibinfo {author} {\bibfnamefont {L.~S.}\ \bibnamefont {Bishop}}, \bibinfo {author} {\bibfnamefont {P.}~\bibnamefont {Nation}}, \bibinfo {author} {\bibfnamefont {A.}~\bibnamefont {Wack}},\ and\ \bibinfo {author} {\bibfnamefont {D.~C.}\ \bibnamefont {McKay}},\ }\bibfield  {title} {\bibinfo {title} {Defining best practices for quantum benchmarks},\ }\href {https://doi.org/10.1109/QCE57702.2023.00084} {\bibfield  {journal} {\bibinfo  {journal} {2023 IEEE International Conference on Quantum Computing and Engineering (QCE)}\ }\textbf {\bibinfo {volume} {01}},\ \bibinfo {pages} {692} (\bibinfo {year} {2023})}\BibitemShut {NoStop}%
\bibitem [{\citenamefont {Magesan}\ \emph {et~al.}(2012)\citenamefont {Magesan}, \citenamefont {Gambetta},\ and\ \citenamefont {Emerson}}]{Magesan2012rb}%
  \BibitemOpen
  \bibfield  {author} {\bibinfo {author} {\bibfnamefont {E.}~\bibnamefont {Magesan}}, \bibinfo {author} {\bibfnamefont {J.~M.}\ \bibnamefont {Gambetta}},\ and\ \bibinfo {author} {\bibfnamefont {J.}~\bibnamefont {Emerson}},\ }\bibfield  {title} {\bibinfo {title} {Characterizing quantum gates via randomized benchmarking},\ }\href {https://doi.org/10.1103/PhysRevA.85.042311} {\bibfield  {journal} {\bibinfo  {journal} {Phys. Rev. A}\ }\textbf {\bibinfo {volume} {85}},\ \bibinfo {pages} {042311} (\bibinfo {year} {2012})}\BibitemShut {NoStop}%
\bibitem [{\citenamefont {Proctor}\ \emph {et~al.}(2022)\citenamefont {Proctor}, \citenamefont {Seritan}, \citenamefont {Rudinger}, \citenamefont {Nielsen}, \citenamefont {Blume-Kohout},\ and\ \citenamefont {Young}}]{Proctor2022mirrorRB}%
  \BibitemOpen
  \bibfield  {author} {\bibinfo {author} {\bibfnamefont {T.}~\bibnamefont {Proctor}}, \bibinfo {author} {\bibfnamefont {S.}~\bibnamefont {Seritan}}, \bibinfo {author} {\bibfnamefont {K.}~\bibnamefont {Rudinger}}, \bibinfo {author} {\bibfnamefont {E.}~\bibnamefont {Nielsen}}, \bibinfo {author} {\bibfnamefont {R.}~\bibnamefont {Blume-Kohout}},\ and\ \bibinfo {author} {\bibfnamefont {K.}~\bibnamefont {Young}},\ }\bibfield  {title} {\bibinfo {title} {Scalable randomized benchmarking of quantum computers using mirror circuits},\ }\href {https://doi.org/10.1103/PhysRevLett.129.150502} {\bibfield  {journal} {\bibinfo  {journal} {Phys. Rev. Lett.}\ }\textbf {\bibinfo {volume} {129}},\ \bibinfo {pages} {150502} (\bibinfo {year} {2022})}\BibitemShut {NoStop}%
\bibitem [{\citenamefont {Hines}\ \emph {et~al.}(2023)\citenamefont {Hines}, \citenamefont {Hothem}, \citenamefont {Blume-Kohout}, \citenamefont {Whaley},\ and\ \citenamefont {Proctor}}]{Hines2023}%
  \BibitemOpen
  \bibfield  {author} {\bibinfo {author} {\bibfnamefont {J.}~\bibnamefont {Hines}}, \bibinfo {author} {\bibfnamefont {D.}~\bibnamefont {Hothem}}, \bibinfo {author} {\bibfnamefont {R.}~\bibnamefont {Blume-Kohout}}, \bibinfo {author} {\bibfnamefont {B.}~\bibnamefont {Whaley}},\ and\ \bibinfo {author} {\bibfnamefont {T.}~\bibnamefont {Proctor}},\ }\bibfield  {title} {\bibinfo {title} {Fully scalable randomized benchmarking without motion reversal},\ }\href {http://arxiv.org/abs/2309.05147} {\bibfield  {journal} {\bibinfo  {journal} {arXiv preprint}\ } (\bibinfo {year} {2023})},\ \Eprint {https://arxiv.org/abs/2309.05147} {arXiv:2309.05147} \BibitemShut {NoStop}%
\bibitem [{\citenamefont {Boixo}\ \emph {et~al.}(2018)\citenamefont {Boixo}, \citenamefont {Isakov}, \citenamefont {Smelyanskiy}, \citenamefont {Babbush}, \citenamefont {Ding}, \citenamefont {Jiang}, \citenamefont {Bremner}, \citenamefont {Martinis},\ and\ \citenamefont {Neven}}]{Boixo2018xeb}%
  \BibitemOpen
  \bibfield  {author} {\bibinfo {author} {\bibfnamefont {S.}~\bibnamefont {Boixo}}, \bibinfo {author} {\bibfnamefont {S.~V.}\ \bibnamefont {Isakov}}, \bibinfo {author} {\bibfnamefont {V.~N.}\ \bibnamefont {Smelyanskiy}}, \bibinfo {author} {\bibfnamefont {R.}~\bibnamefont {Babbush}}, \bibinfo {author} {\bibfnamefont {N.}~\bibnamefont {Ding}}, \bibinfo {author} {\bibfnamefont {Z.}~\bibnamefont {Jiang}}, \bibinfo {author} {\bibfnamefont {M.~J.}\ \bibnamefont {Bremner}}, \bibinfo {author} {\bibfnamefont {J.~M.}\ \bibnamefont {Martinis}},\ and\ \bibinfo {author} {\bibfnamefont {H.}~\bibnamefont {Neven}},\ }\bibfield  {title} {\bibinfo {title} {Characterizing quantum supremacy in near-term devices},\ }\href {https://doi.org/10.1038/s41567-018-0124-x} {\bibfield  {journal} {\bibinfo  {journal} {Nat. Phys.}\ }\textbf {\bibinfo {volume} {14}},\ \bibinfo {pages} {595} (\bibinfo {year} {2018})}\BibitemShut {NoStop}%
\bibitem [{\citenamefont {Cross}\ \emph {et~al.}(2019)\citenamefont {Cross}, \citenamefont {Bishop}, \citenamefont {Sheldon}, \citenamefont {Nation},\ and\ \citenamefont {Gambetta}}]{Cross2019qv}%
  \BibitemOpen
  \bibfield  {author} {\bibinfo {author} {\bibfnamefont {A.~W.}\ \bibnamefont {Cross}}, \bibinfo {author} {\bibfnamefont {L.~S.}\ \bibnamefont {Bishop}}, \bibinfo {author} {\bibfnamefont {S.}~\bibnamefont {Sheldon}}, \bibinfo {author} {\bibfnamefont {P.~D.}\ \bibnamefont {Nation}},\ and\ \bibinfo {author} {\bibfnamefont {J.~M.}\ \bibnamefont {Gambetta}},\ }\bibfield  {title} {\bibinfo {title} {Validating quantum computers using randomized model circuits},\ }\href {https://doi.org/10.1103/PhysRevA.100.032328} {\bibfield  {journal} {\bibinfo  {journal} {Phys. Rev. A}\ }\textbf {\bibinfo {volume} {100}},\ \bibinfo {pages} {032328} (\bibinfo {year} {2019})}\BibitemShut {NoStop}%
\bibitem [{\citenamefont {McKay}\ \emph {et~al.}(2023)\citenamefont {McKay}, \citenamefont {Hincks}, \citenamefont {Pritchett}, \citenamefont {Carroll}, \citenamefont {Govia},\ and\ \citenamefont {Merkel}}]{Mckay2023layerfidelity}%
  \BibitemOpen
  \bibfield  {author} {\bibinfo {author} {\bibfnamefont {D.~C.}\ \bibnamefont {McKay}}, \bibinfo {author} {\bibfnamefont {I.}~\bibnamefont {Hincks}}, \bibinfo {author} {\bibfnamefont {E.~J.}\ \bibnamefont {Pritchett}}, \bibinfo {author} {\bibfnamefont {M.}~\bibnamefont {Carroll}}, \bibinfo {author} {\bibfnamefont {L.~C.~G.}\ \bibnamefont {Govia}},\ and\ \bibinfo {author} {\bibfnamefont {S.~T.}\ \bibnamefont {Merkel}},\ }\bibfield  {title} {\bibinfo {title} {Benchmarking quantum processor performance at scale},\ }\href {http://arxiv.org/abs/2311.05933} {\bibfield  {journal} {\bibinfo  {journal} {arXiv preprint}\ } (\bibinfo {year} {2023})},\ \Eprint {https://arxiv.org/abs/2311.05933} {arXiv:2311.05933} \BibitemShut {NoStop}%
\bibitem [{\citenamefont {Lubinski}\ \emph {et~al.}(2023)\citenamefont {Lubinski}, \citenamefont {Johri}, \citenamefont {Varosy}, \citenamefont {Coleman}, \citenamefont {Zhao}, \citenamefont {Necaise}, \citenamefont {Baldwin}, \citenamefont {Mayer},\ and\ \citenamefont {Proctor}}]{Lubinski2023}%
  \BibitemOpen
  \bibfield  {author} {\bibinfo {author} {\bibfnamefont {T.}~\bibnamefont {Lubinski}}, \bibinfo {author} {\bibfnamefont {S.}~\bibnamefont {Johri}}, \bibinfo {author} {\bibfnamefont {P.}~\bibnamefont {Varosy}}, \bibinfo {author} {\bibfnamefont {J.}~\bibnamefont {Coleman}}, \bibinfo {author} {\bibfnamefont {L.}~\bibnamefont {Zhao}}, \bibinfo {author} {\bibfnamefont {J.}~\bibnamefont {Necaise}}, \bibinfo {author} {\bibfnamefont {C.~H.}\ \bibnamefont {Baldwin}}, \bibinfo {author} {\bibfnamefont {K.}~\bibnamefont {Mayer}},\ and\ \bibinfo {author} {\bibfnamefont {T.}~\bibnamefont {Proctor}},\ }\bibfield  {title} {\bibinfo {title} {Application-oriented performance benchmarks for quantum computing},\ }\href {https://doi.org/10.1109/TQE.2023.3253761} {\bibfield  {journal} {\bibinfo  {journal} {IEEE Transactions on Quantum Engineering}\ }\textbf {\bibinfo {volume} {4}},\ \bibinfo {pages} {1} (\bibinfo {year} {2023})}\BibitemShut {NoStop}%
\bibitem [{\citenamefont {Wu}\ \emph {et~al.}(2023)\citenamefont {Wu}, \citenamefont {Rossi}, \citenamefont {Vicentini}, \citenamefont {Astrakhantsev}, \citenamefont {Becca}, \citenamefont {Cao}, \citenamefont {Carrasquilla}, \citenamefont {Ferrari}, \citenamefont {Georges}, \citenamefont {Hibat-Allah} \emph {et~al.}}]{Wu2023}%
  \BibitemOpen
  \bibfield  {author} {\bibinfo {author} {\bibfnamefont {D.}~\bibnamefont {Wu}}, \bibinfo {author} {\bibfnamefont {R.}~\bibnamefont {Rossi}}, \bibinfo {author} {\bibfnamefont {F.}~\bibnamefont {Vicentini}}, \bibinfo {author} {\bibfnamefont {N.}~\bibnamefont {Astrakhantsev}}, \bibinfo {author} {\bibfnamefont {F.}~\bibnamefont {Becca}}, \bibinfo {author} {\bibfnamefont {X.}~\bibnamefont {Cao}}, \bibinfo {author} {\bibfnamefont {J.}~\bibnamefont {Carrasquilla}}, \bibinfo {author} {\bibfnamefont {F.}~\bibnamefont {Ferrari}}, \bibinfo {author} {\bibfnamefont {A.}~\bibnamefont {Georges}}, \bibinfo {author} {\bibfnamefont {M.}~\bibnamefont {Hibat-Allah}}, \emph {et~al.},\ }\bibfield  {title} {\bibinfo {title} {Variational benchmarks for quantum many-body problems},\ }\href {http://arxiv.org/abs/2302.04919} {\bibfield  {journal} {\bibinfo  {journal} {arXiv preprint}\ } (\bibinfo {year} {2023})},\ \Eprint {https://arxiv.org/abs/2302.04919} {arXiv:2302.04919} \BibitemShut {NoStop}%
\bibitem [{\citenamefont {Santra}\ \emph {et~al.}(2024)\citenamefont {Santra}, \citenamefont {Jendrzejewski}, \citenamefont {Hauke},\ and\ \citenamefont {Egger}}]{Santra2024}%
  \BibitemOpen
  \bibfield  {author} {\bibinfo {author} {\bibfnamefont {G.~C.}\ \bibnamefont {Santra}}, \bibinfo {author} {\bibfnamefont {F.}~\bibnamefont {Jendrzejewski}}, \bibinfo {author} {\bibfnamefont {P.}~\bibnamefont {Hauke}},\ and\ \bibinfo {author} {\bibfnamefont {D.~J.}\ \bibnamefont {Egger}},\ }\bibfield  {title} {\bibinfo {title} {Squeezing and quantum approximate optimization},\ }\href {https://doi.org/10.1103/PhysRevA.109.012413} {\bibfield  {journal} {\bibinfo  {journal} {Phys. Rev. A}\ }\textbf {\bibinfo {volume} {109}},\ \bibinfo {pages} {012413} (\bibinfo {year} {2024})}\BibitemShut {NoStop}%
\bibitem [{\citenamefont {Zhang}\ and\ \citenamefont {Nation}(2023)}]{zhang2023characterizing}%
  \BibitemOpen
  \bibfield  {author} {\bibinfo {author} {\bibfnamefont {V.}~\bibnamefont {Zhang}}\ and\ \bibinfo {author} {\bibfnamefont {P.~D.}\ \bibnamefont {Nation}},\ }\bibfield  {title} {\bibinfo {title} {Characterizing quantum processors using discrete time crystals},\ }\href {http://arxiv.org/abs/2301.07625} {\bibfield  {journal} {\bibinfo  {journal} {arXiv}\ } (\bibinfo {year} {2023})},\ \Eprint {https://arxiv.org/abs/2301.07625} {arXiv:2301.07625 [quant-ph]} \BibitemShut {NoStop}%
\bibitem [{\citenamefont {Farhi}\ \emph {et~al.}(2014)\citenamefont {Farhi}, \citenamefont {Goldstone},\ and\ \citenamefont {Gutmann}}]{farhi2014quantum}%
  \BibitemOpen
  \bibfield  {author} {\bibinfo {author} {\bibfnamefont {E.}~\bibnamefont {Farhi}}, \bibinfo {author} {\bibfnamefont {J.}~\bibnamefont {Goldstone}},\ and\ \bibinfo {author} {\bibfnamefont {S.}~\bibnamefont {Gutmann}},\ }\bibfield  {title} {\bibinfo {title} {A quantum approximate optimization algorithm},\ }\href {http://arxiv.org/abs/1411.4028} {\bibfield  {journal} {\bibinfo  {journal} {arXiv}\ } (\bibinfo {year} {2014})},\ \Eprint {https://arxiv.org/abs/1411.4028} {arXiv:1411.4028 [quant-ph]} \BibitemShut {NoStop}%
\bibitem [{\citenamefont {Zhou}\ \emph {et~al.}(2020)\citenamefont {Zhou}, \citenamefont {Wang}, \citenamefont {Choi}, \citenamefont {Pichler},\ and\ \citenamefont {Lukin}}]{Zhou2020}%
  \BibitemOpen
  \bibfield  {author} {\bibinfo {author} {\bibfnamefont {L.}~\bibnamefont {Zhou}}, \bibinfo {author} {\bibfnamefont {S.-T.}\ \bibnamefont {Wang}}, \bibinfo {author} {\bibfnamefont {S.}~\bibnamefont {Choi}}, \bibinfo {author} {\bibfnamefont {H.}~\bibnamefont {Pichler}},\ and\ \bibinfo {author} {\bibfnamefont {M.~D.}\ \bibnamefont {Lukin}},\ }\bibfield  {title} {\bibinfo {title} {Quantum approximate optimization algorithm: Performance, mechanism, and implementation on near-term devices},\ }\href {https://doi.org/10.1103/PhysRevX.10.021067} {\bibfield  {journal} {\bibinfo  {journal} {Phys. Rev. X}\ }\textbf {\bibinfo {volume} {10}},\ \bibinfo {pages} {021067} (\bibinfo {year} {2020})}\BibitemShut {NoStop}%
\bibitem [{\citenamefont {Farhi}\ \emph {et~al.}(2020)\citenamefont {Farhi}, \citenamefont {Gamarnik},\ and\ \citenamefont {Gutmann}}]{farhi2020quantum}%
  \BibitemOpen
  \bibfield  {author} {\bibinfo {author} {\bibfnamefont {E.}~\bibnamefont {Farhi}}, \bibinfo {author} {\bibfnamefont {D.}~\bibnamefont {Gamarnik}},\ and\ \bibinfo {author} {\bibfnamefont {S.}~\bibnamefont {Gutmann}},\ }\bibfield  {title} {\bibinfo {title} {The quantum approximate optimization algorithm needs to see the whole graph: A typical case},\ }\href {http://arxiv.org/abs/2004.09002} {\bibfield  {journal} {\bibinfo  {journal} {arXiv}\ } (\bibinfo {year} {2020})},\ \Eprint {https://arxiv.org/abs/2004.09002} {arXiv:2004.09002 [quant-ph]} \BibitemShut {NoStop}%
\bibitem [{\citenamefont {Heim}\ \emph {et~al.}(2015)\citenamefont {Heim}, \citenamefont {Rønnow}, \citenamefont {Isakov},\ and\ \citenamefont {Troyer}}]{Heim2015spinGlass}%
  \BibitemOpen
  \bibfield  {author} {\bibinfo {author} {\bibfnamefont {B.}~\bibnamefont {Heim}}, \bibinfo {author} {\bibfnamefont {T.~F.}\ \bibnamefont {Rønnow}}, \bibinfo {author} {\bibfnamefont {S.~V.}\ \bibnamefont {Isakov}},\ and\ \bibinfo {author} {\bibfnamefont {M.}~\bibnamefont {Troyer}},\ }\bibfield  {title} {\bibinfo {title} {Quantum versus classical annealing of ising spin glasses},\ }\href {https://doi.org/10.1126/science.aaa4170} {\bibfield  {journal} {\bibinfo  {journal} {Science}\ }\textbf {\bibinfo {volume} {348}},\ \bibinfo {pages} {215} (\bibinfo {year} {2015})}\BibitemShut {NoStop}%
\bibitem [{\citenamefont {Young}\ \emph {et~al.}(2010)\citenamefont {Young}, \citenamefont {Knysh},\ and\ \citenamefont {Smelyanskiy}}]{young2010first}%
  \BibitemOpen
  \bibfield  {author} {\bibinfo {author} {\bibfnamefont {A.~P.}\ \bibnamefont {Young}}, \bibinfo {author} {\bibfnamefont {S.}~\bibnamefont {Knysh}},\ and\ \bibinfo {author} {\bibfnamefont {V.~N.}\ \bibnamefont {Smelyanskiy}},\ }\bibfield  {title} {\bibinfo {title} {First-order phase transition in the quantum adiabatic algorithm},\ }\href {https://doi.org/10.1103/PhysRevLett.104.020502} {\bibfield  {journal} {\bibinfo  {journal} {Phys. Rev. Lett.}\ }\textbf {\bibinfo {volume} {104}},\ \bibinfo {pages} {020502} (\bibinfo {year} {2010})}\BibitemShut {NoStop}%
\bibitem [{\citenamefont {J\"org}\ \emph {et~al.}(2010)\citenamefont {J\"org}, \citenamefont {Krzakala}, \citenamefont {Semerjian},\ and\ \citenamefont {Zamponi}}]{PhysRevLett.104.207206}%
  \BibitemOpen
  \bibfield  {author} {\bibinfo {author} {\bibfnamefont {T.}~\bibnamefont {J\"org}}, \bibinfo {author} {\bibfnamefont {F.}~\bibnamefont {Krzakala}}, \bibinfo {author} {\bibfnamefont {G.}~\bibnamefont {Semerjian}},\ and\ \bibinfo {author} {\bibfnamefont {F.}~\bibnamefont {Zamponi}},\ }\bibfield  {title} {\bibinfo {title} {First-order transitions and the performance of quantum algorithms in random optimization problems},\ }\href {https://doi.org/10.1103/PhysRevLett.104.207206} {\bibfield  {journal} {\bibinfo  {journal} {Phys. Rev. Lett.}\ }\textbf {\bibinfo {volume} {104}},\ \bibinfo {pages} {207206} (\bibinfo {year} {2010})}\BibitemShut {NoStop}%
\bibitem [{\citenamefont {Schmitt}\ \emph {et~al.}(2022)\citenamefont {Schmitt}, \citenamefont {Rams}, \citenamefont {Dziarmaga}, \citenamefont {Heyl},\ and\ \citenamefont {Zurek}}]{Schmitt2022}%
  \BibitemOpen
  \bibfield  {author} {\bibinfo {author} {\bibfnamefont {M.}~\bibnamefont {Schmitt}}, \bibinfo {author} {\bibfnamefont {M.~M.}\ \bibnamefont {Rams}}, \bibinfo {author} {\bibfnamefont {J.}~\bibnamefont {Dziarmaga}}, \bibinfo {author} {\bibfnamefont {M.}~\bibnamefont {Heyl}},\ and\ \bibinfo {author} {\bibfnamefont {W.~H.}\ \bibnamefont {Zurek}},\ }\bibfield  {title} {\bibinfo {title} {Quantum phase transition dynamics in the two-dimensional transverse-field ising model},\ }\href {https://doi.org/10.1126/sciadv.abl6850} {\bibfield  {journal} {\bibinfo  {journal} {Sci. Adv.}\ }\textbf {\bibinfo {volume} {8}},\ \bibinfo {pages} {6850} (\bibinfo {year} {2022})}\BibitemShut {NoStop}%
\bibitem [{\citenamefont {Chandran}\ \emph {et~al.}(2012)\citenamefont {Chandran}, \citenamefont {Erez}, \citenamefont {Gubser},\ and\ \citenamefont {Sondhi}}]{chandran2012kibble}%
  \BibitemOpen
  \bibfield  {author} {\bibinfo {author} {\bibfnamefont {A.}~\bibnamefont {Chandran}}, \bibinfo {author} {\bibfnamefont {A.}~\bibnamefont {Erez}}, \bibinfo {author} {\bibfnamefont {S.~S.}\ \bibnamefont {Gubser}},\ and\ \bibinfo {author} {\bibfnamefont {S.~L.}\ \bibnamefont {Sondhi}},\ }\bibfield  {title} {\bibinfo {title} {Kibble-zurek problem: Universality and the scaling limit},\ }\href {https://doi.org/10.1103/PhysRevB.86.064304} {\bibfield  {journal} {\bibinfo  {journal} {Phys. Rev. B}\ }\textbf {\bibinfo {volume} {86}},\ \bibinfo {pages} {064304} (\bibinfo {year} {2012})}\BibitemShut {NoStop}%
\bibitem [{\citenamefont {Morita}\ and\ \citenamefont {Nishimori}(2008)}]{Morita2008}%
  \BibitemOpen
  \bibfield  {author} {\bibinfo {author} {\bibfnamefont {S.}~\bibnamefont {Morita}}\ and\ \bibinfo {author} {\bibfnamefont {H.}~\bibnamefont {Nishimori}},\ }\bibfield  {title} {\bibinfo {title} {Mathematical foundation of quantum annealing},\ }\href {https://doi.org/10.1063/1.2995837} {\bibfield  {journal} {\bibinfo  {journal} {J. Math. Phys.}\ }\textbf {\bibinfo {volume} {49}},\ \bibinfo {pages} {125210} (\bibinfo {year} {2008})}\BibitemShut {NoStop}%
\bibitem [{\citenamefont {Caneva}\ \emph {et~al.}(2008)\citenamefont {Caneva}, \citenamefont {Fazio},\ and\ \citenamefont {Santoro}}]{Caneva2008}%
  \BibitemOpen
  \bibfield  {author} {\bibinfo {author} {\bibfnamefont {T.}~\bibnamefont {Caneva}}, \bibinfo {author} {\bibfnamefont {R.}~\bibnamefont {Fazio}},\ and\ \bibinfo {author} {\bibfnamefont {G.~E.}\ \bibnamefont {Santoro}},\ }\bibfield  {title} {\bibinfo {title} {Adiabatic quantum dynamics of the lipkin-meshkov-glick model},\ }\href {https://doi.org/10.1103/PhysRevB.78.104426} {\bibfield  {journal} {\bibinfo  {journal} {Phys. Rev. B}\ }\textbf {\bibinfo {volume} {78}},\ \bibinfo {pages} {104426} (\bibinfo {year} {2008})}\BibitemShut {NoStop}%
\bibitem [{\citenamefont {Poulin}\ \emph {et~al.}(2011)\citenamefont {Poulin}, \citenamefont {Qarry}, \citenamefont {Somma},\ and\ \citenamefont {Verstraete}}]{Poulin2011}%
  \BibitemOpen
  \bibfield  {author} {\bibinfo {author} {\bibfnamefont {D.}~\bibnamefont {Poulin}}, \bibinfo {author} {\bibfnamefont {A.}~\bibnamefont {Qarry}}, \bibinfo {author} {\bibfnamefont {R.}~\bibnamefont {Somma}},\ and\ \bibinfo {author} {\bibfnamefont {F.}~\bibnamefont {Verstraete}},\ }\bibfield  {title} {\bibinfo {title} {Quantum simulation of time-dependent hamiltonians and the convenient illusion of hilbert space},\ }\href {https://doi.org/10.1103/PhysRevLett.106.170501} {\bibfield  {journal} {\bibinfo  {journal} {Phys. Rev. Lett.}\ }\textbf {\bibinfo {volume} {106}},\ \bibinfo {pages} {170501} (\bibinfo {year} {2011})}\BibitemShut {NoStop}%
\bibitem [{\citenamefont {Childs}\ \emph {et~al.}(2021)\citenamefont {Childs}, \citenamefont {Su}, \citenamefont {Tran}, \citenamefont {Wiebe},\ and\ \citenamefont {Zhu}}]{Childs2021theoryTrotterError}%
  \BibitemOpen
  \bibfield  {author} {\bibinfo {author} {\bibfnamefont {A.~M.}\ \bibnamefont {Childs}}, \bibinfo {author} {\bibfnamefont {Y.}~\bibnamefont {Su}}, \bibinfo {author} {\bibfnamefont {M.~C.}\ \bibnamefont {Tran}}, \bibinfo {author} {\bibfnamefont {N.}~\bibnamefont {Wiebe}},\ and\ \bibinfo {author} {\bibfnamefont {S.}~\bibnamefont {Zhu}},\ }\bibfield  {title} {\bibinfo {title} {Theory of trotter error with commutator scaling},\ }\href {https://doi.org/10.1103/PhysRevX.11.011020} {\bibfield  {journal} {\bibinfo  {journal} {Phys. Rev. X}\ }\textbf {\bibinfo {volume} {11}},\ \bibinfo {pages} {011020} (\bibinfo {year} {2021})}\BibitemShut {NoStop}%
\bibitem [{\citenamefont {Rajput}\ \emph {et~al.}(2022)\citenamefont {Rajput}, \citenamefont {Roggero},\ and\ \citenamefont {Wiebe}}]{Rajput2021}%
  \BibitemOpen
  \bibfield  {author} {\bibinfo {author} {\bibfnamefont {A.}~\bibnamefont {Rajput}}, \bibinfo {author} {\bibfnamefont {A.}~\bibnamefont {Roggero}},\ and\ \bibinfo {author} {\bibfnamefont {N.}~\bibnamefont {Wiebe}},\ }\bibfield  {title} {\bibinfo {title} {Hybridized methods for quantum simulation in the interaction picture},\ }\href {https://doi.org/10.22331/q-2022-08-17-780} {\bibfield  {journal} {\bibinfo  {journal} {Quantum}\ }\textbf {\bibinfo {volume} {6}},\ \bibinfo {pages} {780} (\bibinfo {year} {2022})}\BibitemShut {NoStop}%
\bibitem [{\citenamefont {Dutta}\ \emph {et~al.}(2016)\citenamefont {Dutta}, \citenamefont {Rahmani},\ and\ \citenamefont {del Campo}}]{Dutta2016anti}%
  \BibitemOpen
  \bibfield  {author} {\bibinfo {author} {\bibfnamefont {A.}~\bibnamefont {Dutta}}, \bibinfo {author} {\bibfnamefont {A.}~\bibnamefont {Rahmani}},\ and\ \bibinfo {author} {\bibfnamefont {A.}~\bibnamefont {del Campo}},\ }\bibfield  {title} {\bibinfo {title} {Anti-kibble-zurek behavior in crossing the quantum critical point of a thermally isolated system driven by a noisy control field},\ }\href {https://doi.org/10.1103/PhysRevLett.117.080402} {\bibfield  {journal} {\bibinfo  {journal} {Phys. Rev. Lett.}\ }\textbf {\bibinfo {volume} {117}},\ \bibinfo {pages} {080402} (\bibinfo {year} {2016})}\BibitemShut {NoStop}%
\bibitem [{\citenamefont {Arceci}\ \emph {et~al.}(2018)\citenamefont {Arceci}, \citenamefont {Barbarino}, \citenamefont {Rossini},\ and\ \citenamefont {Santoro}}]{Arceci2018}%
  \BibitemOpen
  \bibfield  {author} {\bibinfo {author} {\bibfnamefont {L.}~\bibnamefont {Arceci}}, \bibinfo {author} {\bibfnamefont {S.}~\bibnamefont {Barbarino}}, \bibinfo {author} {\bibfnamefont {D.}~\bibnamefont {Rossini}},\ and\ \bibinfo {author} {\bibfnamefont {G.~E.}\ \bibnamefont {Santoro}},\ }\bibfield  {title} {\bibinfo {title} {Optimal working point in dissipative quantum annealing},\ }\href {https://doi.org/10.1103/PhysRevB.98.064307} {\bibfield  {journal} {\bibinfo  {journal} {Phys. Rev. B}\ }\textbf {\bibinfo {volume} {98}},\ \bibinfo {pages} {064307} (\bibinfo {year} {2018})}\BibitemShut {NoStop}%
\bibitem [{\citenamefont {Mayo}\ \emph {et~al.}(2021)\citenamefont {Mayo}, \citenamefont {Fan}, \citenamefont {Chern},\ and\ \citenamefont {del Campo}}]{Mayo2021}%
  \BibitemOpen
  \bibfield  {author} {\bibinfo {author} {\bibfnamefont {J.~J.}\ \bibnamefont {Mayo}}, \bibinfo {author} {\bibfnamefont {Z.}~\bibnamefont {Fan}}, \bibinfo {author} {\bibfnamefont {G.-W.}\ \bibnamefont {Chern}},\ and\ \bibinfo {author} {\bibfnamefont {A.}~\bibnamefont {del Campo}},\ }\bibfield  {title} {\bibinfo {title} {Distribution of kinks in an ising ferromagnet after annealing and the generalized kibble-zurek mechanism},\ }\href {https://doi.org/10.1103/PhysRevResearch.3.033150} {\bibfield  {journal} {\bibinfo  {journal} {Phys. Rev. Res.}\ }\textbf {\bibinfo {volume} {3}},\ \bibinfo {pages} {033150} (\bibinfo {year} {2021})}\BibitemShut {NoStop}%
\bibitem [{\citenamefont {Boixo}\ \emph {et~al.}(2014)\citenamefont {Boixo}, \citenamefont {R{\o}nnow}, \citenamefont {Isakov}, \citenamefont {Wang}, \citenamefont {Wecker}, \citenamefont {Lidar}, \citenamefont {Martinis},\ and\ \citenamefont {Troyer}}]{boixo2014evidence}%
  \BibitemOpen
  \bibfield  {author} {\bibinfo {author} {\bibfnamefont {S.}~\bibnamefont {Boixo}}, \bibinfo {author} {\bibfnamefont {T.~F.}\ \bibnamefont {R{\o}nnow}}, \bibinfo {author} {\bibfnamefont {S.~V.}\ \bibnamefont {Isakov}}, \bibinfo {author} {\bibfnamefont {Z.}~\bibnamefont {Wang}}, \bibinfo {author} {\bibfnamefont {D.}~\bibnamefont {Wecker}}, \bibinfo {author} {\bibfnamefont {D.~A.}\ \bibnamefont {Lidar}}, \bibinfo {author} {\bibfnamefont {J.~M.}\ \bibnamefont {Martinis}},\ and\ \bibinfo {author} {\bibfnamefont {M.}~\bibnamefont {Troyer}},\ }\bibfield  {title} {\bibinfo {title} {Evidence for quantum annealing with more than one hundred qubits},\ }\href {https://doi.org/10.1038/nphys2900} {\bibfield  {journal} {\bibinfo  {journal} {Nature physics}\ }\textbf {\bibinfo {volume} {10}},\ \bibinfo {pages} {218} (\bibinfo {year} {2014})}\BibitemShut {NoStop}%
\bibitem [{\citenamefont {Denchev}\ \emph {et~al.}(2016)\citenamefont {Denchev}, \citenamefont {Boixo}, \citenamefont {Isakov}, \citenamefont {Ding}, \citenamefont {Babbush}, \citenamefont {Smelyanskiy}, \citenamefont {Martinis},\ and\ \citenamefont {Neven}}]{PhysRevX.6.031015}%
  \BibitemOpen
  \bibfield  {author} {\bibinfo {author} {\bibfnamefont {V.~S.}\ \bibnamefont {Denchev}}, \bibinfo {author} {\bibfnamefont {S.}~\bibnamefont {Boixo}}, \bibinfo {author} {\bibfnamefont {S.~V.}\ \bibnamefont {Isakov}}, \bibinfo {author} {\bibfnamefont {N.}~\bibnamefont {Ding}}, \bibinfo {author} {\bibfnamefont {R.}~\bibnamefont {Babbush}}, \bibinfo {author} {\bibfnamefont {V.}~\bibnamefont {Smelyanskiy}}, \bibinfo {author} {\bibfnamefont {J.}~\bibnamefont {Martinis}},\ and\ \bibinfo {author} {\bibfnamefont {H.}~\bibnamefont {Neven}},\ }\bibfield  {title} {\bibinfo {title} {What is the computational value of finite-range tunneling?},\ }\href {https://doi.org/10.1103/PhysRevX.6.031015} {\bibfield  {journal} {\bibinfo  {journal} {Phys. Rev. X}\ }\textbf {\bibinfo {volume} {6}},\ \bibinfo {pages} {031015} (\bibinfo {year} {2016})}\BibitemShut {NoStop}%
\bibitem [{\citenamefont {Isakov}\ \emph {et~al.}(2016)\citenamefont {Isakov}, \citenamefont {Mazzola}, \citenamefont {Smelyanskiy}, \citenamefont {Jiang}, \citenamefont {Boixo}, \citenamefont {Neven},\ and\ \citenamefont {Troyer}}]{isakov2016understanding}%
  \BibitemOpen
  \bibfield  {author} {\bibinfo {author} {\bibfnamefont {S.~V.}\ \bibnamefont {Isakov}}, \bibinfo {author} {\bibfnamefont {G.}~\bibnamefont {Mazzola}}, \bibinfo {author} {\bibfnamefont {V.~N.}\ \bibnamefont {Smelyanskiy}}, \bibinfo {author} {\bibfnamefont {Z.}~\bibnamefont {Jiang}}, \bibinfo {author} {\bibfnamefont {S.}~\bibnamefont {Boixo}}, \bibinfo {author} {\bibfnamefont {H.}~\bibnamefont {Neven}},\ and\ \bibinfo {author} {\bibfnamefont {M.}~\bibnamefont {Troyer}},\ }\bibfield  {title} {\bibinfo {title} {Understanding quantum tunneling through quantum monte carlo simulations},\ }\href@noop {} {\bibfield  {journal} {\bibinfo  {journal} {Physical review letters}\ }\textbf {\bibinfo {volume} {117}},\ \bibinfo {pages} {180402} (\bibinfo {year} {2016})}\BibitemShut {NoStop}%
\bibitem [{\citenamefont {Dziarmaga}\ and\ \citenamefont {Rams}(2022)}]{Dziarmaga2022kinkcorrelations}%
  \BibitemOpen
  \bibfield  {author} {\bibinfo {author} {\bibfnamefont {J.}~\bibnamefont {Dziarmaga}}\ and\ \bibinfo {author} {\bibfnamefont {M.~M.}\ \bibnamefont {Rams}},\ }\bibfield  {title} {\bibinfo {title} {Kink correlations, domain-size distribution, and emptiness formation probability after a kibble-zurek quench in the quantum ising chain},\ }\href {https://doi.org/10.1103/PhysRevB.106.014309} {\bibfield  {journal} {\bibinfo  {journal} {Phys. Rev. B}\ }\textbf {\bibinfo {volume} {106}},\ \bibinfo {pages} {014309} (\bibinfo {year} {2022})}\BibitemShut {NoStop}%
\bibitem [{\citenamefont {Motta}\ and\ \citenamefont {Rice}(2022)}]{Motta2022}%
  \BibitemOpen
  \bibfield  {author} {\bibinfo {author} {\bibfnamefont {M.}~\bibnamefont {Motta}}\ and\ \bibinfo {author} {\bibfnamefont {J.~E.}\ \bibnamefont {Rice}},\ }\bibfield  {title} {\bibinfo {title} {Emerging quantum computing algorithms for quantum chemistry},\ }\href {https://doi.org/10.1002/wcms.1580} {\bibfield  {journal} {\bibinfo  {journal} {WIREs Comput Mol Sci.}\ }\textbf {\bibinfo {volume} {12}},\ \bibinfo {pages} {e1580} (\bibinfo {year} {2022})}\BibitemShut {NoStop}%
\bibitem [{\citenamefont {Layden}\ \emph {et~al.}(2023)\citenamefont {Layden}, \citenamefont {Mazzola}, \citenamefont {Mishmash}, \citenamefont {Motta}, \citenamefont {Wocjan}, \citenamefont {Kim},\ and\ \citenamefont {Sheldon}}]{layden2023quantum}%
  \BibitemOpen
  \bibfield  {author} {\bibinfo {author} {\bibfnamefont {D.}~\bibnamefont {Layden}}, \bibinfo {author} {\bibfnamefont {G.}~\bibnamefont {Mazzola}}, \bibinfo {author} {\bibfnamefont {R.~V.}\ \bibnamefont {Mishmash}}, \bibinfo {author} {\bibfnamefont {M.}~\bibnamefont {Motta}}, \bibinfo {author} {\bibfnamefont {P.}~\bibnamefont {Wocjan}}, \bibinfo {author} {\bibfnamefont {J.-S.}\ \bibnamefont {Kim}},\ and\ \bibinfo {author} {\bibfnamefont {S.}~\bibnamefont {Sheldon}},\ }\bibfield  {title} {\bibinfo {title} {Quantum-enhanced markov chain monte carlo},\ }\href {https://doi.org/10.1038/s41586-023-06095-4} {\bibfield  {journal} {\bibinfo  {journal} {Nature}\ }\textbf {\bibinfo {volume} {619}},\ \bibinfo {pages} {282} (\bibinfo {year} {2023})}\BibitemShut {NoStop}%
\bibitem [{\citenamefont {{Qiskit contributors}}(2023)}]{Qiskit}%
  \BibitemOpen
  \bibfield  {author} {\bibinfo {author} {\bibnamefont {{Qiskit contributors}}},\ }\href {https://doi.org/10.5281/zenodo.2573505} {\bibinfo {title} {Qiskit: An open-source framework for quantum computing}} (\bibinfo {year} {2023})\BibitemShut {NoStop}%
\bibitem [{IBM(2021)}]{IBMQuantumPlatform}%
  \BibitemOpen
  \href {https://quantum.ibm.com/} {\bibinfo {title} {{IBM Quantum}}},\ \bibinfo {howpublished} {\url{https://quantum-computing.ibm.com/}} (\bibinfo {year} {2021})\BibitemShut {NoStop}%
\bibitem [{\citenamefont {Rigetti}\ and\ \citenamefont {Devoret}(2010)}]{Rigetti2010ecrGate}%
  \BibitemOpen
  \bibfield  {author} {\bibinfo {author} {\bibfnamefont {C.}~\bibnamefont {Rigetti}}\ and\ \bibinfo {author} {\bibfnamefont {M.}~\bibnamefont {Devoret}},\ }\bibfield  {title} {\bibinfo {title} {Fully microwave-tunable universal gates in superconducting qubits with linear couplings and fixed transition frequencies},\ }\href {https://doi.org/10.1103/PhysRevB.81.134507} {\bibfield  {journal} {\bibinfo  {journal} {Phys. Rev. B}\ }\textbf {\bibinfo {volume} {81}},\ \bibinfo {pages} {134507} (\bibinfo {year} {2010})}\BibitemShut {NoStop}%
\bibitem [{\citenamefont {Sheldon}\ \emph {et~al.}(2016)\citenamefont {Sheldon}, \citenamefont {Magesan}, \citenamefont {Chow},\ and\ \citenamefont {Gambetta}}]{Sheldon2016ecrGate}%
  \BibitemOpen
  \bibfield  {author} {\bibinfo {author} {\bibfnamefont {S.}~\bibnamefont {Sheldon}}, \bibinfo {author} {\bibfnamefont {E.}~\bibnamefont {Magesan}}, \bibinfo {author} {\bibfnamefont {J.~M.}\ \bibnamefont {Chow}},\ and\ \bibinfo {author} {\bibfnamefont {J.~M.}\ \bibnamefont {Gambetta}},\ }\bibfield  {title} {\bibinfo {title} {Procedure for systematically tuning up cross-talk in the cross-resonance gate},\ }\href {https://doi.org/10.1103/PhysRevA.93.060302} {\bibfield  {journal} {\bibinfo  {journal} {Phys. Rev. A}\ }\textbf {\bibinfo {volume} {93}},\ \bibinfo {pages} {060302} (\bibinfo {year} {2016})}\BibitemShut {NoStop}%
\bibitem [{\citenamefont {McKay}\ \emph {et~al.}(2016)\citenamefont {McKay}, \citenamefont {Filipp}, \citenamefont {Mezzacapo}, \citenamefont {Magesan}, \citenamefont {Chow},\ and\ \citenamefont {Gambetta}}]{Mckay2016universal}%
  \BibitemOpen
  \bibfield  {author} {\bibinfo {author} {\bibfnamefont {D.~C.}\ \bibnamefont {McKay}}, \bibinfo {author} {\bibfnamefont {S.}~\bibnamefont {Filipp}}, \bibinfo {author} {\bibfnamefont {A.}~\bibnamefont {Mezzacapo}}, \bibinfo {author} {\bibfnamefont {E.}~\bibnamefont {Magesan}}, \bibinfo {author} {\bibfnamefont {J.~M.}\ \bibnamefont {Chow}},\ and\ \bibinfo {author} {\bibfnamefont {J.~M.}\ \bibnamefont {Gambetta}},\ }\bibfield  {title} {\bibinfo {title} {Universal gate for fixed-frequency qubits via a tunable bus},\ }\href {https://doi.org/10.1103/PhysRevApplied.6.064007} {\bibfield  {journal} {\bibinfo  {journal} {Phys. Rev. Appl.}\ }\textbf {\bibinfo {volume} {6}},\ \bibinfo {pages} {064007} (\bibinfo {year} {2016})}\BibitemShut {NoStop}%
\bibitem [{\citenamefont {Ganzhorn}\ \emph {et~al.}(2020)\citenamefont {Ganzhorn}, \citenamefont {Salis}, \citenamefont {Egger}, \citenamefont {Fuhrer}, \citenamefont {Mergenthaler}, \citenamefont {M\"uller}, \citenamefont {M\"uller}, \citenamefont {Paredes}, \citenamefont {Pechal}, \citenamefont {Werninghaus},\ and\ \citenamefont {Filipp}}]{Ganzhorn2020benchmarking}%
  \BibitemOpen
  \bibfield  {author} {\bibinfo {author} {\bibfnamefont {M.}~\bibnamefont {Ganzhorn}}, \bibinfo {author} {\bibfnamefont {G.}~\bibnamefont {Salis}}, \bibinfo {author} {\bibfnamefont {D.~J.}\ \bibnamefont {Egger}}, \bibinfo {author} {\bibfnamefont {A.}~\bibnamefont {Fuhrer}}, \bibinfo {author} {\bibfnamefont {M.}~\bibnamefont {Mergenthaler}}, \bibinfo {author} {\bibfnamefont {C.}~\bibnamefont {M\"uller}}, \bibinfo {author} {\bibfnamefont {P.}~\bibnamefont {M\"uller}}, \bibinfo {author} {\bibfnamefont {S.}~\bibnamefont {Paredes}}, \bibinfo {author} {\bibfnamefont {M.}~\bibnamefont {Pechal}}, \bibinfo {author} {\bibfnamefont {M.}~\bibnamefont {Werninghaus}},\ and\ \bibinfo {author} {\bibfnamefont {S.}~\bibnamefont {Filipp}},\ }\bibfield  {title} {\bibinfo {title} {Benchmarking the noise sensitivity of different parametric two-qubit gates in a single superconducting quantum computing platform},\ }\href {https://doi.org/10.1103/PhysRevResearch.2.033447} {\bibfield  {journal} {\bibinfo  {journal} {Phys. Rev. Res.}\
  }\textbf {\bibinfo {volume} {2}},\ \bibinfo {pages} {033447} (\bibinfo {year} {2020})}\BibitemShut {NoStop}%
\bibitem [{\citenamefont {van~den Berg}\ \emph {et~al.}(2022)\citenamefont {van~den Berg}, \citenamefont {Minev},\ and\ \citenamefont {Temme}}]{Berg2022trex}%
  \BibitemOpen
  \bibfield  {author} {\bibinfo {author} {\bibfnamefont {E.}~\bibnamefont {van~den Berg}}, \bibinfo {author} {\bibfnamefont {Z.~K.}\ \bibnamefont {Minev}},\ and\ \bibinfo {author} {\bibfnamefont {K.}~\bibnamefont {Temme}},\ }\bibfield  {title} {\bibinfo {title} {Model-free readout-error mitigation for quantum expectation values},\ }\href {https://doi.org/10.1103/PhysRevA.105.032620} {\bibfield  {journal} {\bibinfo  {journal} {Phys. Rev. A}\ }\textbf {\bibinfo {volume} {105}},\ \bibinfo {pages} {032620} (\bibinfo {year} {2022})}\BibitemShut {NoStop}%
\bibitem [{\citenamefont {Nation}\ \emph {et~al.}(2021)\citenamefont {Nation}, \citenamefont {Kang}, \citenamefont {Sundaresan},\ and\ \citenamefont {Gambetta}}]{Nation2021m3}%
  \BibitemOpen
  \bibfield  {author} {\bibinfo {author} {\bibfnamefont {P.~D.}\ \bibnamefont {Nation}}, \bibinfo {author} {\bibfnamefont {H.}~\bibnamefont {Kang}}, \bibinfo {author} {\bibfnamefont {N.}~\bibnamefont {Sundaresan}},\ and\ \bibinfo {author} {\bibfnamefont {J.~M.}\ \bibnamefont {Gambetta}},\ }\bibfield  {title} {\bibinfo {title} {Scalable mitigation of measurement errors on quantum computers},\ }\href {https://doi.org/10.1103/PRXQuantum.2.040326} {\bibfield  {journal} {\bibinfo  {journal} {PRX Quantum}\ }\textbf {\bibinfo {volume} {2}},\ \bibinfo {pages} {040326} (\bibinfo {year} {2021})}\BibitemShut {NoStop}%
\bibitem [{\citenamefont {Barron}\ \emph {et~al.}(2023)\citenamefont {Barron}, \citenamefont {Egger}, \citenamefont {Pelofske}, \citenamefont {Bärtschi}, \citenamefont {Eidenbenz}, \citenamefont {Lehmkuehler},\ and\ \citenamefont {Woerner}}]{Barron2023qaoa}%
  \BibitemOpen
  \bibfield  {author} {\bibinfo {author} {\bibfnamefont {S.~V.}\ \bibnamefont {Barron}}, \bibinfo {author} {\bibfnamefont {D.~J.}\ \bibnamefont {Egger}}, \bibinfo {author} {\bibfnamefont {E.}~\bibnamefont {Pelofske}}, \bibinfo {author} {\bibfnamefont {A.}~\bibnamefont {Bärtschi}}, \bibinfo {author} {\bibfnamefont {S.}~\bibnamefont {Eidenbenz}}, \bibinfo {author} {\bibfnamefont {M.}~\bibnamefont {Lehmkuehler}},\ and\ \bibinfo {author} {\bibfnamefont {S.}~\bibnamefont {Woerner}},\ }\bibfield  {title} {\bibinfo {title} {Provable bounds for noise-free expectation values computed from noisy samples},\ }\href {http://arxiv.org/abs/2312.00733} {\bibfield  {journal} {\bibinfo  {journal} {arXiv preprint}\ } (\bibinfo {year} {2023})},\ \Eprint {https://arxiv.org/abs/2312.00733} {arXiv:2312.00733} \BibitemShut {NoStop}%
\bibitem [{\citenamefont {Earnest}\ \emph {et~al.}(2021)\citenamefont {Earnest}, \citenamefont {Tornow},\ and\ \citenamefont {Egger}}]{Earnest2023pulse}%
  \BibitemOpen
  \bibfield  {author} {\bibinfo {author} {\bibfnamefont {N.}~\bibnamefont {Earnest}}, \bibinfo {author} {\bibfnamefont {C.}~\bibnamefont {Tornow}},\ and\ \bibinfo {author} {\bibfnamefont {D.~J.}\ \bibnamefont {Egger}},\ }\bibfield  {title} {\bibinfo {title} {Pulse-efficient circuit transpilation for quantum applications on cross-resonance-based hardware},\ }\href {https://doi.org/10.1103/PhysRevResearch.3.043088} {\bibfield  {journal} {\bibinfo  {journal} {Phys. Rev. Res.}\ }\textbf {\bibinfo {volume} {3}},\ \bibinfo {pages} {043088} (\bibinfo {year} {2021})}\BibitemShut {NoStop}%
\bibitem [{\citenamefont {Nowak}\ and\ \citenamefont {Dziarmaga}(2021)}]{Nowak2021quantum}%
  \BibitemOpen
  \bibfield  {author} {\bibinfo {author} {\bibfnamefont {R.~J.}\ \bibnamefont {Nowak}}\ and\ \bibinfo {author} {\bibfnamefont {J.}~\bibnamefont {Dziarmaga}},\ }\bibfield  {title} {\bibinfo {title} {Quantum kibble-zurek mechanism: Kink correlations after a quench in the quantum ising chain},\ }\href {https://doi.org/10.1103/PhysRevB.104.075448} {\bibfield  {journal} {\bibinfo  {journal} {Phys. Rev. B}\ }\textbf {\bibinfo {volume} {104}},\ \bibinfo {pages} {075448} (\bibinfo {year} {2021})}\BibitemShut {NoStop}%
\bibitem [{\citenamefont {Vazquez}\ \emph {et~al.}(2023)\citenamefont {Vazquez}, \citenamefont {Egger}, \citenamefont {Ochsner},\ and\ \citenamefont {Woerner}}]{Vazquez2023wellConditioned}%
  \BibitemOpen
  \bibfield  {author} {\bibinfo {author} {\bibfnamefont {A.~C.}\ \bibnamefont {Vazquez}}, \bibinfo {author} {\bibfnamefont {D.~J.}\ \bibnamefont {Egger}}, \bibinfo {author} {\bibfnamefont {D.}~\bibnamefont {Ochsner}},\ and\ \bibinfo {author} {\bibfnamefont {S.}~\bibnamefont {Woerner}},\ }\bibfield  {title} {\bibinfo {title} {Well-conditioned multi-product formulas for hardware-friendly hamiltonian simulation},\ }\href {https://doi.org/10.22331/q-2023-07-25-1067} {\bibfield  {journal} {\bibinfo  {journal} {Quantum}\ }\textbf {\bibinfo {volume} {7}},\ \bibinfo {pages} {1067} (\bibinfo {year} {2023})}\BibitemShut {NoStop}%
\bibitem [{\citenamefont {Astrakhantsev}\ \emph {et~al.}(2023)\citenamefont {Astrakhantsev}, \citenamefont {Mazzola}, \citenamefont {Tavernelli},\ and\ \citenamefont {Carleo}}]{astrakhantsev2023phenomenological}%
  \BibitemOpen
  \bibfield  {author} {\bibinfo {author} {\bibfnamefont {N.}~\bibnamefont {Astrakhantsev}}, \bibinfo {author} {\bibfnamefont {G.}~\bibnamefont {Mazzola}}, \bibinfo {author} {\bibfnamefont {I.}~\bibnamefont {Tavernelli}},\ and\ \bibinfo {author} {\bibfnamefont {G.}~\bibnamefont {Carleo}},\ }\bibfield  {title} {\bibinfo {title} {Phenomenological theory of variational quantum ground-state preparation},\ }\href {https://doi.org/10.1103/PhysRevResearch.5.033225} {\bibfield  {journal} {\bibinfo  {journal} {Phys. Rev. Res.}\ }\textbf {\bibinfo {volume} {5}},\ \bibinfo {pages} {033225} (\bibinfo {year} {2023})}\BibitemShut {NoStop}%
\bibitem [{\citenamefont {Havl{\'\i}{\v{c}}ek}\ \emph {et~al.}(2019)\citenamefont {Havl{\'\i}{\v{c}}ek}, \citenamefont {C{\'o}rcoles}, \citenamefont {Temme}, \citenamefont {Harrow}, \citenamefont {Kandala}, \citenamefont {Chow},\ and\ \citenamefont {Gambetta}}]{havlivcek2019supervised}%
  \BibitemOpen
  \bibfield  {author} {\bibinfo {author} {\bibfnamefont {V.}~\bibnamefont {Havl{\'\i}{\v{c}}ek}}, \bibinfo {author} {\bibfnamefont {A.~D.}\ \bibnamefont {C{\'o}rcoles}}, \bibinfo {author} {\bibfnamefont {K.}~\bibnamefont {Temme}}, \bibinfo {author} {\bibfnamefont {A.~W.}\ \bibnamefont {Harrow}}, \bibinfo {author} {\bibfnamefont {A.}~\bibnamefont {Kandala}}, \bibinfo {author} {\bibfnamefont {J.~M.}\ \bibnamefont {Chow}},\ and\ \bibinfo {author} {\bibfnamefont {J.~M.}\ \bibnamefont {Gambetta}},\ }\bibfield  {title} {\bibinfo {title} {Supervised learning with quantum-enhanced feature spaces},\ }\href {https://doi.org/10.1038/s41586-019-0980-2} {\bibfield  {journal} {\bibinfo  {journal} {Nature}\ }\textbf {\bibinfo {volume} {567}},\ \bibinfo {pages} {209} (\bibinfo {year} {2019})}\BibitemShut {NoStop}%
\bibitem [{\citenamefont {Melo}\ \emph {et~al.}(2023)\citenamefont {Melo}, \citenamefont {Earnest-Noble},\ and\ \citenamefont {Tacchino}}]{Melo2023pulseefficient}%
  \BibitemOpen
  \bibfield  {author} {\bibinfo {author} {\bibfnamefont {A.}~\bibnamefont {Melo}}, \bibinfo {author} {\bibfnamefont {N.}~\bibnamefont {Earnest-Noble}},\ and\ \bibinfo {author} {\bibfnamefont {F.}~\bibnamefont {Tacchino}},\ }\bibfield  {title} {\bibinfo {title} {Pulse-efficient quantum machine learning},\ }\href {https://doi.org/10.22331/q-2023-10-09-1130} {\bibfield  {journal} {\bibinfo  {journal} {{Quantum}}\ }\textbf {\bibinfo {volume} {7}},\ \bibinfo {pages} {1130} (\bibinfo {year} {2023})}\BibitemShut {NoStop}%
\bibitem [{\citenamefont {Abbas}\ \emph {et~al.}(2021)\citenamefont {Abbas}, \citenamefont {Sutter}, \citenamefont {Zoufal}, \citenamefont {Lucchi}, \citenamefont {Figalli},\ and\ \citenamefont {Woerner}}]{abbas2021power}%
  \BibitemOpen
  \bibfield  {author} {\bibinfo {author} {\bibfnamefont {A.}~\bibnamefont {Abbas}}, \bibinfo {author} {\bibfnamefont {D.}~\bibnamefont {Sutter}}, \bibinfo {author} {\bibfnamefont {C.}~\bibnamefont {Zoufal}}, \bibinfo {author} {\bibfnamefont {A.}~\bibnamefont {Lucchi}}, \bibinfo {author} {\bibfnamefont {A.}~\bibnamefont {Figalli}},\ and\ \bibinfo {author} {\bibfnamefont {S.}~\bibnamefont {Woerner}},\ }\bibfield  {title} {\bibinfo {title} {The power of quantum neural networks},\ }\href@noop {} {\bibfield  {journal} {\bibinfo  {journal} {Nature Computational Science}\ }\textbf {\bibinfo {volume} {1}},\ \bibinfo {pages} {403} (\bibinfo {year} {2021})}\BibitemShut {NoStop}%
\bibitem [{\citenamefont {R{\o}nnow}\ \emph {et~al.}(2014)\citenamefont {R{\o}nnow}, \citenamefont {Wang}, \citenamefont {Job}, \citenamefont {Boixo}, \citenamefont {Isakov}, \citenamefont {Wecker}, \citenamefont {Martinis}, \citenamefont {Lidar},\ and\ \citenamefont {Troyer}}]{ronnow2014defining}%
  \BibitemOpen
  \bibfield  {author} {\bibinfo {author} {\bibfnamefont {T.~F.}\ \bibnamefont {R{\o}nnow}}, \bibinfo {author} {\bibfnamefont {Z.}~\bibnamefont {Wang}}, \bibinfo {author} {\bibfnamefont {J.}~\bibnamefont {Job}}, \bibinfo {author} {\bibfnamefont {S.}~\bibnamefont {Boixo}}, \bibinfo {author} {\bibfnamefont {S.~V.}\ \bibnamefont {Isakov}}, \bibinfo {author} {\bibfnamefont {D.}~\bibnamefont {Wecker}}, \bibinfo {author} {\bibfnamefont {J.~M.}\ \bibnamefont {Martinis}}, \bibinfo {author} {\bibfnamefont {D.~A.}\ \bibnamefont {Lidar}},\ and\ \bibinfo {author} {\bibfnamefont {M.}~\bibnamefont {Troyer}},\ }\bibfield  {title} {\bibinfo {title} {Defining and detecting quantum speedup},\ }\href {https://doi.org/10.1126/science.1252319} {\bibfield  {journal} {\bibinfo  {journal} {Science}\ }\textbf {\bibinfo {volume} {345}},\ \bibinfo {pages} {420} (\bibinfo {year} {2014})}\BibitemShut {NoStop}%
\bibitem [{\citenamefont {Albash}\ and\ \citenamefont {Lidar}(2018{\natexlab{b}})}]{Albash2018scalingAdvantageAnnealing}%
  \BibitemOpen
  \bibfield  {author} {\bibinfo {author} {\bibfnamefont {T.}~\bibnamefont {Albash}}\ and\ \bibinfo {author} {\bibfnamefont {D.~A.}\ \bibnamefont {Lidar}},\ }\bibfield  {title} {\bibinfo {title} {Demonstration of a scaling advantage for a quantum annealer over simulated annealing},\ }\href {https://doi.org/10.1103/PhysRevX.8.031016} {\bibfield  {journal} {\bibinfo  {journal} {Phys. Rev. X}\ }\textbf {\bibinfo {volume} {8}},\ \bibinfo {pages} {031016} (\bibinfo {year} {2018}{\natexlab{b}})}\BibitemShut {NoStop}%
\bibitem [{\citenamefont {Mc~Keever}\ and\ \citenamefont {Lubasch}(2024)}]{keever2023adiabatic}%
  \BibitemOpen
  \bibfield  {author} {\bibinfo {author} {\bibfnamefont {C.}~\bibnamefont {Mc~Keever}}\ and\ \bibinfo {author} {\bibfnamefont {M.}~\bibnamefont {Lubasch}},\ }\bibfield  {title} {\bibinfo {title} {Towards adiabatic quantum computing using compressed quantum circuits},\ }\href {https://doi.org/10.1103/PRXQuantum.5.020362} {\bibfield  {journal} {\bibinfo  {journal} {PRX Quantum}\ }\textbf {\bibinfo {volume} {5}},\ \bibinfo {pages} {020362} (\bibinfo {year} {2024})}\BibitemShut {NoStop}%
\bibitem [{\citenamefont {Sack}\ and\ \citenamefont {Egger}(2024)}]{Sack2024qaoa}%
  \BibitemOpen
  \bibfield  {author} {\bibinfo {author} {\bibfnamefont {S.~H.}\ \bibnamefont {Sack}}\ and\ \bibinfo {author} {\bibfnamefont {D.~J.}\ \bibnamefont {Egger}},\ }\bibfield  {title} {\bibinfo {title} {Large-scale quantum approximate optimization on nonplanar graphs with machine learning noise mitigation},\ }\href {https://doi.org/10.1103/PhysRevResearch.6.013223} {\bibfield  {journal} {\bibinfo  {journal} {Phys. Rev. Res.}\ }\textbf {\bibinfo {volume} {6}},\ \bibinfo {pages} {013223} (\bibinfo {year} {2024})}\BibitemShut {NoStop}%
\bibitem [{\citenamefont {Hadfield}\ \emph {et~al.}(2019)\citenamefont {Hadfield}, \citenamefont {Wang}, \citenamefont {O'Gorman}, \citenamefont {Rieffel}, \citenamefont {Venturelli},\ and\ \citenamefont {Biswas}}]{Hadfield2019}%
  \BibitemOpen
  \bibfield  {author} {\bibinfo {author} {\bibfnamefont {S.}~\bibnamefont {Hadfield}}, \bibinfo {author} {\bibfnamefont {Z.}~\bibnamefont {Wang}}, \bibinfo {author} {\bibfnamefont {B.}~\bibnamefont {O'Gorman}}, \bibinfo {author} {\bibfnamefont {E.}~\bibnamefont {Rieffel}}, \bibinfo {author} {\bibfnamefont {D.}~\bibnamefont {Venturelli}},\ and\ \bibinfo {author} {\bibfnamefont {R.}~\bibnamefont {Biswas}},\ }\bibfield  {title} {\bibinfo {title} {From the quantum approximate optimization algorithm to a quantum alternating operator ansatz},\ }\href {https://doi.org/10.3390/a12020034} {\bibfield  {journal} {\bibinfo  {journal} {Algorithms}\ }\textbf {\bibinfo {volume} {12}},\ \bibinfo {pages} {34} (\bibinfo {year} {2019})}\BibitemShut {NoStop}%
\bibitem [{\citenamefont {Scriva}\ \emph {et~al.}(2024)\citenamefont {Scriva}, \citenamefont {Astrakhantsev}, \citenamefont {Pilati},\ and\ \citenamefont {Mazzola}}]{Scriva2024qaoa}%
  \BibitemOpen
  \bibfield  {author} {\bibinfo {author} {\bibfnamefont {G.}~\bibnamefont {Scriva}}, \bibinfo {author} {\bibfnamefont {N.}~\bibnamefont {Astrakhantsev}}, \bibinfo {author} {\bibfnamefont {S.}~\bibnamefont {Pilati}},\ and\ \bibinfo {author} {\bibfnamefont {G.}~\bibnamefont {Mazzola}},\ }\bibfield  {title} {\bibinfo {title} {Challenges of variational quantum optimization with measurement shot noise},\ }\href {https://doi.org/10.1103/PhysRevA.109.032408} {\bibfield  {journal} {\bibinfo  {journal} {Phys. Rev. A}\ }\textbf {\bibinfo {volume} {109}},\ \bibinfo {pages} {032408} (\bibinfo {year} {2024})}\BibitemShut {NoStop}%
\bibitem [{\citenamefont {Sack}\ and\ \citenamefont {Serbyn}(2021)}]{Sack2019qaoa}%
  \BibitemOpen
  \bibfield  {author} {\bibinfo {author} {\bibfnamefont {S.~H.}\ \bibnamefont {Sack}}\ and\ \bibinfo {author} {\bibfnamefont {M.}~\bibnamefont {Serbyn}},\ }\bibfield  {title} {\bibinfo {title} {Quantum annealing initialization of the quantum approximate optimization algorithm},\ }\href {https://doi.org/10.22331/q-2021-07-01-491} {\bibfield  {journal} {\bibinfo  {journal} {Quantum}\ }\textbf {\bibinfo {volume} {5}},\ \bibinfo {pages} {491} (\bibinfo {year} {2021})}\BibitemShut {NoStop}%
\bibitem [{\citenamefont {Barends}\ \emph {et~al.}(2016)\citenamefont {Barends}, \citenamefont {Shabani}, \citenamefont {Lamata}, \citenamefont {Kelly}, \citenamefont {Mezzacapo}, \citenamefont {Heras}, \citenamefont {Babbush}, \citenamefont {Fowler}, \citenamefont {Campbell}, \citenamefont {Chen} \emph {et~al.}}]{barends2016digitized}%
  \BibitemOpen
  \bibfield  {author} {\bibinfo {author} {\bibfnamefont {R.}~\bibnamefont {Barends}}, \bibinfo {author} {\bibfnamefont {A.}~\bibnamefont {Shabani}}, \bibinfo {author} {\bibfnamefont {L.}~\bibnamefont {Lamata}}, \bibinfo {author} {\bibfnamefont {J.}~\bibnamefont {Kelly}}, \bibinfo {author} {\bibfnamefont {A.}~\bibnamefont {Mezzacapo}}, \bibinfo {author} {\bibfnamefont {U.~L.}\ \bibnamefont {Heras}}, \bibinfo {author} {\bibfnamefont {R.}~\bibnamefont {Babbush}}, \bibinfo {author} {\bibfnamefont {A.~G.}\ \bibnamefont {Fowler}}, \bibinfo {author} {\bibfnamefont {B.}~\bibnamefont {Campbell}}, \bibinfo {author} {\bibfnamefont {Y.}~\bibnamefont {Chen}}, \emph {et~al.},\ }\bibfield  {title} {\bibinfo {title} {Digitized adiabatic quantum computing with a superconducting circuit},\ }\href {https://doi.org/10.1038/nature17658} {\bibfield  {journal} {\bibinfo  {journal} {Nature}\ }\textbf {\bibinfo {volume} {534}},\ \bibinfo {pages} {222} (\bibinfo {year} {2016})}\BibitemShut {NoStop}%
\bibitem [{\citenamefont {Santoro}\ \emph {et~al.}(2002)\citenamefont {Santoro}, \citenamefont {Marton{\'a}k}, \citenamefont {Tosatti},\ and\ \citenamefont {Car}}]{santoro2002theory}%
  \BibitemOpen
  \bibfield  {author} {\bibinfo {author} {\bibfnamefont {G.~E.}\ \bibnamefont {Santoro}}, \bibinfo {author} {\bibfnamefont {R.}~\bibnamefont {Marton{\'a}k}}, \bibinfo {author} {\bibfnamefont {E.}~\bibnamefont {Tosatti}},\ and\ \bibinfo {author} {\bibfnamefont {R.}~\bibnamefont {Car}},\ }\bibfield  {title} {\bibinfo {title} {Theory of quantum annealing of an ising spin glass},\ }\href {https://doi.org/10.1126/science.1068774} {\bibfield  {journal} {\bibinfo  {journal} {Science}\ }\textbf {\bibinfo {volume} {295}},\ \bibinfo {pages} {2427} (\bibinfo {year} {2002})}\BibitemShut {NoStop}%
\bibitem [{cpl()}]{cplex}%
  \BibitemOpen
  \href@noop {} {\bibinfo {title} {{IBM ILOG CPLEX Optimizer}}}\BibitemShut {NoStop}%
\bibitem [{Note1()}]{Note1}%
  \BibitemOpen
  \bibinfo {note} {\protect \texttt {ibm\protect \_auckland}{} is one of IBM's by now retired 27-qubit Falcon chips}\BibitemShut {NoStop}%
\bibitem [{\citenamefont {Ebadi}\ \emph {et~al.}(2022)\citenamefont {Ebadi}, \citenamefont {Keesling}, \citenamefont {Cain}, \citenamefont {Wang}, \citenamefont {Levine}, \citenamefont {Bluvstein}, \citenamefont {Semeghini}, \citenamefont {Omran}, \citenamefont {Liu}, \citenamefont {Samajdar} \emph {et~al.}}]{ebadi2022quantum}%
  \BibitemOpen
  \bibfield  {author} {\bibinfo {author} {\bibfnamefont {S.}~\bibnamefont {Ebadi}}, \bibinfo {author} {\bibfnamefont {A.}~\bibnamefont {Keesling}}, \bibinfo {author} {\bibfnamefont {M.}~\bibnamefont {Cain}}, \bibinfo {author} {\bibfnamefont {T.~T.}\ \bibnamefont {Wang}}, \bibinfo {author} {\bibfnamefont {H.}~\bibnamefont {Levine}}, \bibinfo {author} {\bibfnamefont {D.}~\bibnamefont {Bluvstein}}, \bibinfo {author} {\bibfnamefont {G.}~\bibnamefont {Semeghini}}, \bibinfo {author} {\bibfnamefont {A.}~\bibnamefont {Omran}}, \bibinfo {author} {\bibfnamefont {J.-G.}\ \bibnamefont {Liu}}, \bibinfo {author} {\bibfnamefont {R.}~\bibnamefont {Samajdar}}, \emph {et~al.},\ }\bibfield  {title} {\bibinfo {title} {Quantum optimization of maximum independent set using rydberg atom arrays},\ }\href {https://doi.org/10.1126/science.abo6587} {\bibfield  {journal} {\bibinfo  {journal} {Science}\ }\textbf {\bibinfo {volume} {376}},\ \bibinfo {pages} {1209} (\bibinfo {year} {2022})}\BibitemShut {NoStop}%
\bibitem [{\citenamefont {Weidenfeller}\ \emph {et~al.}(2022)\citenamefont {Weidenfeller}, \citenamefont {Valor}, \citenamefont {Gacon}, \citenamefont {Tornow}, \citenamefont {Bello}, \citenamefont {Woerner},\ and\ \citenamefont {Egger}}]{Weidenfeller2022scalingofquantum}%
  \BibitemOpen
  \bibfield  {author} {\bibinfo {author} {\bibfnamefont {J.}~\bibnamefont {Weidenfeller}}, \bibinfo {author} {\bibfnamefont {L.~C.}\ \bibnamefont {Valor}}, \bibinfo {author} {\bibfnamefont {J.}~\bibnamefont {Gacon}}, \bibinfo {author} {\bibfnamefont {C.}~\bibnamefont {Tornow}}, \bibinfo {author} {\bibfnamefont {L.}~\bibnamefont {Bello}}, \bibinfo {author} {\bibfnamefont {S.}~\bibnamefont {Woerner}},\ and\ \bibinfo {author} {\bibfnamefont {D.~J.}\ \bibnamefont {Egger}},\ }\bibfield  {title} {\bibinfo {title} {Scaling of the quantum approximate optimization algorithm on superconducting qubit based hardware},\ }\href {https://doi.org/10.22331/q-2022-12-07-870} {\bibfield  {journal} {\bibinfo  {journal} {{Quantum}}\ }\textbf {\bibinfo {volume} {6}},\ \bibinfo {pages} {870} (\bibinfo {year} {2022})}\BibitemShut {NoStop}%
\bibitem [{\citenamefont {Mazzola}(2024)}]{mazzola2024quantum}%
  \BibitemOpen
  \bibfield  {author} {\bibinfo {author} {\bibfnamefont {G.}~\bibnamefont {Mazzola}},\ }\bibfield  {title} {\bibinfo {title} {Quantum computing for chemistry and physics applications from a monte carlo perspective},\ }\href {https://doi.org/10.1063/5.0173591} {\bibfield  {journal} {\bibinfo  {journal} {J. Chem. Phys.}\ }\textbf {\bibinfo {volume} {160}},\ \bibinfo {pages} {010901} (\bibinfo {year} {2024})}\BibitemShut {NoStop}%
\bibitem [{\citenamefont {Farrell}\ \emph {et~al.}(2024)\citenamefont {Farrell}, \citenamefont {Illa}, \citenamefont {Ciavarella},\ and\ \citenamefont {Savage}}]{farrell2023scalable}%
  \BibitemOpen
  \bibfield  {author} {\bibinfo {author} {\bibfnamefont {R.~C.}\ \bibnamefont {Farrell}}, \bibinfo {author} {\bibfnamefont {M.}~\bibnamefont {Illa}}, \bibinfo {author} {\bibfnamefont {A.~N.}\ \bibnamefont {Ciavarella}},\ and\ \bibinfo {author} {\bibfnamefont {M.~J.}\ \bibnamefont {Savage}},\ }\bibfield  {title} {\bibinfo {title} {Scalable circuits for preparing ground states on digital quantum computers: The schwinger model vacuum on 100 qubits},\ }\href {https://doi.org/10.1103/PRXQuantum.5.020315} {\bibfield  {journal} {\bibinfo  {journal} {PRX Quantum}\ }\textbf {\bibinfo {volume} {5}},\ \bibinfo {pages} {020315} (\bibinfo {year} {2024})}\BibitemShut {NoStop}%
\bibitem [{\citenamefont {Pelofske}\ \emph {et~al.}(2023)\citenamefont {Pelofske}, \citenamefont {Bärtschi}, \citenamefont {Cincio}, \citenamefont {Golden},\ and\ \citenamefont {Eidenbenz}}]{pelofske2023scaling}%
  \BibitemOpen
  \bibfield  {author} {\bibinfo {author} {\bibfnamefont {E.}~\bibnamefont {Pelofske}}, \bibinfo {author} {\bibfnamefont {A.}~\bibnamefont {Bärtschi}}, \bibinfo {author} {\bibfnamefont {L.}~\bibnamefont {Cincio}}, \bibinfo {author} {\bibfnamefont {J.}~\bibnamefont {Golden}},\ and\ \bibinfo {author} {\bibfnamefont {S.}~\bibnamefont {Eidenbenz}},\ }\bibfield  {title} {\bibinfo {title} {Scaling whole-chip qaoa for higher-order ising spin glass models on heavy-hex graphs},\ }\href {http://arxiv.org/abs/2312.00997} {\bibfield  {journal} {\bibinfo  {journal} {arXiv}\ } (\bibinfo {year} {2023})},\ \Eprint {https://arxiv.org/abs/2312.00997} {arXiv:2312.00997 [quant-ph]} \BibitemShut {NoStop}%
\bibitem [{\citenamefont {Chowdhury}\ \emph {et~al.}(2024)\citenamefont {Chowdhury}, \citenamefont {Yu}, \citenamefont {Shamim}, \citenamefont {Kabir},\ and\ \citenamefont {Sufian}}]{chowdhury2024enhancing}%
  \BibitemOpen
  \bibfield  {author} {\bibinfo {author} {\bibfnamefont {T.~A.}\ \bibnamefont {Chowdhury}}, \bibinfo {author} {\bibfnamefont {K.}~\bibnamefont {Yu}}, \bibinfo {author} {\bibfnamefont {M.~A.}\ \bibnamefont {Shamim}}, \bibinfo {author} {\bibfnamefont {M.~L.}\ \bibnamefont {Kabir}},\ and\ \bibinfo {author} {\bibfnamefont {R.~S.}\ \bibnamefont {Sufian}},\ }\bibfield  {title} {\bibinfo {title} {Enhancing quantum utility: simulating large-scale quantum spin chains on superconducting quantum computers},\ }\href {http://arxiv.org/abs/2312.12427} {\bibfield  {journal} {\bibinfo  {journal} {arXiv}\ } (\bibinfo {year} {2024})},\ \Eprint {https://arxiv.org/abs/2312.12427} {arXiv:2312.12427 [quant-ph]} \BibitemShut {NoStop}%
\bibitem [{\citenamefont {Rehfeldt}\ \emph {et~al.}(2023)\citenamefont {Rehfeldt}, \citenamefont {Koch},\ and\ \citenamefont {Shinano}}]{Rehfeldt2023}%
  \BibitemOpen
  \bibfield  {author} {\bibinfo {author} {\bibfnamefont {D.}~\bibnamefont {Rehfeldt}}, \bibinfo {author} {\bibfnamefont {T.}~\bibnamefont {Koch}},\ and\ \bibinfo {author} {\bibfnamefont {Y.}~\bibnamefont {Shinano}},\ }\bibfield  {title} {\bibinfo {title} {Faster exact solution of sparse maxcut and qubo problems},\ }\href {https://doi.org/10.1007/s12532-023-00236-6} {\bibfield  {journal} {\bibinfo  {journal} {Mathematical Programming Computation}\ }\textbf {\bibinfo {volume} {15}},\ \bibinfo {pages} {445} (\bibinfo {year} {2023})}\BibitemShut {NoStop}%
\bibitem [{\citenamefont {Egger}\ \emph {et~al.}(2021)\citenamefont {Egger}, \citenamefont {Mare{\v{c}}ek},\ and\ \citenamefont {Woerner}}]{Egger2021warmstart}%
  \BibitemOpen
  \bibfield  {author} {\bibinfo {author} {\bibfnamefont {D.~J.}\ \bibnamefont {Egger}}, \bibinfo {author} {\bibfnamefont {J.}~\bibnamefont {Mare{\v{c}}ek}},\ and\ \bibinfo {author} {\bibfnamefont {S.}~\bibnamefont {Woerner}},\ }\bibfield  {title} {\bibinfo {title} {Warm-starting quantum optimization},\ }\href {https://doi.org/10.22331/q-2021-06-17-479} {\bibfield  {journal} {\bibinfo  {journal} {{Quantum}}\ }\textbf {\bibinfo {volume} {5}},\ \bibinfo {pages} {479} (\bibinfo {year} {2021})}\BibitemShut {NoStop}%
\bibitem [{\citenamefont {Tate}\ \emph {et~al.}(2023)\citenamefont {Tate}, \citenamefont {Farhadi}, \citenamefont {Herold}, \citenamefont {Mohler},\ and\ \citenamefont {Gupta}}]{Tate2023warmstart}%
  \BibitemOpen
  \bibfield  {author} {\bibinfo {author} {\bibfnamefont {R.}~\bibnamefont {Tate}}, \bibinfo {author} {\bibfnamefont {M.}~\bibnamefont {Farhadi}}, \bibinfo {author} {\bibfnamefont {C.}~\bibnamefont {Herold}}, \bibinfo {author} {\bibfnamefont {G.}~\bibnamefont {Mohler}},\ and\ \bibinfo {author} {\bibfnamefont {S.}~\bibnamefont {Gupta}},\ }\bibfield  {title} {\bibinfo {title} {{Bridging Classical and Quantum with SDP initialized warm-starts for QAOA}},\ }\href {https://doi.org/10.1145/3549554} {\bibfield  {journal} {\bibinfo  {journal} {ACM Transactions on Quantum Computing}\ }\textbf {\bibinfo {volume} {4}} (\bibinfo {year} {2023})}\BibitemShut {NoStop}%
\bibitem [{\citenamefont {Chandarana}\ \emph {et~al.}(2022)\citenamefont {Chandarana}, \citenamefont {Hegade}, \citenamefont {Paul}, \citenamefont {Albarr\'an-Arriagada}, \citenamefont {Solano}, \citenamefont {del Campo},\ and\ \citenamefont {Chen}}]{Chandarana2022Counterdiabatic}%
  \BibitemOpen
  \bibfield  {author} {\bibinfo {author} {\bibfnamefont {P.}~\bibnamefont {Chandarana}}, \bibinfo {author} {\bibfnamefont {N.~N.}\ \bibnamefont {Hegade}}, \bibinfo {author} {\bibfnamefont {K.}~\bibnamefont {Paul}}, \bibinfo {author} {\bibfnamefont {F.}~\bibnamefont {Albarr\'an-Arriagada}}, \bibinfo {author} {\bibfnamefont {E.}~\bibnamefont {Solano}}, \bibinfo {author} {\bibfnamefont {A.}~\bibnamefont {del Campo}},\ and\ \bibinfo {author} {\bibfnamefont {X.}~\bibnamefont {Chen}},\ }\bibfield  {title} {\bibinfo {title} {Digitized-counterdiabatic quantum approximate optimization algorithm},\ }\href {https://doi.org/10.1103/PhysRevResearch.4.013141} {\bibfield  {journal} {\bibinfo  {journal} {Phys. Rev. Res.}\ }\textbf {\bibinfo {volume} {4}},\ \bibinfo {pages} {013141} (\bibinfo {year} {2022})}\BibitemShut {NoStop}%
\bibitem [{\citenamefont {Ezzell}\ \emph {et~al.}(2023)\citenamefont {Ezzell}, \citenamefont {Pokharel}, \citenamefont {Tewala}, \citenamefont {Quiroz},\ and\ \citenamefont {Lidar}}]{Ezzell2023dd}%
  \BibitemOpen
  \bibfield  {author} {\bibinfo {author} {\bibfnamefont {N.}~\bibnamefont {Ezzell}}, \bibinfo {author} {\bibfnamefont {B.}~\bibnamefont {Pokharel}}, \bibinfo {author} {\bibfnamefont {L.}~\bibnamefont {Tewala}}, \bibinfo {author} {\bibfnamefont {G.}~\bibnamefont {Quiroz}},\ and\ \bibinfo {author} {\bibfnamefont {D.~A.}\ \bibnamefont {Lidar}},\ }\bibfield  {title} {\bibinfo {title} {Dynamical decoupling for superconducting qubits: A performance survey},\ }\href {https://doi.org/10.1103/PhysRevApplied.20.064027} {\bibfield  {journal} {\bibinfo  {journal} {Phys. Rev. Appl.}\ }\textbf {\bibinfo {volume} {20}},\ \bibinfo {pages} {064027} (\bibinfo {year} {2023})}\BibitemShut {NoStop}%
\bibitem [{\citenamefont {Cincio}\ \emph {et~al.}(2007)\citenamefont {Cincio}, \citenamefont {Dziarmaga}, \citenamefont {Rams},\ and\ \citenamefont {Zurek}}]{Cincio2007entropy}%
  \BibitemOpen
  \bibfield  {author} {\bibinfo {author} {\bibfnamefont {L.}~\bibnamefont {Cincio}}, \bibinfo {author} {\bibfnamefont {J.}~\bibnamefont {Dziarmaga}}, \bibinfo {author} {\bibfnamefont {M.~M.}\ \bibnamefont {Rams}},\ and\ \bibinfo {author} {\bibfnamefont {W.~H.}\ \bibnamefont {Zurek}},\ }\bibfield  {title} {\bibinfo {title} {Entropy of entanglement and correlations induced by a quench: Dynamics of a quantum phase transition in the quantum ising model},\ }\href {https://doi.org/10.1103/PhysRevA.75.052321} {\bibfield  {journal} {\bibinfo  {journal} {Phys. Rev. A}\ }\textbf {\bibinfo {volume} {75}},\ \bibinfo {pages} {052321} (\bibinfo {year} {2007})}\BibitemShut {NoStop}%
\bibitem [{\citenamefont {Roychowdhury}\ \emph {et~al.}(2021)\citenamefont {Roychowdhury}, \citenamefont {Moessner},\ and\ \citenamefont {Das}}]{Roychowdhury2021}%
  \BibitemOpen
  \bibfield  {author} {\bibinfo {author} {\bibfnamefont {K.}~\bibnamefont {Roychowdhury}}, \bibinfo {author} {\bibfnamefont {R.}~\bibnamefont {Moessner}},\ and\ \bibinfo {author} {\bibfnamefont {A.}~\bibnamefont {Das}},\ }\bibfield  {title} {\bibinfo {title} {Dynamics and correlations at a quantum phase transition beyond kibble-zurek},\ }\href {https://doi.org/10.1103/PhysRevB.104.014406} {\bibfield  {journal} {\bibinfo  {journal} {Phys. Rev. B}\ }\textbf {\bibinfo {volume} {104}},\ \bibinfo {pages} {014406} (\bibinfo {year} {2021})}\BibitemShut {NoStop}%
\end{thebibliography}%

\end{document}